%% file: sz_likelihood_paper.tex
\documentclass[traditabstract]{aa} 
\usepackage{epsfig}
\usepackage{graphicx}
\usepackage{epstopdf}
\usepackage{color}  
\usepackage{array}   
\usepackage{units}  
\usepackage{natbib}  
\usepackage{multirow}   
\usepackage{amsmath,amssymb}  
\usepackage[english]{babel}
\usepackage[latin1]{inputenc}
\usepackage[T1]{fontenc}
\usepackage{longtable,lscape}
\usepackage{footnote}

\usepackage{txfonts}
\usepackage[toc,page]{appendix} 
\usepackage{color}

\input Planck.tex

\def\aap{A\&A}%
\def\reff@jnl#1{{#1\/}}
\def\apj{\reff@jnl{ApJ}}       
\def\apjs{\reff@jnl{ApJS}}     
\def\aaps{\reff@jnl{A\&AS}}    
\def\mnras{\reff@jnl{MNRAS}}   
\def\prd{\reff@jnl{Phys.\ Rev.\ D}}    

\newcommand{\beq}{\begin{equation}}
\newcommand{\eeq}{\end{equation}}

\newcommand{\be}{\begin{equation}}
\newcommand{\ee}{\end{equation}}
\newcommand{\bea}{\begin{eq}}
\newcommand{\eea}{\end{equation}}

{\begin{enumerate}\setlength{\itemsep}{0mm}}%
{\end{enumerate}}
{\begin{enumerate}\setlength{\itemsep}{0mm}}%
{\end{enumerate}}

\newcommand{\bc}{\begin{center}}
\newcommand{\ec}{\end{center}}
\newcommand{\bi}{\begin{itemize}}
\newcommand{\ei}{\end{itemize}}
\newcommand{\ben}{\begin{enumerate}}
\newcommand{\een}{\end{enumerate}}

\newfont{\gwpfont}{cmssq8 scaled 1000}

\def\xmm{{\it XMM-Newton}}

\def\lesssim{\mathrel{\hbox{\rlap{\hbox{\lower4pt\hbox{$\sim$}}}\hbox{$<$}}}}
\def\gtrsim{\mathrel{\hbox{\rlap{\hbox{\lower4pt\hbox{$\sim$}}}\hbox{$>$}}}}

\newcommand{\propsim}{\lower 3pt \hbox{$\, \buildrel {\textstyle
     \propto}\over {\textstyle \sim}\,$}}

\def\xmm{{\it XMM-Newton}}

\def\rosat{{\it ROSAT}}

\def\lesssim{\mathrel{\hbox{\rlap{\hbox{\lower4pt\hbox{$\sim$}}}\hbox{$<$}}}}
\def\gtrsim{\mathrel{\hbox{\rlap{\hbox{\lower4pt\hbox{$\sim$}}}\hbox{$>$}}}}

\begin{document}

\title{{\it The Good, the Bad and the Ugly}:\\ Statistical quality assessment
  of SZ detections} \author{N. Aghanim\inst{1}, G. Hurier\inst{1}, J.-M. Diego\inst{2},  M. Douspis\inst{1}, J. Macias-Perez\inst{3}, E. Pointecouteau\inst{4}, B. Comis\inst{3}, M. Arnaud\inst{5},\\ L. Montier\inst{4}}

\institute{Institut d'Astrophysique Spatiale, CNRS (UMR8617) Universit\'{e} Paris-Sud 11, B\^{a}timent 121, Orsay, France
\and 
Instituto de F\'{\i}sica de Cantabria (CSIC-Universidad de Cantabria), Avda. de los Castros s/n, Santander, Spain
\and
Laboratoire de Physique Subatomique et de Cosmologie, Universit\'{e} Joseph Fourier Grenoble I, CNRS/IN2P3, Institut National Polytechnique de Grenoble, 53 rue des Martyrs, 38026 Grenoble cedex, France
\and
CNRS, IRAP, 9 Av. colonel Roche, BP 44346, F-31028 Toulouse cedex 4, France
\and 
Laboratoire AIM, IRFU/Service d'Astrophysique - CEA/DSM - CNRS - Universit\'{e} Paris Diderot, B\^{a}t. 709, CEA-Saclay, F-91191 Gif-sur-Yvette Cedex, France
}

\abstract {We examine three approaches to the problem of source
  classification in catalogues. Our goal is to determine the
  confidence with which the elements in these catalogues can be
  distinguished in populations on the basis of their spectral energy
  distribution (SED).  Our analysis is based on the projection of the
  measurements onto a comprehensive SED model of the main signals in
  the considered range of frequencies. We first first consider
  likelihood analysis, which half way between supervised and
  unsupervised methods. Next, we investigate an unsupervised
  clustering technique. Finally, we consider a supervised classifier
  based on Artificial Neural Networks. We illustrate the approach and
  results using catalogues from various surveys. i.e., X-Rays (MCXC),
  optical (SDSS) and millimetric (\Planck\ Sunyaev-Zeldovich (SZ)). We
  show that the results from the statistical classifications of the
  three methods are in very good agreement with each others, although
  the supervised neural network-based classification shows better
  performances allowing the best separation into populations of
  reliable and unreliable sources in catalogues. The latest method was
  applied to the SZ sources detected by the \Planck\ satellite. It led
  to a classification assessing and thereby agreeing with the
  reliability assessment published in the \Planck\ SZ catalogue. Our
  method could easily be applied to catalogues from future large
  survey such as SRG/eROSITA and {\it Euclid}.

}


\authorrunning{N. Aghanim, G. Hurier, et al.}
\titlerunning{Statistical quality assessment of SZ detections}

\maketitle


\section{Introduction}

Astronomy and cosmology are witnessing a transition from specific point
observations to larger and larger astronomical surveys covering large
fractions of the sky, as large as the whole sky in some cases. In this
context, the need for reliable classification tools are important to
assess the quality and the confidence in detected sources. For
example, this is a crucial point for future surveys like SRG/eROSITA
\citep[see e.g.,][]{mer12} or {\it
  Euclid}\footnote{{http://www.euclid-ec.org/}} that expect to
detect or order of $6\times10^4$ to $9\times10^4$ clusters of
galaxies.  In these experiments, a purity of 90 to 80\% of the
catalogues of clusters would translate into a few thousands of false
detections. These large numbers may pose serious issues for the
cosmological interpretation of the number counts. They will also put a
heavy load on the ground-based telescopes since the follow up
observations will need to mitigate the large number of such false
sources. In such a context, an assessment of the quality factor for
the detections or even better a classification of the detected
clusters in terms of their reliability will be a key information.

In the present article, we address the topic of multivariate tools for
sample classification applying machine learning techniques that are commonly
used various scientific domains such as sociology, genetic classification, cosmology, spectroscopy,
etc. Two distinct approaches can be used: 
Supervised and unsupervised learning. 
The difference relies on the utilization of hypothesis for supervised learning.
Unsupervised learning can be used when no a-priori on the potential classes are known. \\

The traditional method for detecting structure within a population is
some form of exploratory technique such as Principal Components
Analysis (PCA). Such methods do not use prior information on the
classification of the candidate populations. Another unsupervised method
commonly used is the clustering technique \citep[see
  e.g.,][]{har75,har79}. It consists of the search for the nearest
neighbors in a canonic space and thus permits to classify automatically
unknown populations in relation with a reference population.  Such a
method was used since the eighties in different domains ranging from
apiculture \citep[e.g.,][]{tom71,cor75} to planetary science
\citep[e.g.,][]{for13}. Clustering and in particular Voronoi
tessellation is also used in astronomy to model and reproduce the
cosmic web \citep[e.g.,][]{she04}. Of the second class of
classification methods, i.e. supervised methods, the most commonly
used is the Artificial Neural Networks (ANN) \citep[][and references therein]{swa14}. 
ANN are algorithms that mimic the learning abilities of brains; they have been successfully 
used in the analysis of dataset from many scientific domains \citep{reb97,bri11}.\\

In our study, We illustrate the use and effects of the statistical
classification techniques on the recently published catalogue of
Sunyaev-Zeldovich (SZ) sources \citep{planck2013-p29} which contains
both confirmed galaxy clusters and candidate clusters.  We consider a
likelihood analysis, half way between supervised and unsupervised
methods. Next, we investigate a clustering technique. Finally, we
consider an Artificial Neural Networks. Our aim is to assess whether
an ensemble of sources detected through the SZ signal can be
distinguishable on the basis of their spectral energy distribution.

The article is organised as follows: We describe the data in
Sect. \ref{data}, we then present both the SED model and the
associated fitted parameters in Sects. \ref{SED} and \ref{d-sed}. We
describe the different classification methods used in the study in
Sect. \ref{QF} and discuss the results in Sect. \ref{disc}. We
summarise our findings and conclusions in Sect. \ref{ccl}.


\section{Data}\label{data}

For our study, we use catalogues and samples of sources including
clusters of galaxies detected in the X-rays and in the optical and in
SZ. We also use catalogues of radio and IR point sources as well as
galactic cold sources. We use the \Planck\ frequency maps. We finally
construct a test set on 2000 random positions over the sky. \\

Namely, we use the Meta-Catalogue of X-ray detected
Clusters of galaxies \citep[MCXC,][and reference therein]{pif11}. It
is a compilation constructed from the publicly available \rosat\ All
Sky Survey-based and serendipitous cluster catalogues, as well as the
{\it Einstein} Medium Sensitivity Survey.  It includes only clusters
with available redshift information in the original catalogues which
yields an ensemble of 1789 clusters.

Furthermore, we use a catalogue of clusters extracted from the Sloan
Digital Sky Survey (SDSS, \citealt{yor00}) data, the WHL12 catalogue
\citep[132,\,684 objects,][]{wen12}. It provides an estimated
richness. We apply a cut in richness, $N_{200}$, of 50 to exclude low mass
systems and groups that have no significant SZ signal.

Finally, we use the \Planck\ SZ source catalogue \citep[PSZ1
  hereafter, see ][]{planck2013-p29}. It consists of 1227 sources
detected through their SZ effect in the \Planck\ frequency maps. The
catalogue contains a large fraction of galaxy clusters but it also
contains un-confirmed cluster candidates including low reliability
ones. 

We also use catalogues of sources detected in the radio, at 30~GHz,
and in the infra-red (IR), at 353~GHz, both are extracted from the
\Planck\ Catalogue of Compact Sources (PCCS)
\citep{planck2013-p05}. We use a catalogue of cold Galactic (CG)
sources \citep[see ][]{planck2011-7.7b,planck2011-7.7a} detected in
the \Planck\ channel maps following \citep{mon10}. We construct a
catalogue of false SZ detections which consists of the major sources
of contamination identified in \citet{planck2011_d} and
\citet{planck2013-p29}. Namely, we take 100 sources, outside the mask
used for the \Planck\ SZ detection, from each the the GS sources
catalogue, the IR sources at 353~GHz and radio sources at 30~GHz. The
obtained sample of false SZ detection is representative of unreliable
SZ sources.

In order to compute the spectral energy distribution (SED) of the
considered sources, we use the \Planck\ channel maps from 70 to
857~GHz.  Each map is set to a resolution of 13 arcmin, i.e. the
lowest resolution associated with 70~GHz channel. This allows us to
access to the emission in the radio domain (belo 100~GHz) without
decreasing the resolution too much.


\section{SED fitting} \label{SED}

In the context of multivariate classification, it is impossible from
the statistical point of view to model the interplay of all the
physical variables and observational parameters. We therefore need to
resort to some dimensionality reduction approaches prior any
classification. There are standard dimensionality reduction techniques
like PCA or Independent Component Analysis (ICA) \citep{swa14} that do not
include any pre-knowledge of the physical variables/processes. In the
following, we rather choose to reduce the dimensionality by decomposing
the signal in the form of a spectral energy distribution (SED).  

It is beyond the scope of our study to model the SED taking into
account all the contributions to the signal. We rather focus on the
astrophysical emissions that affect the most the SZ detection in
multi-frequency experiments. This was discussed in both
\citep{planck11d} and \citep{planck2013-p29}. From 70 to 857~GHz,
several astrophysical sources contribute to the measured signal:
Diffuse galactic free-free, synchrotron, and thermal dust
\citep{planck,planck2011-7.3} emissions; anomalous microwave emission
\citep[AME,][]{planck2011-7.2}; molecular Galactic emissions
\citep[mainly $^{12}$CO in the 100, 217, and 353~GHz
  bands,][]{planckco}; emission from Galactic and extragalactic point
sources \citep[radio and infrared
  sources,][]{planck2012-VII,planck2011-1.10}; CIB
\citep[][]{planck11b}; zodiacal light emission
\citep{2013arXiv1303.5074P}; and thermal Sunyaev-Zeldovich effect
\citep{SZ} in clusters of galaxies.

Therefore, we model the SED taking into account five components: the
tSZ effect neglecting relativistic corrections, the CMB signal and the
CO emission. We also add an effective IR component representing the
contamination by dust emission, CG sources and CIB fluctuations; and
an effective radio component accounting for diffuse radio and
synchrotron emission and radio sources.

The flux in each channel, i.e. frequency, is then written as:
\begin{align}
F_\nu =& A_{\rm SZ} F_{\rm SZ}(\nu) + A_{\rm CMB} F_{\rm CMB}(\nu) +
A_{\rm IR} F_{\rm IR}(\nu) \nonumber \\
&+ A_{\rm RAD} F_{\rm RAD}(\nu) + A_{\rm CO}
F_{\rm CO}(\nu) + N(\nu),
\label{eq:model}
\end{align}
where, $F_{\rm SZ}(\nu)$, $F_{\rm CMB}(\nu)$, $F_{\rm IR}(\nu)$,
$F_{\rm RAD}(\nu)$, and $F_{\rm CO}(\nu)$ are the spectra of SZ, CMB,
IR, radio, and CO emissions, $A_{\rm SZ}$, $A_{\rm CMB}$, $A_{\rm
  IR}$, $A_{\rm RAD}$, and $A_{\rm CO}$ the corresponding
amplitudes. $N(\nu)$ is the instrumental noise.\\ 

For $F_{\rm IR}(\nu)$, we consider a modified black-body spectrum with
temperature $T_d = 17$~K and index $\beta_d = 1.6$. This assumption is
representative of the dust properties at high galactic latitudes. The
contribution from CIB fluctuations affects the flux measurement but is
not a major contamination from the point of view of the detection,
i.e. spurious sources. For $F_{\rm RAD}(\nu)$, we consider a power law
emission, $\nu^{\alpha_{\rm r}}$, with index $\alpha_{\rm r} = -0.7$
in intensity units representative of the average property of the radio
emission.

We compute the flux, at the position of each source of the catalogues
described in Sect. \ref{data}, with aperture photometry. We set the
aperture to 10~arcmin; the background level is estimated in an annulus
between 20 to 50~arcmin.  We have checked that varying the size of the
aperture, from 5 to 15 arcmin, does not affect the results. Larger
apertures obviously capture more contamination from the background.

Each derived spectrum, $F_\nu$, is fitted assuming the model in
Eq. \ref{eq:model}, where we fit for $A_{\rm SZ}$, $A_{\rm CMB}$,
$A_{\rm IR}$, $A_{\rm RAD}$, and $A_{\rm CO}$ through a linear fit of
the form,
\begin{equation}
\mathbf{A} = ({\cal F}^T {\cal C}^{-1}_{N} {\cal F})^{-1} {\cal F}^T
       {\cal C}^{-1}_{N} F_\nu,
\end{equation}
with the mixing matrix ${\cal F}^T$, the instrumental noise covariance
matrix ${\cal C}_{N}$, and $\mathbf{A}$ a vector containing the fitted
parameters. In this approach, ${\cal C}_{N}$ only accounts for the
instrumental noise and we implicitly assume that the five components
considered in the model reproduce the astrophysical signal in the
data.

\begin{figure}[!th]
\center
\includegraphics[width=9cm]{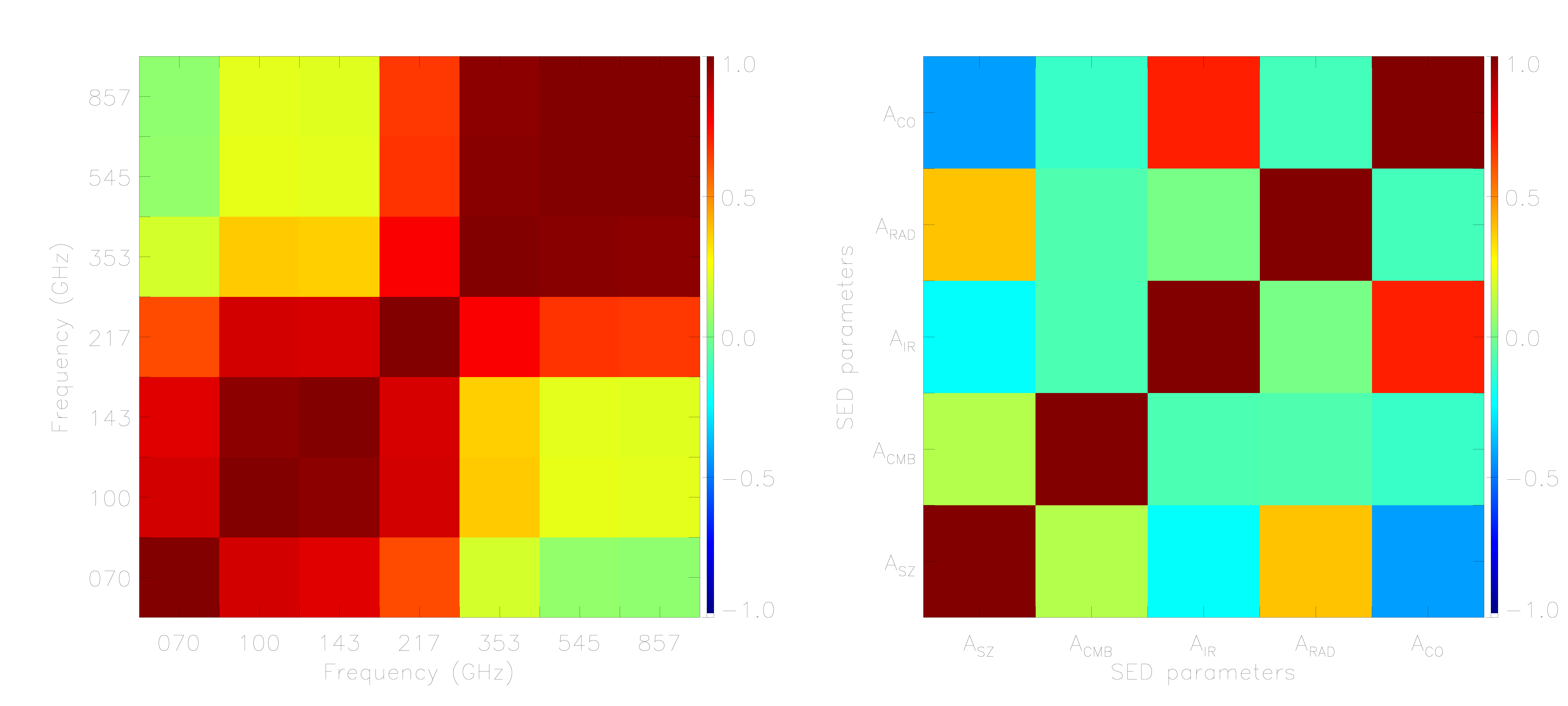}
\caption{Left panel : correlation matrix of the measured fluxes from
  30 to 857~GHz estimated on 2000 random positions over the sky. Right
  panel : correlation matrix of fitted SED parameters from the same
  positions.}
\label{cormat}
\end{figure}

This efficiency of the dimensionality reduction is illustrated in
Fig.~\ref{cormat} (right panel), where we show the SED fitted
parameters correlation matrix, compared to the correlation matrix of
measured fluxes from 30 to 857 GHz (left panel).  We observe that we
have a high degree of correlation between frequencies, especially at
low frequency due to the CMB component (< 217~GHz), and at higher
frequency (> 217 GHz) due to the thermal due component. By contrast,
in the SED parameter space, we observed that the correlation matrix is
almost diagonal, except a spatial correlation between thermal dust and
CO emission.\\
  
\section{Distribution of the fitted SED parameters}\label{d-sed}

We start by fitting the amplitudes of the different components in the
SED, namely $A_{\rm SZ}$, $A_{\rm CMB}$, $A_{\rm IR}$, $A_{\rm RAD}$,
and $A_{\rm CO}$ at the positions of each source in the catalogue and
samples described in Sect. \ref{data}. We examine the distribution of
the fitted SED parameters and display, for each catalogue and sample,
both the distributions and the correlation between fitted parameters.
We also perform the same fitting at 2000 random positions in the
sky.

\begin{figure}[!th]
\center
\includegraphics[width=8cm]{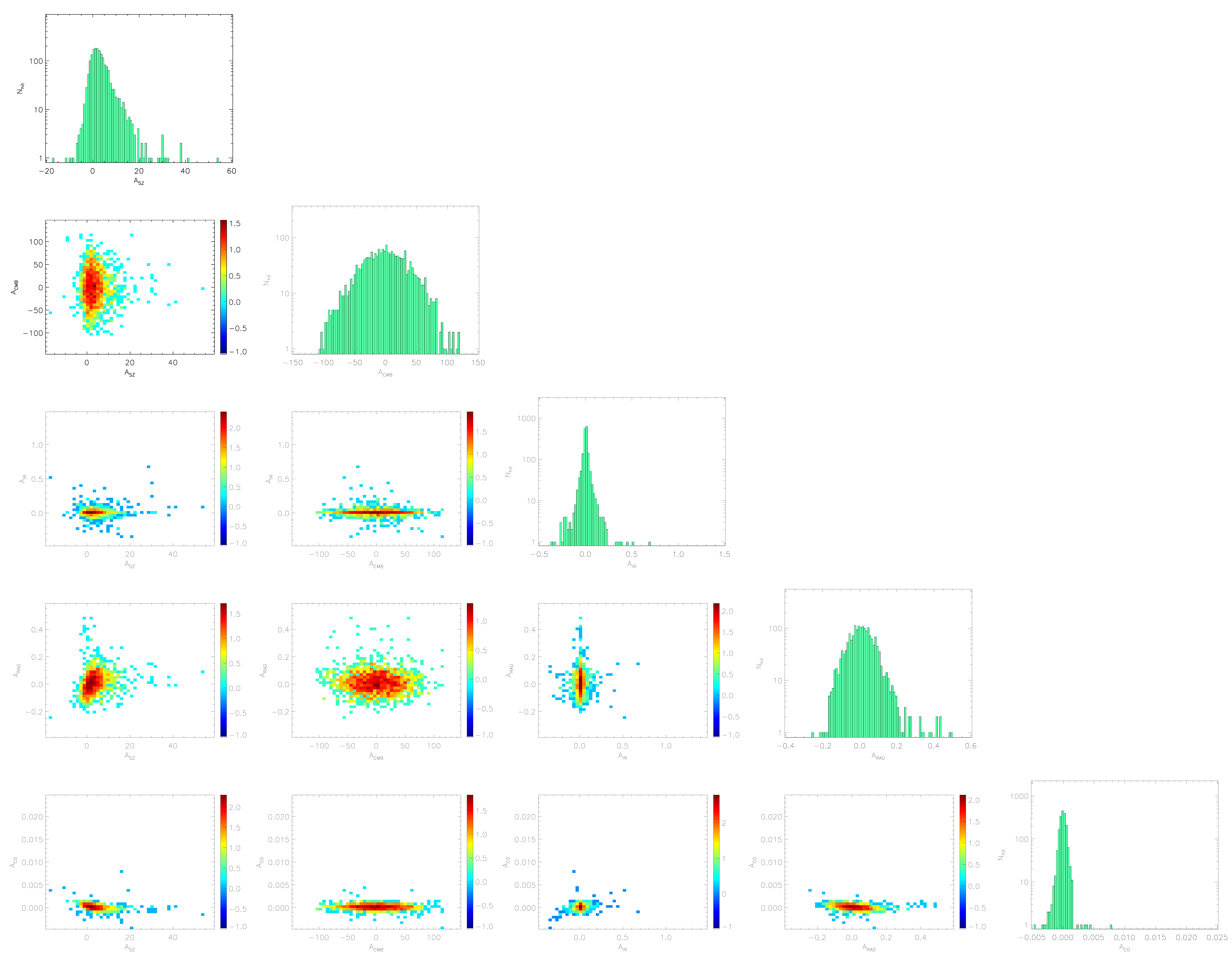}
\caption{Diagonal: The distribution of, from top to bottom, $A_{\rm
    SZ}$, $A_{\rm CMB}$, $A_{\rm IR}$, $A_{\rm RAD}$, and $A_{\rm CO}$
  at the position of MCXC galaxy clusters from \citet{pif11}. Off
  diagonal: The 2-D histograms that present the correlation between
  parameters.}
\label{mcxc}
\end{figure}

\subsection{X-ray clusters from MCXC}
In Fig.~\ref{mcxc} we present the derived distribution for each
amplitude fitted at the positions of MCXC galaxy clusters. Besides
some negative values, due to statistical noise, we observe and
asymmetric distribution with positive values for $A_{\rm SZ}$, as
expected for galaxy clusters. $A_{\rm CMB}$ presents a Gaussian
distribution, with a dispersion given by the amplitude of primordial
CMB fluctuations. $A_{\rm IR}$ presents a Cauchy distribution centered
on zero. $A_{\rm RAD}$ has a Gaussian distribution, except for a few
outliers associated with contamination clusters from radio-loud Active
Galactic Nuclei (AGN) (e.g., Perseus, Virgo). $A_{\rm CO}$ presents a
Gaussian distribution centered on zero. \\

We observe a positive correlation between $A_{\rm SZ}$ and $A_{\rm
  RAD}$. Indeed, radio contamination mimics a tSZ effect at
frequencies below 217~GHz, and thus an apparent increase on the tSZ
flux can be compensated by an increase of radio emission amplitude,
leading to the observed degeneracy. $A_{\rm RAD}$ and $A_{\rm CO}$ are
anti-correlated as both components induce an excess of emission at
100~GHz.  We notice that $A_{\rm CMB}$ and $A_{\rm IR}$ are not
correlated with the other fitted parameters.

\subsection{Optical clusters from SDSS}

\begin{figure}[!th]
\center
\includegraphics[width=8cm]{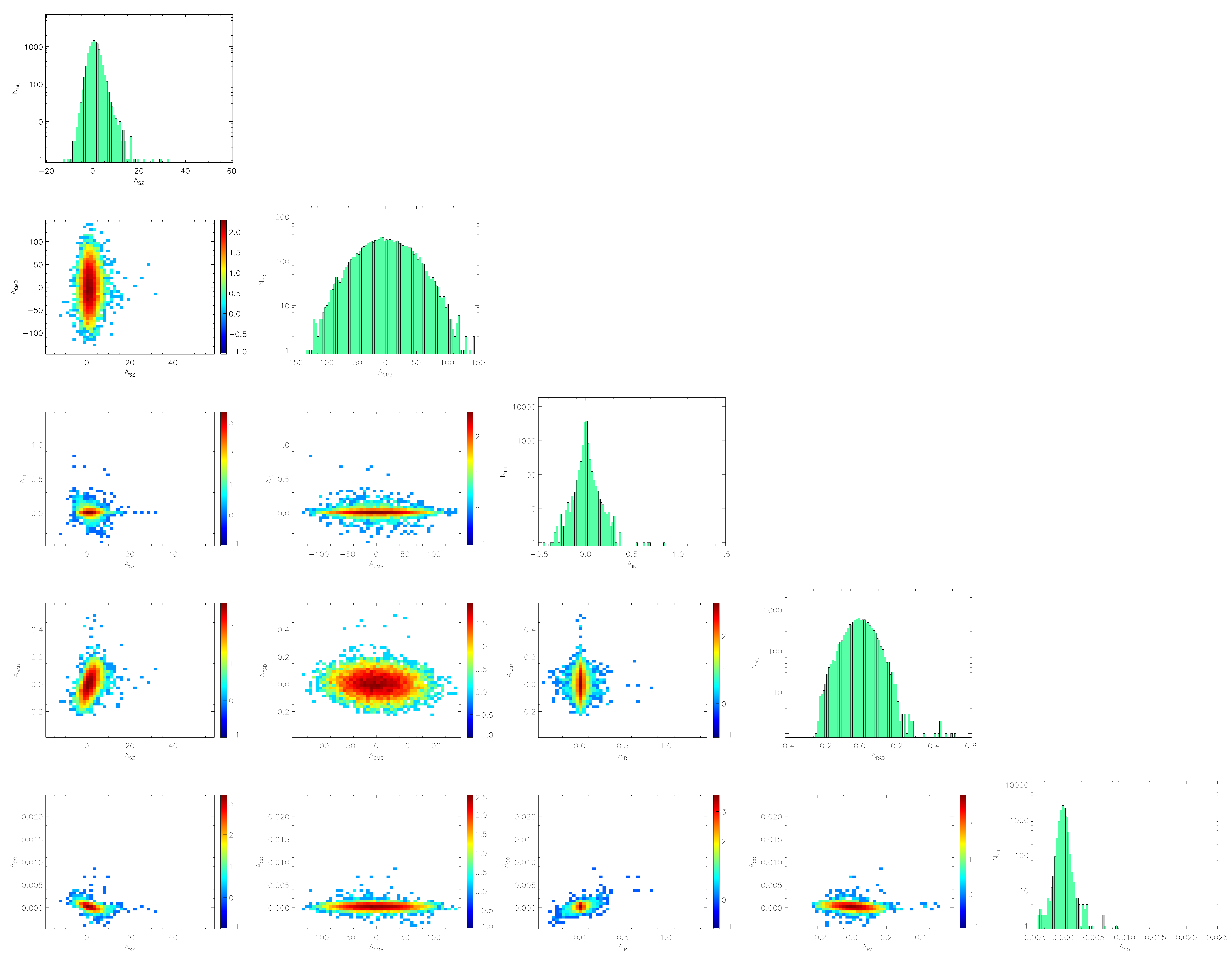}
\caption{Same as  Fig. \ref{mcxc} for the SDSS galaxy clusters
  with richness above 50 from WHL catalogue \citep{wen12}.}
\label{sdss}
\end{figure}

We now fit the amplitudes of the SED components at the position of
optical clusters selected from the \citet[hereafter WHL][]{wen12}
catalogue. In Fig.~\ref{sdss}, we present the derived values for
$A_{\rm SZ}$, $A_{\rm CMB}$, $A_{\rm IR}$, $A_{\rm RAD}$, and $A_{\rm
  CO}$. The distributions of amplitudes are similar to the those of
the MCXC clusters. They show the asymmetric distribution for $A_{\rm
  SZ}$ and symmetric Gaussian or Cauchy distributions for the
distributions of the other amplitudes.

\subsection{SZ clusters from PSZ1}

The distributions of amplitudes of the fitted SED in the direction of
861 confirmed galaxy clusters from the PSZ1 catalogue share the same
characteristics as the two other cluster catalogues detected the in
X-rays or the optical.

\begin{figure}[!th]
\center
\includegraphics[width=8cm]{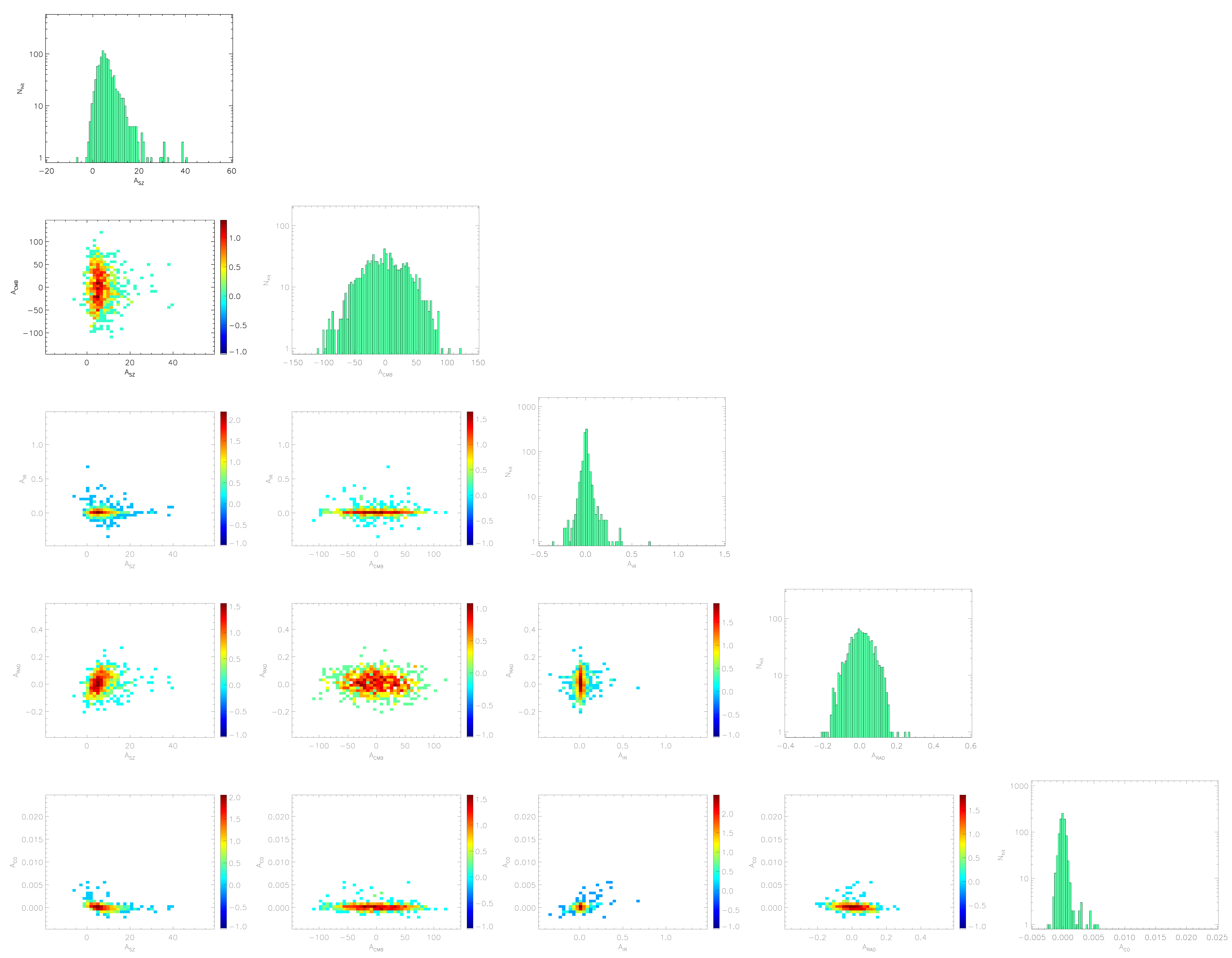}
\caption{same as  Fig. \ref{mcxc} for the 861 confirmed SZ galaxy clusters
  from the PSZ1 catalogue \citep{planck2013-p29}.}
\label{psz1}
\end{figure}

\subsection{Radio, IR and CG sources}
We now examine the cases of sources emitting in the radio and in the
IR that are not galaxy clusters. We focus on three cases which
represent cases of spurious detections that affect the cluster
extraction as described in \citet{planck11d} and in
\citet{planck2013-p29}. Namely, we fit for the SED in the direction of
IR and radio sources from the PCCS catalogue, taken at 353 and 30~GHz
respectively and we also consider GCS from \Planck. All of the sources
are taken outside a galactic mask leaving 85\% of the sky.

\begin{figure}[!th]
\center
\includegraphics[width=8cm]{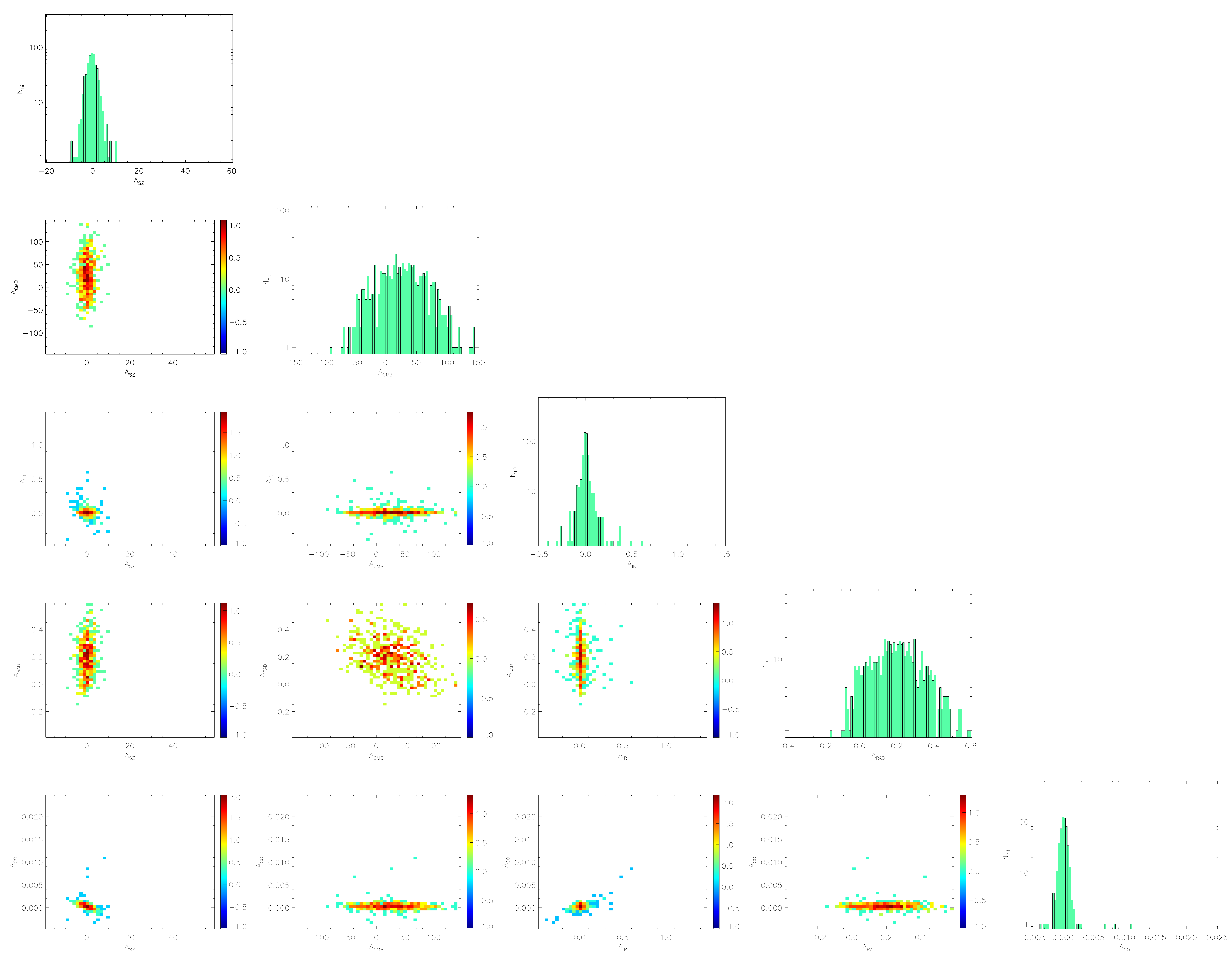}
\caption{Same as  Fig. \ref{mcxc} for the sources detected in the 30~GHz
  channel of {\it Planck}.}
\label{30GHz}
\end{figure}

\begin{figure}[!th]
\center
\includegraphics[width=8cm]{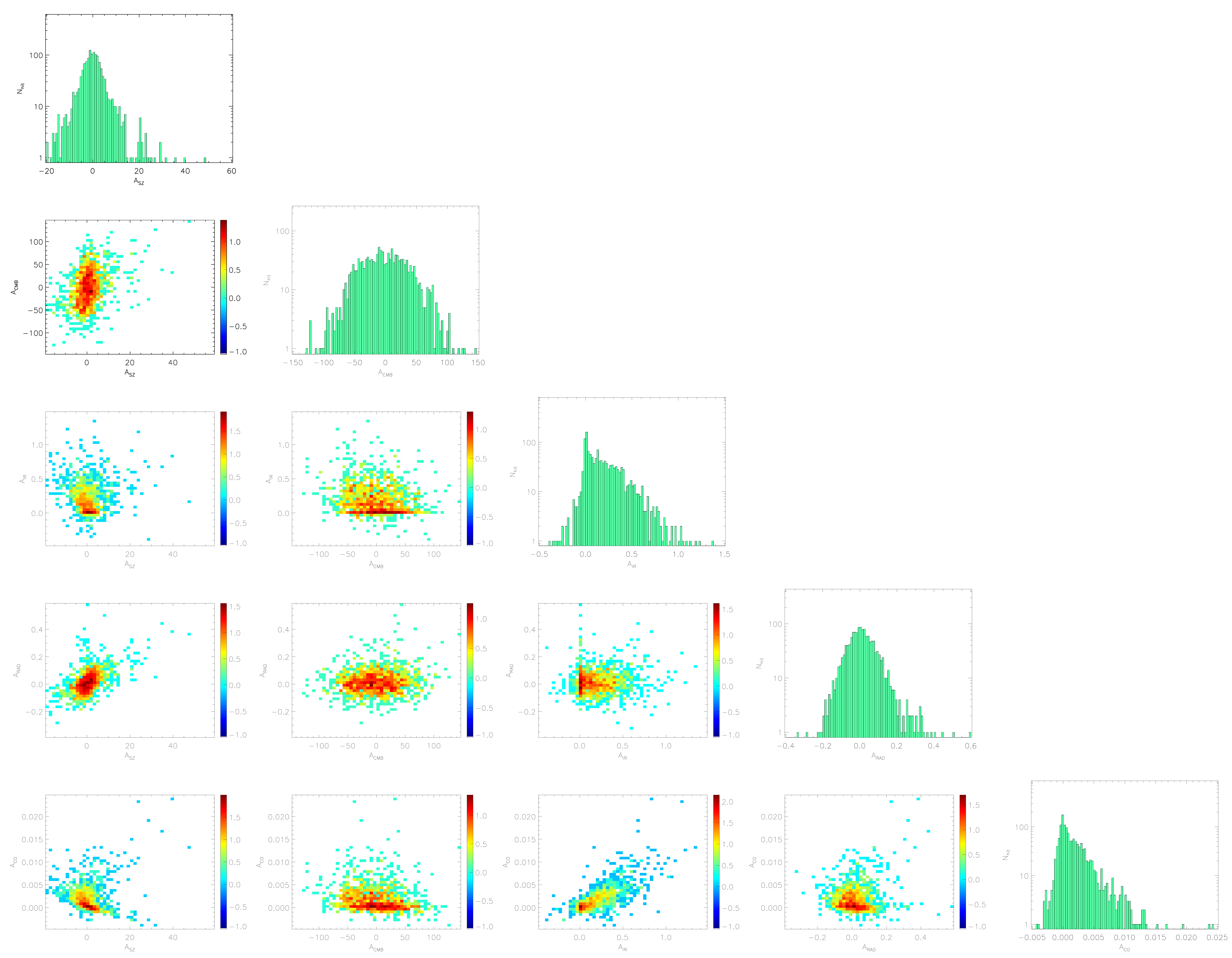}
\caption{Same as  Fig. \ref{mcxc} for sources detected in the 353~GHz
  channel of {\it Planck}. }
\label{857GHz}
\end{figure}

\begin{figure}[!th]
\center
\includegraphics[width=8cm]{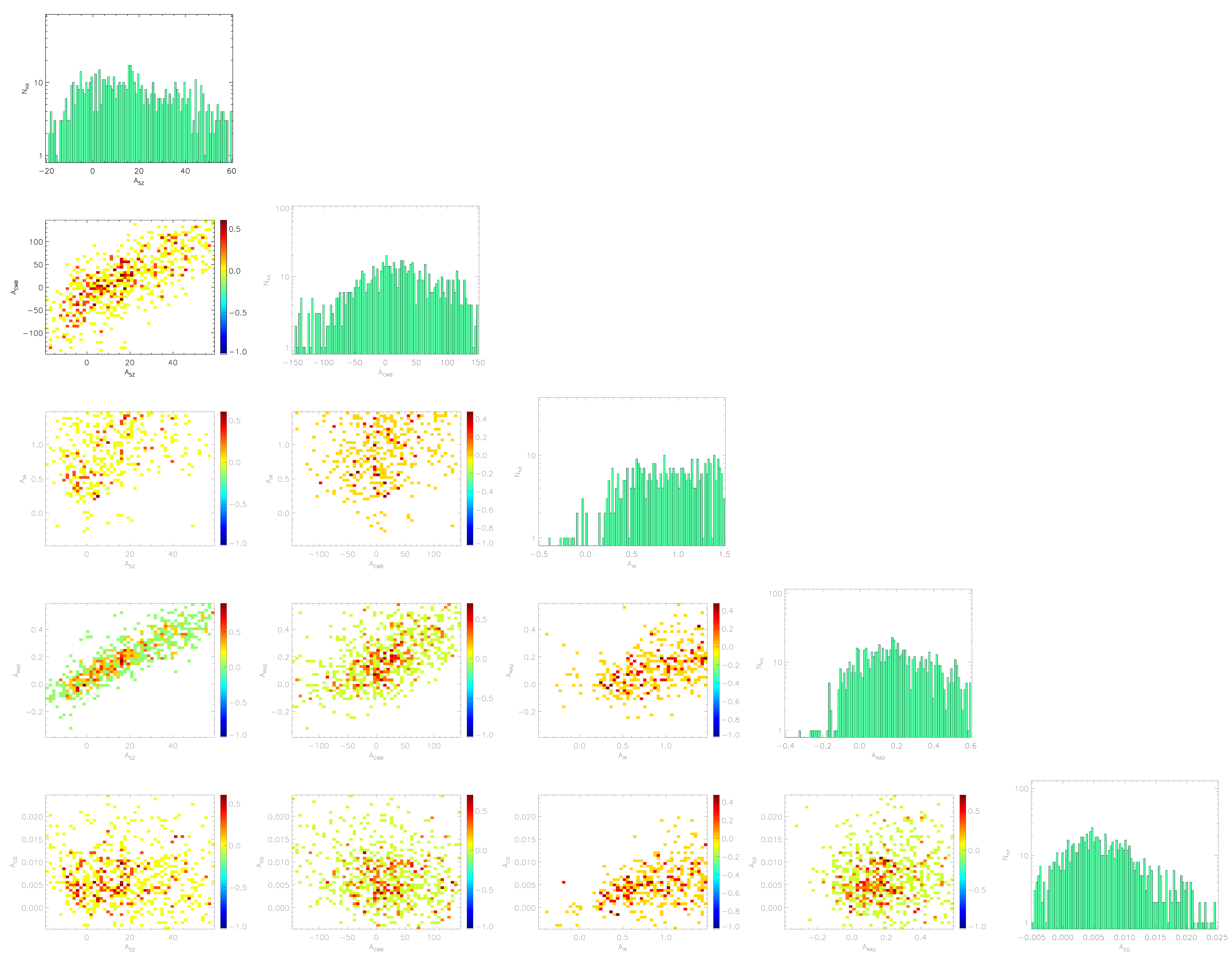}
\caption{Same as  Fig. \ref{mcxc} for CG sources from \Planck. }
\label{ECC}
\end{figure}

The derived values are presented in Figs.~\ref{30GHz}, \ref{857GHz}
and \ref{ECC}. The distributions of fitted SED amplitudes are very
different form what they look like in the case of actual galaxy
clusters. For the radio sources, we note that IR and CMB distributions
are similar to those of the cluster catalogs and that the tSZ
amplitude distribution is more symmetric as compared to the case of
galaxy clusters.  For the IR sources detected at 353~GHz , the
distribution of all amplitudes are ``pathological''.  The IR emission
contaminates all the components including CMB and tSZ. For the CG
sources, the distributions are much less compact. The $A_{\rm CMB}$
distribution is mostly symmetric. The $A_{\rm SZ}$ is symmetric and
extending to very high values, unrealistic for clusters of
galaxies. The distributions of $A_{\rm IR}$, $A_{\rm RAD}$, and
$A_{\rm CO}$ are mostly asymmetric extending to large values similarly
to the IR and radio-source cases.

\subsection{Random positions}

We perform the same SED fitting in random positions on the
sky outside the mask. The distributions of the fitted
amplitudes all show symmetric behavior and do not extend to high
values for any of the components considered here.

\begin{figure}[!th]
\center
\includegraphics[width=8cm]{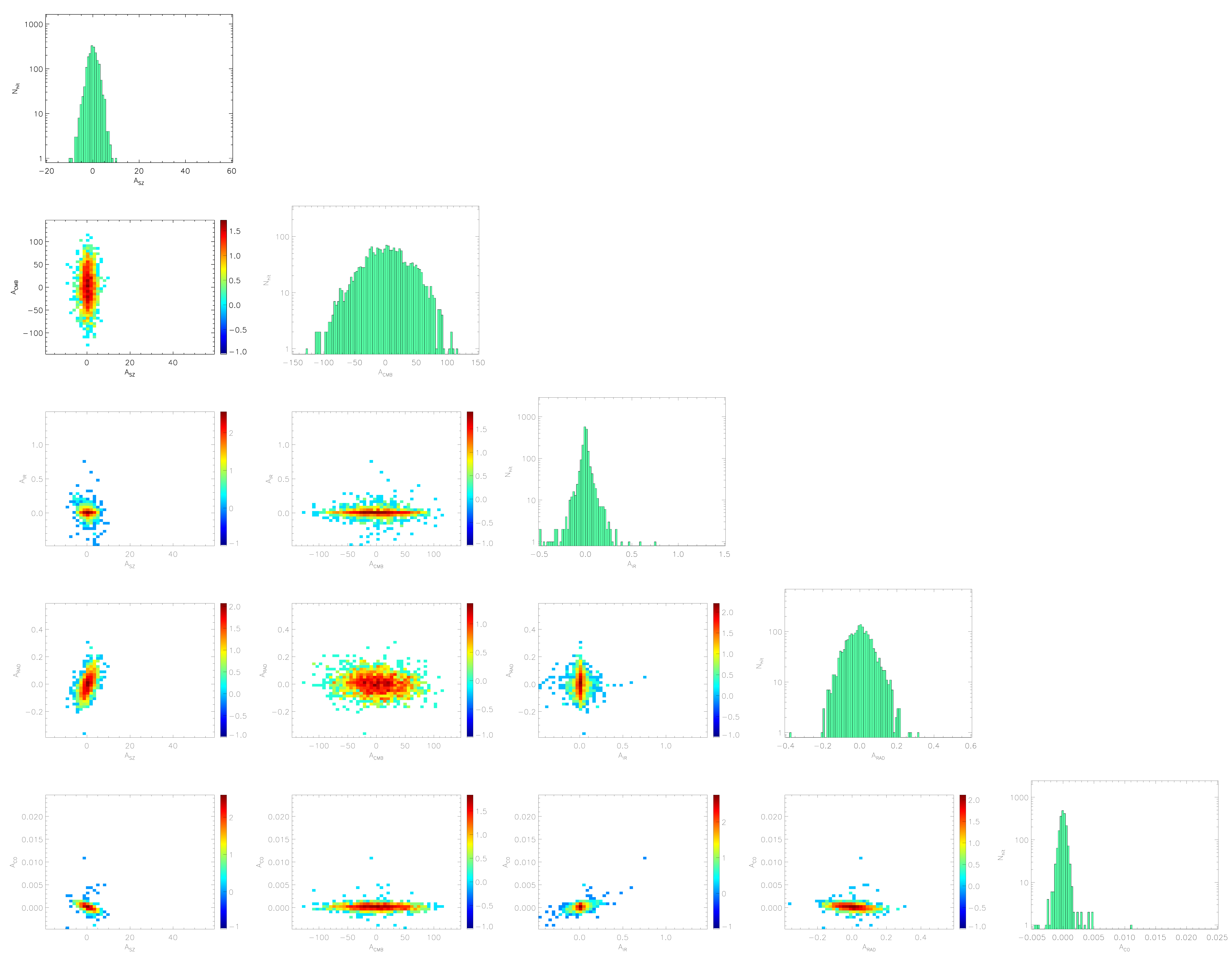}
\caption{Same as Fig. \ref{mcxc} at random positions on the sky. }
\label{jk}
\end{figure}

\subsection{PSZ1 sources}

All the results presented above are either obtained for random
positions on the sky or for catalogues and samples of actual clusters
of galaxies or IR/radio or CG sources. The PSZ1 corresponds to a
catalogue of sources detected through their tSZ effect. As such, it
contains bona fide clusters of galaxies, 861 in total, but it also
contains tSZ sources of different degrees of reliability including
false detections.

\begin{figure}[!th]
\center
\includegraphics[width=8cm]{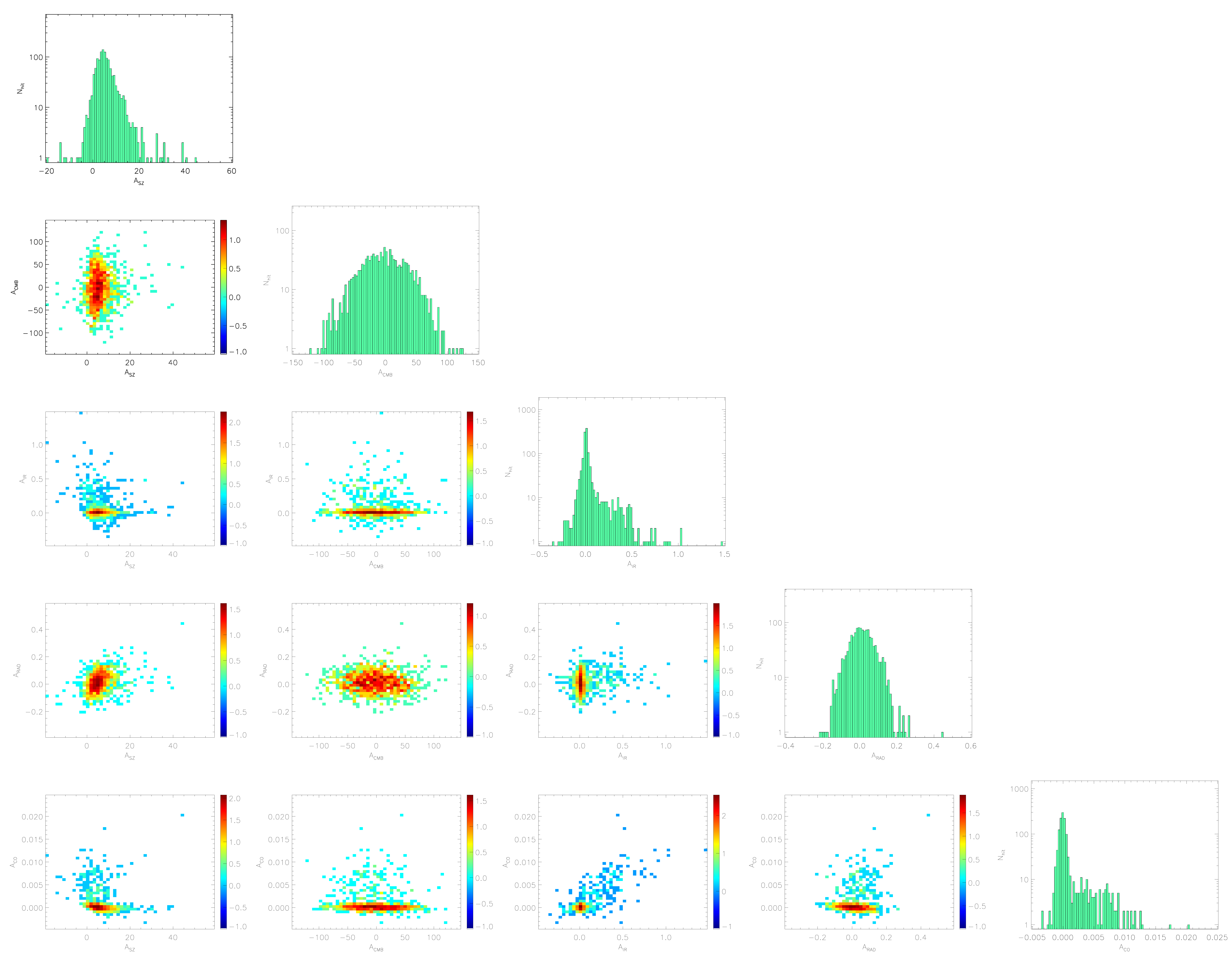}
\caption{Same as Fig. \ref{mcxc} for the PSZ1 sources. }
\label{PSZ1}
\end{figure}

We examine the distribution of the fitted SED amplitudes of all the
PSZ1 sources. The derived values for $A_{\rm SZ}$, $A_{\rm CMB}$,
$A_{\rm IR}$, $A_{\rm RAD}$, and $A_{\rm CO}$ in the direction of PSZ1
sources are shown in Fig.~\ref{PSZ1}. We note that the distribution of
$A_{\rm SZ}$ is similar to that of the clusters from MCXC and
SDSS-based samples, i.e. asymmetric and extending to positive values.
However, and contrary to the case of pure cluster samples of MCXC and
SDSS, we observe a clear excess of IR and CO emissions. This is
exhibited through the bimodal behavior of the distribution extending
to high $A_{\rm IR}$ and $A_{\rm CO}$ values.  We also note a strong
correlation of $A_{\rm IR}$ and $A_{\rm CO}$ that we explained by the fact
that a combination of IR and CO contamination mimics an offset SZ
spectral distortion.

\begin{figure}[!th]
\center
\includegraphics[width=8cm]{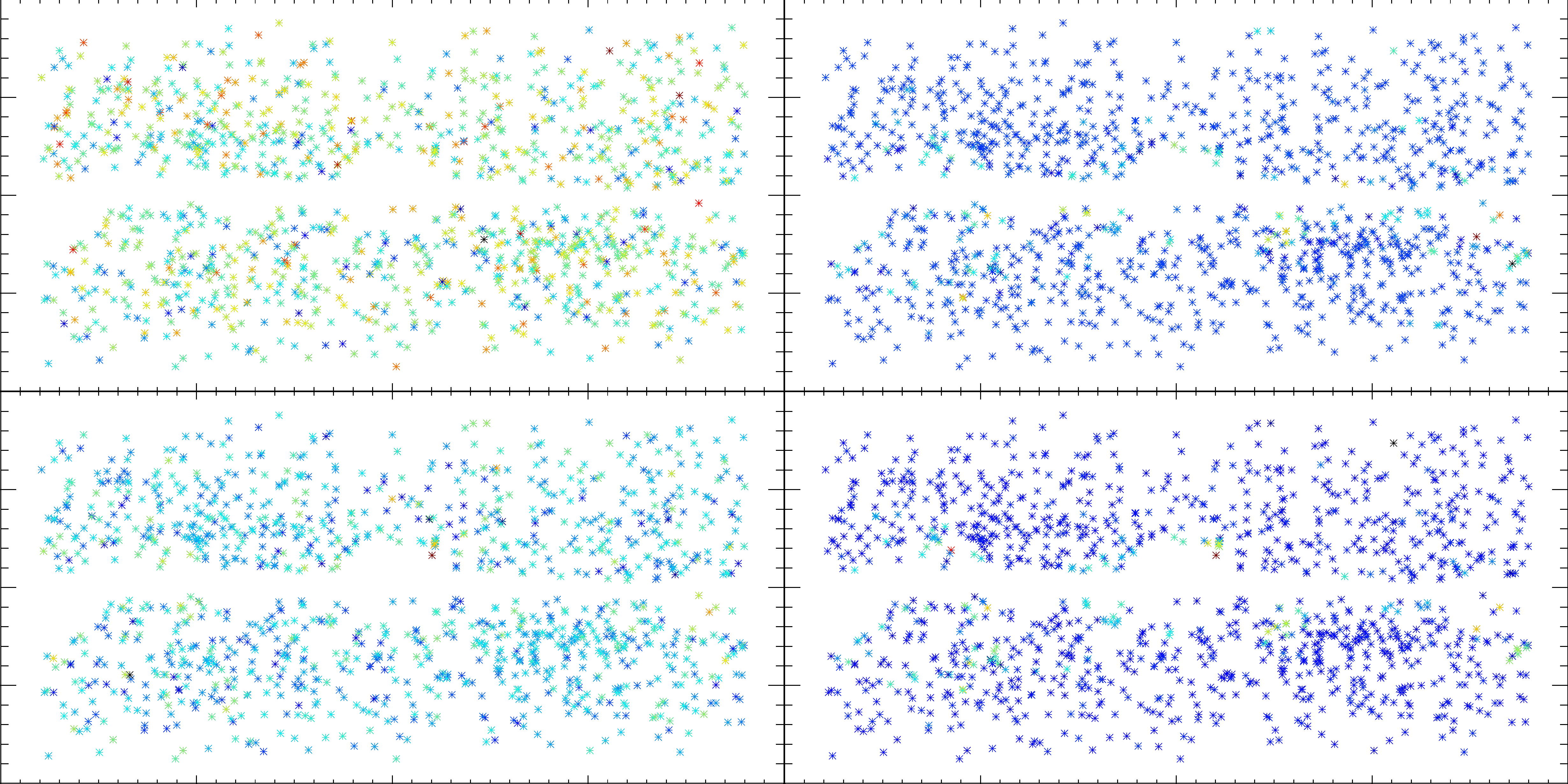}
\caption{Amplitude of $A_{\rm CMB}$, $A_{\rm IR}$, $A_{\rm RAD}$,
  and $A_{\rm CO}$ as a function of the position on the sky for PSZ1 sources.}
\label{PSZ1_sky}
\end{figure}

In Fig.~\ref{PSZ1_sky}, we show the distribution over the sky of the
amplitudes $A_{\rm CMB}$, $A_{\rm IR}$, $A_{\rm RAD}$, and $A_{\rm
  CO}$. As expected, the distribution over the sky of CMB amplitudes
does not show any particular trend of feature. It simply corresponds a
Gaussian background. The distribution over the sky of the radio
amplitudes is also rather Gaussian. As for the IR and the CO
distributions, we clearly see that the contamination is, as expected,
strongly correlated with the galactic emission in the galactic plane
and the molecular clouds. For the $A_{\rm IR}$ we also note son
contamination at higher galactic latitudes.

\section{Classification and SZ quality assessment}\label{QF}

From the analysis of the SED fitted parameters $A_{\rm SZ}$, $A_{\rm
  CMB}$, $A_{\rm IR}$, $A_{\rm RAD}$, and $A_{\rm CO}$, we note that
X-ray, optical and SZ bona fide clusters show distributions of
parameters consistent with no or low contamination. By contrast, the
distribution of fitted SED parameters of the PSZ1sources show some
contamination both by IR and by CO emission. We thus construct quality
assessments of tSZ detections based on the characteristics of SEDs.
We use three different techniques, to assess the quality of the tSZ
detection and thus separate PSZ1 sources into two categories reliable
and unreliable.

As an intermediate step, we define a phenomenological quality
assessment, hereafter penalty factor $Q_{\rm P}$, based on the data
themselves. It does not rely on a model of the SED but rather on the
empirical assessments provided in the PSZ1 which define decreasing
reliability classes 1, 2, and 3. Since the average spectrum of the
class 3 sources of PSZ1 show a clear excess of IR emission, we
restrict to a parametrisation of the IR contamination.  The penalty
factor is defined as:
\begin{equation}
Q_{\rm P} = \sum^{857}_{\nu = 353} \left(1-\frac{F_{\nu}}{F'_{\rm IR} } \right).
\end{equation}
With $F'_{\rm IR}$ set to 2.8$\times 10^{-4}$, 1.6$\times 10^{-3}$,
2.4$\times 10^{-2}$, and 1.9 K in CMB units from 353 to 857~GHz and
$\sigma_\nu = F'_{\rm IR}/3$. As previously, $F_{\nu}$ is estimated
through aperture photometry in an aperture of 7 arcmin.  The amplitude
of the IR component corresponds to the average amplitude, at each
frequency, of the class 3 sources in PSZ1 which represent the typical
low reliability sources as defined from empirical assessments. This
estimator does not require error bars on the fluxes to derive the IR
amplitude. The linear behaviour of $Q_{\rm P}$ penalises cases that
exceed the $F_{\rm IR}$.

\begin{figure}[!th]
\center
\includegraphics[width=8cm]{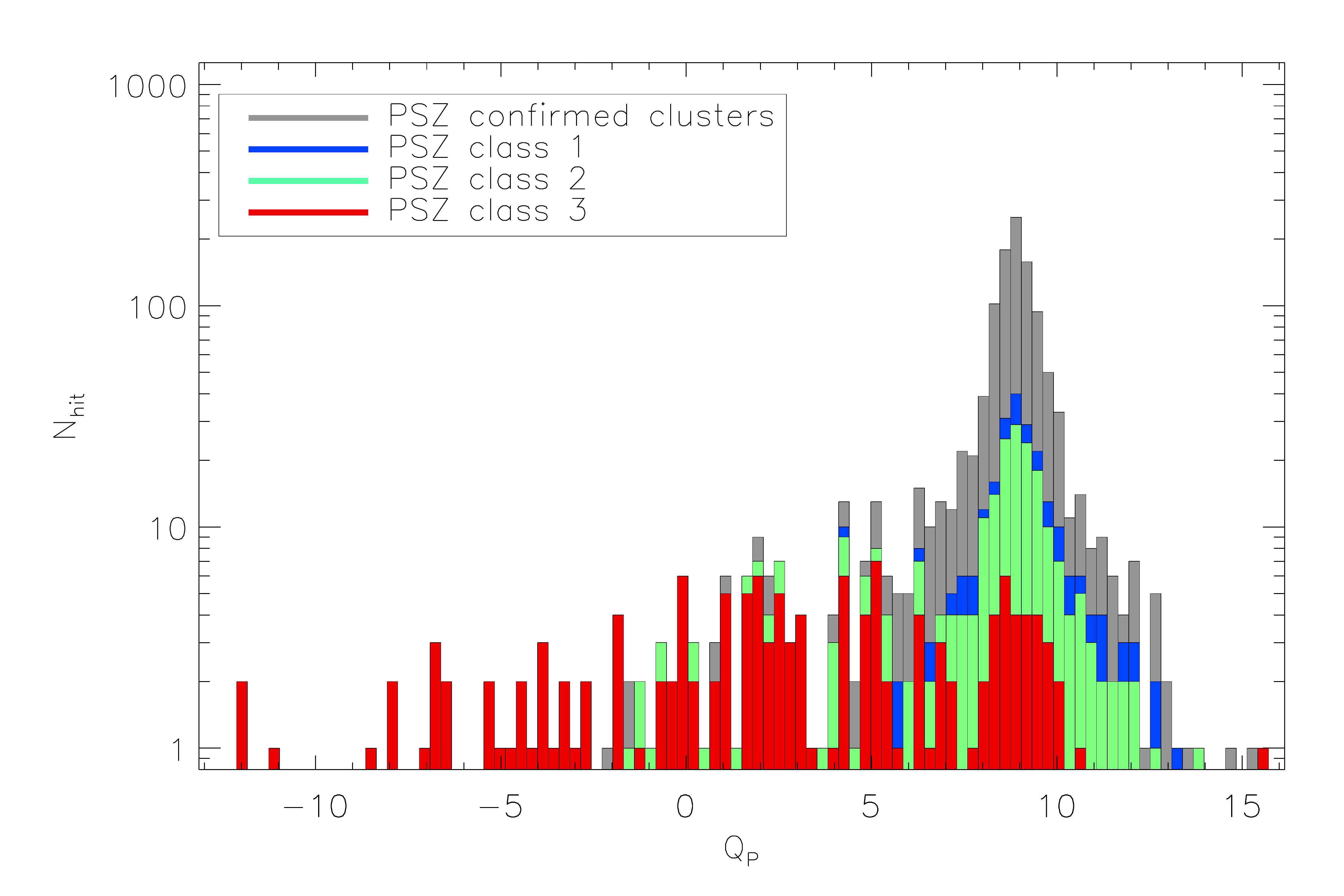}
\caption{Piled-up distribution of the penalty factor for the PSZ1
  catalog. Grey, blue, green and red are for confirmed clusters, class
  1, 2 and 3 sources respectively.}
\label{QUAL_CHEMA_PSZ}
\end{figure}

\subsection{Clustering-based quality assessment}

The clustering algorithm is an unsupervised machine learning method
often presented as assigning objects to the nearest cluster by
distance. There are several choices for the distance: Euclidian,
Manhatan, or generalised distance with the Mahalanobis metric. The
number of clusters $n$ is supplied as an input parameter.

We perform the classification of the sources, in $n$
populations/clusters, using a standard {\it k-means} clustering
\citep{har75,har79} considering a Euclidian metric for the parameter
space. In a first step we define the distance, $d_{\rm cont}$, in the
SED amplitude space from the zero contamination level is defined as
\begin{equation}
d_{\rm cont} = \sqrt{\left(\frac{A_{\rm CMB}}{\sigma_{\rm
      CMB}}\right)^2 + \left(\frac{A_{\rm IR}}{\sigma_{\rm
      IR}}\right)^2 + \left(\frac{A_{\rm RAD}}{\sigma_{\rm
      RAD}}\right)^2 + \left(\frac{A_{\rm CO}}{\sigma_{\rm
      CO}}\right)^2},
\end{equation}
where $\sigma_{\rm CMB}$, $\sigma_{\rm IR}$, $\sigma_{\rm RAD}$, and
$\sigma_{\rm CO}$ are the standard deviations of CMB, IR, radio, and
CO amplitude distributions. 
 
\begin{figure}[!th]
\center
\includegraphics[width=8cm]{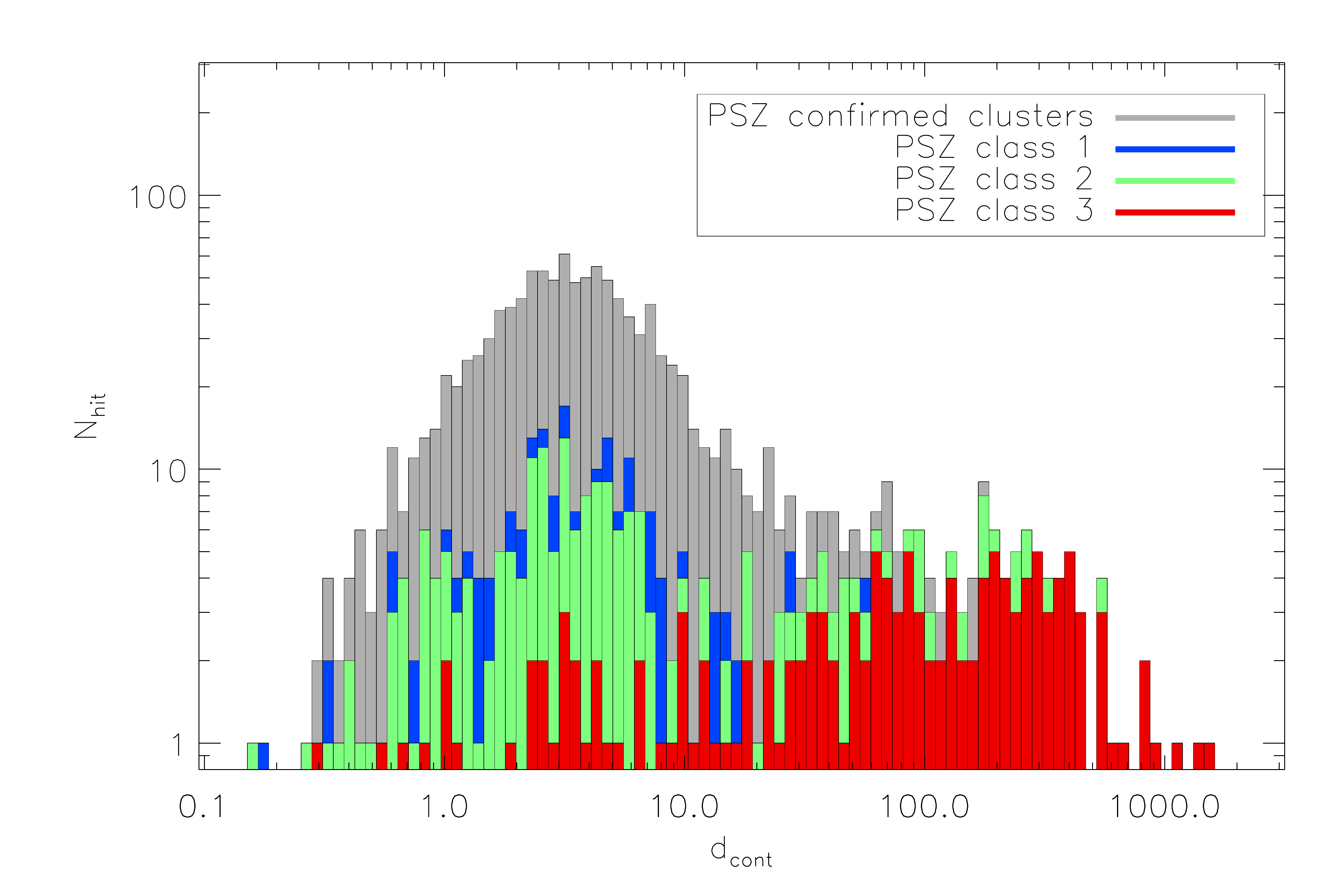}
\caption{Piled-up distribution of the distance, $d_{\rm cont}$, in the
  clustering algorithm for PSZ1 sources. Grey is for confirmed
  clusters, blue is for class 1 sources, green for class 2 sources,
  and red is for class 3 sources.}
\label{CLUS_DIST}
\end{figure}

In Fig.~\ref{CLUS_DIST}, we present the piled-up distribution of the
distance, $d_{\rm cont}$, for the PSZ1 sources. 
We observe that the confirmed clusters (grey), class
1 (blue), and class 2 (green) candidates present similar
distributions, whereas class 3 objects (red) show larger values for
$d_{\rm cont}$. This illustrates the presence of distinct populations of objects 
in the PSZ1 sample.\\

Then, we apply the the {\it k-means} algorithm to the PSZ1 sources.
One drawback the {\it k-means} approach
is that, it requires that each cluster of population is symmetric and
has the same extension with respect to the metric. This implies that
we need a rather large number of populations. Moreover, an
inappropriate choice of $n$ may yield to poor results. That is why, when
performing {\it k-means}, it is important to run diagnostic checks.  
We have tested the clustering techniques considering from $n=2$ to 6
populations of sources. Below $n=3$, the distribution of fitted
amplitudes; $A_{\rm SZ}$, $A_{\rm CMB}$, $A_{\rm IR}$, $A_{\rm RAD}$,
and $A_{\rm CO}$; showed residual contamination.

\begin{figure}[!th]
\center
\includegraphics[width=8cm]{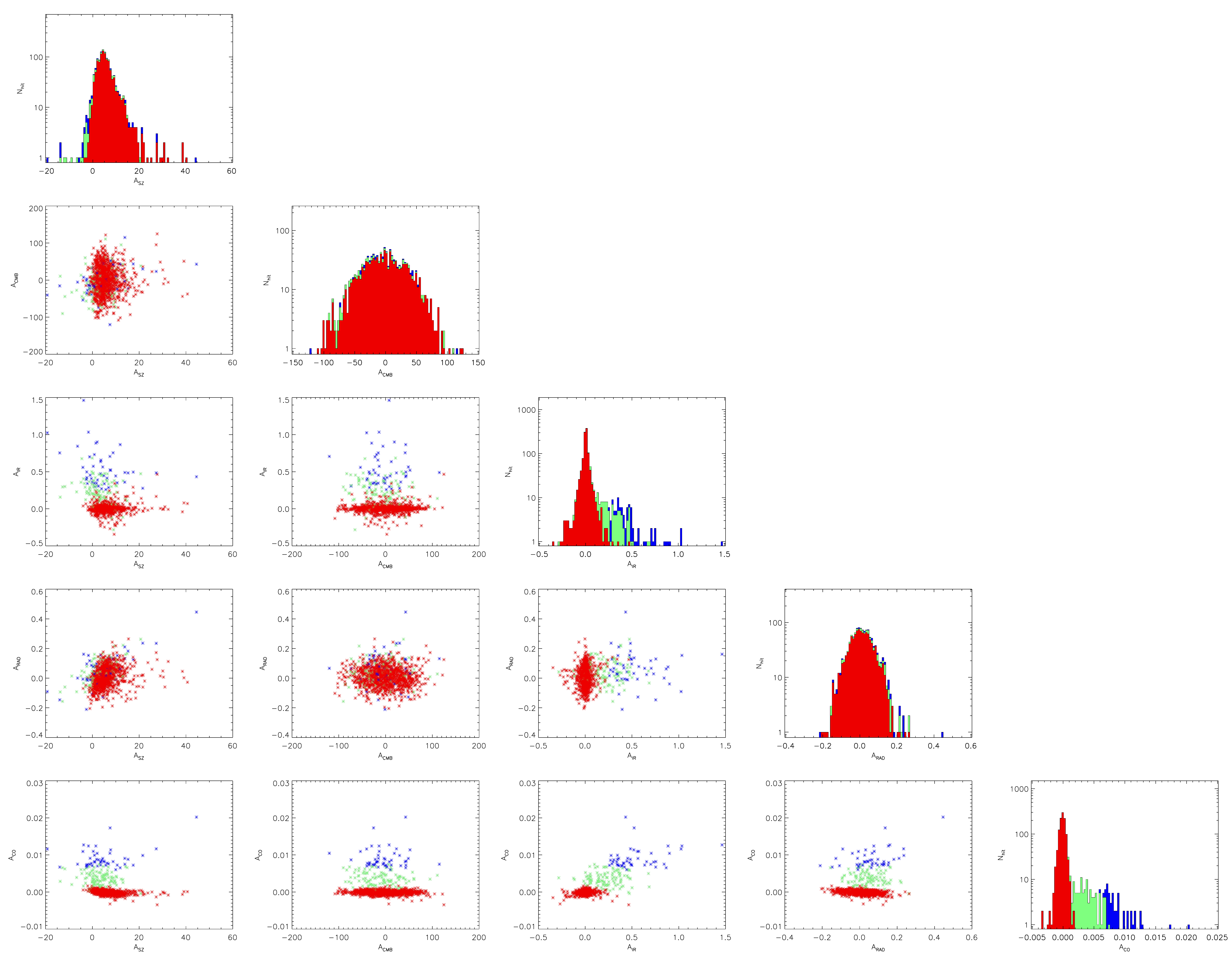}
\caption{Same as Fig. \ref{mcxc} for the PSZ1 sources. In red we show
  the highest-quality sources, in yellow, cyan and blue the sources are
  displayed with decreasing quality.}
\label{CLUST_DISTRIB}
\end{figure}

We present the results for three populations in
Fig.~\ref{CLUST_DISTRIB}.  We found that, above three populations, the
results in terms of the distributions of fitted SED parameters were
unchanged. We show the results of the clustering algorithm
classification for the three populations (in red the highest-quality
sources, in yellow, cyan and blue the sources with decreasing
quality). We note that the population of good/reliable sources shows
little signs of contamination and shares the same characteristics as
the true clusters (see Figs.~\ref{mcxc}, \ref{sdss} and
\ref{psz1}). In particular in the $A_{\rm IR}$/$A_{\rm CO}$ plane, we
observe a good separation between the three populations.

The clustering algorithm separates the low-quality and higher quality
candidates and clusters of PSZ1 catalogue. However, the separation of
populations is not optimal as the two populations of reliable and
unreliable sources shows a large overlap.


\subsection{Likelihood-based quality assessment}

In this second approach half way between supervised and unsupervised
methods, we base the assessment of the SZ detection on the Likelihood
of the contamination.  We thus define a quality factor, $Q_{\rm L}$,
as the product of SED parameter distributions estimated in random
positions over the sky

\begin{equation}
Q_{\rm L} = G_{\rm CMB}(A_{\rm CMB}) C_{\rm IR}(A_{\rm IR}) G_{\rm RAD}(A_{\rm
  RAD}) G_{\rm CO}(A_{\rm CO}),
\end{equation}
with $G_{\rm CMB}$, $C_{\rm IR}$, $G_{\rm RAD}$, and $G_{\rm CO}$ the
distributions of the fitted SED parameters $A_{\rm CMB}$, $A_{\rm
  IR}$, $A_{\rm RAD}$, and $A_{\rm CO}$. $G$ stands for Gaussian
distribution, $A\, {\rm exp}(-(x-m)^2/2\sigma^2)$, and $C$ for Cauchy
distribution, $A/(1+(x-m)^2/\sigma^2)$, distributions.  We show in
Table \ref{distrib} the results of the adjustments for the random
positions in the sky. We also conservatively set that high quality tSZ
detections correspond to a $6\,\sigma$ limit which translates into
$Q_{\rm L}\sim1.5\,10{-8}$.

\begin{table}
\center
\label{distrib}
\caption{Best fitting parameters for $A_{\rm CMB}$, $A_{\rm IR}$,
  $A_{\rm RAD}$, and $A_{\rm CO}$ distributions for random
  positions in the sky.}
\begin{tabular}{|c|c|c|c|c|}
\hline
 & Distrib. & A & m & $\sigma$ \\
\hline
$A_{\rm CMB}$ & Gaussian & 49.6 & 1.12 & 35.5 \\
\hline
$A_{\rm IR}$ & Cauchy & 506.8 & -24.0 $10^{-5}$ & 16.2 $10^{-4}$ \\
\hline
$A_{\rm RAD}$ & Gaussian & 273.4 & -2.78 $10^{-3}$ & 6.71 $10^{-2}$ \\
\hline
$A_{\rm CO}$ & Gaussian & 271.1 & -1.90 $10^{-5}$ & 4.92 $10^{-4}$ \\
\hline
\end{tabular}
\end{table}

\begin{figure}[!th]
\center
\includegraphics[width=8cm]{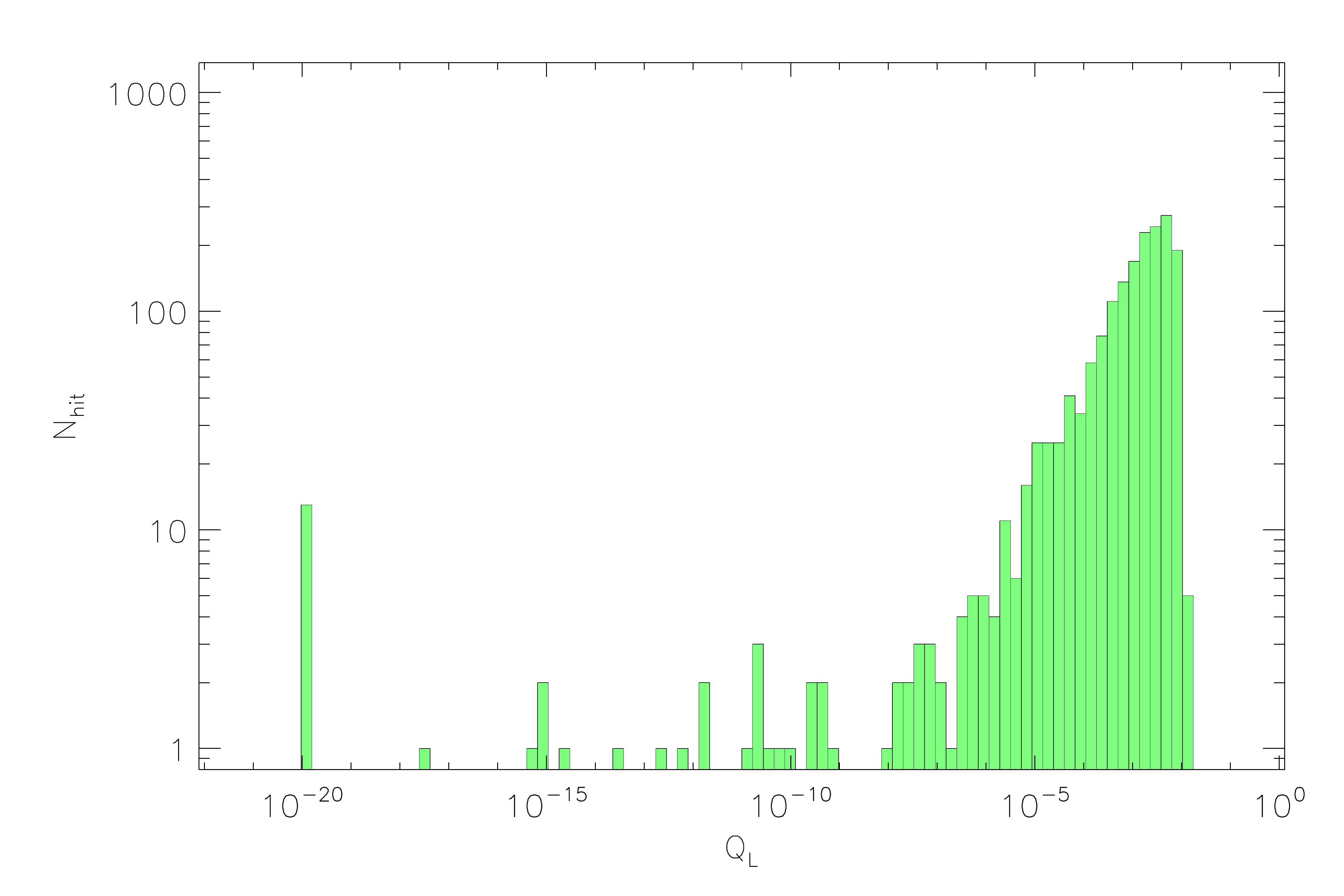}
\caption{Distribution of the quality factor $Q_{\rm L}$ for MCXC cluster of
  galaxies.}
\label{QUAL_mcxc}
\end{figure}

\begin{figure}[!th]
\center
\includegraphics[width=8cm]{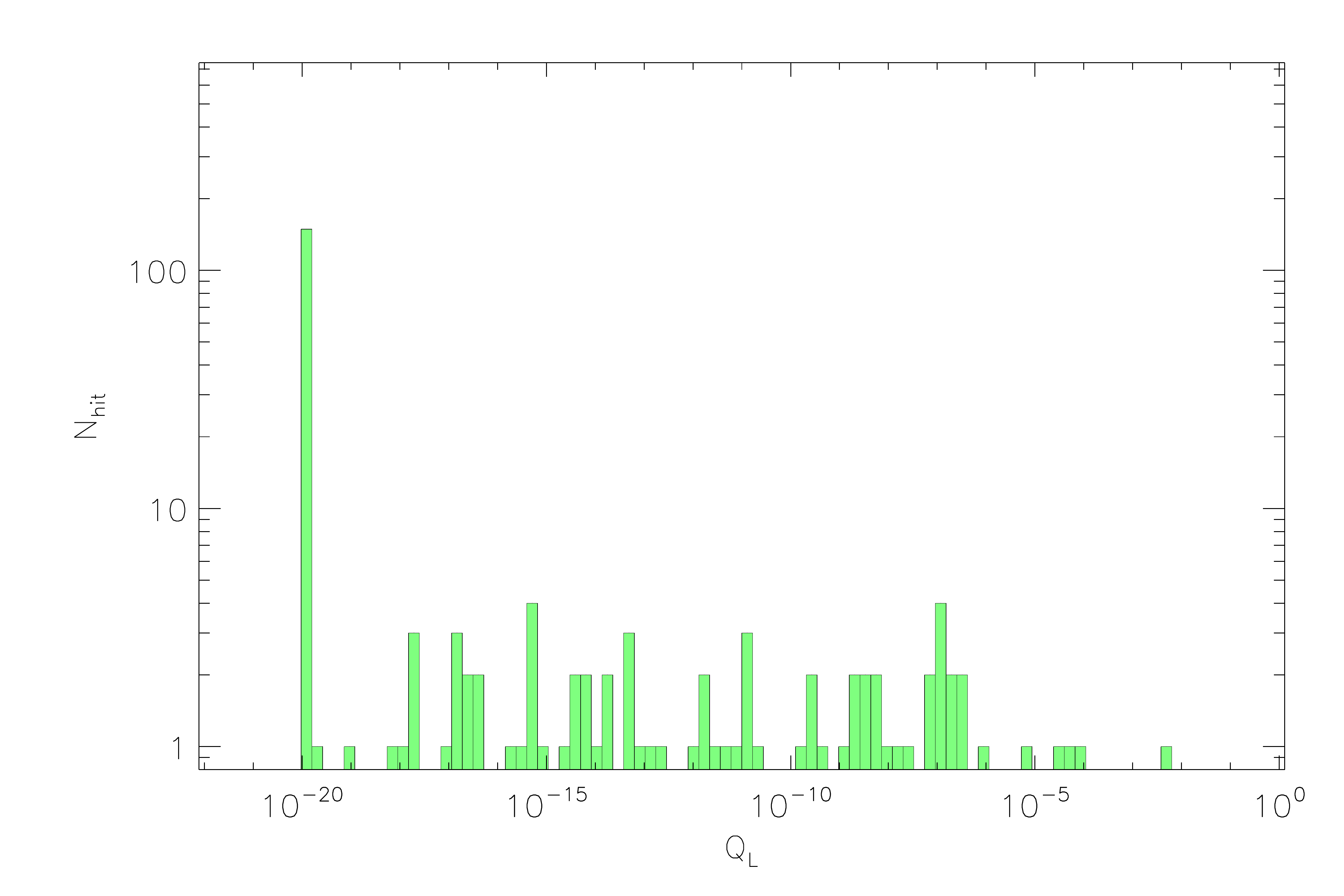}
\caption{Distribution of the quality factor $Q_{\rm L}$ for the sample of
  false detections.}
\label{QUAL_false}
\end{figure}

We first show in Fig.~\ref{QUAL_mcxc} the distribution of $Q_{\rm L}$ for
the MCXC clusters. We note that the vast majority of these clusters
fall above the quality factor of $Q_{\rm L}=1.5\,10^{-8}$. A small number of
clusters from the MCXC have quality factors lower than the cut. They
correspond to clusters exhibiting important contamination from AGN and
radio sources. We also show in Fig. \ref{QUAL_false} the distribution
of $Q_{\rm L}$ for the sample of false detections defined in
Sect. \ref{data}. Only a handful of false detections lay above the
quality factor cut.

\begin{figure}[!th]
\center
\includegraphics[width=8cm]{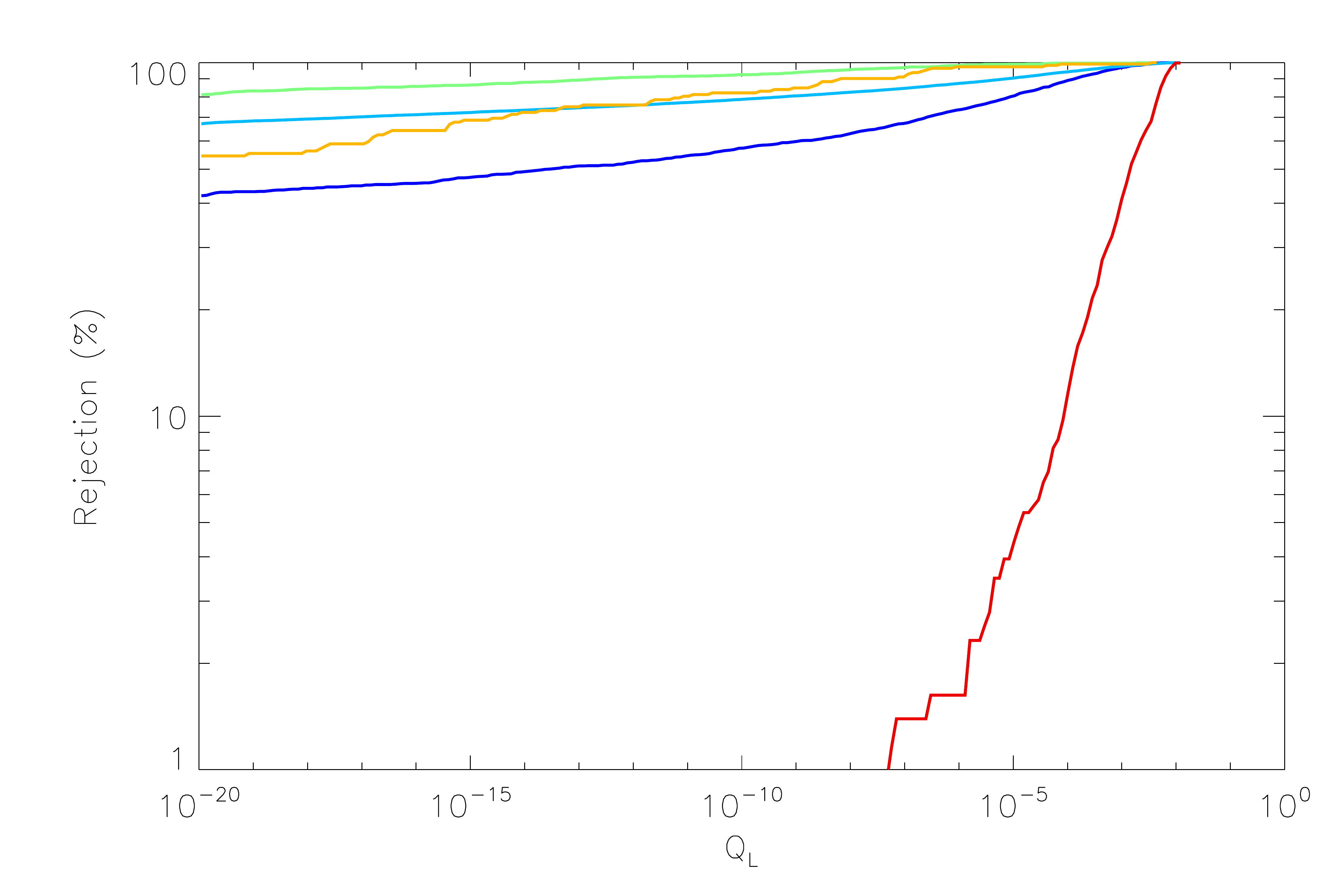}
\caption{For the confirmed clusters of the PSZ1 catalogue: Fraction of
  rejected sources as a function of the quality factor $Q_{\rm L}$
  cut. In red and orange are the true clusters and false detections
  respectively. We also display in blue, cyan and green the radio, IR
  and CG sources.}
\label{rej_like_test}
\end{figure}

In Fig. \ref{rej_like_test} we show the percentage of sources
rejected by the quality factor cut $Q_{\rm L}=1.5\,10^{-8}$. We see that
true confirmed clusters (red line) are not rejected. Applying the
quality factor to the false sources (orange line), allows us to reject
about 10\% of the false detection. We see that the efficiency of the
rejection if the contaminating sources were radio sources at 30~GHz
(blue line), IR sources at 353~GHz (cyan line) or CG sources (green
line) differ.

\begin{figure}[!th]
\center
\includegraphics[width=8cm]{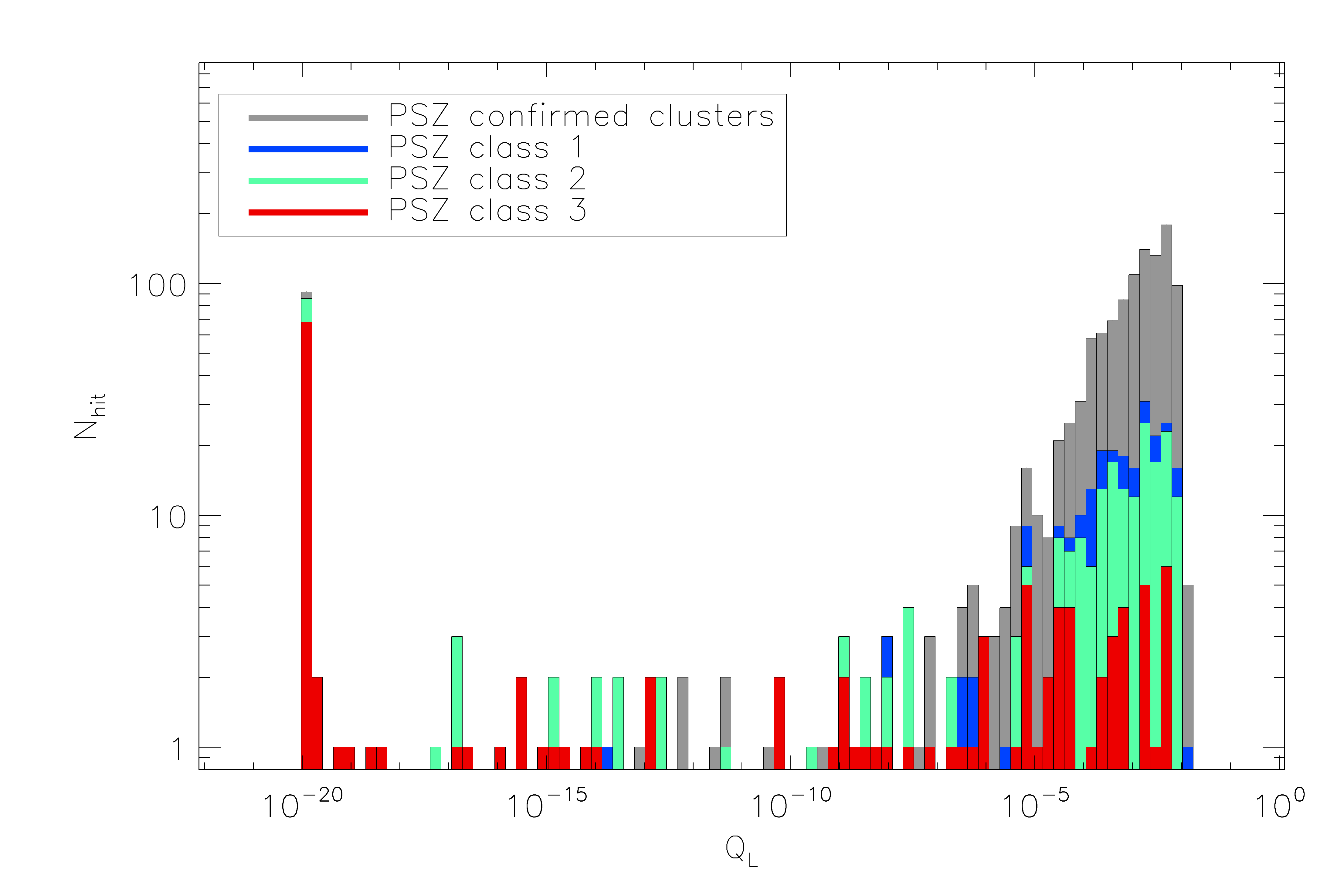}
\caption{Distribution of the quality factor for PSZ1 sources. In grey
  for confirmed clusters, in blue, green and red class 1 , 2 and 3
  sources respectively.}
\label{QUAL1}
\end{figure}

We now show Fig.~\ref{QUAL1} the piled up histograms of $Q_{\rm L}$ for the
PSZ1 sources. We display the class 1, 2, 3 together with the confirmed
clusters in blue, green, red, and grey respectively.  We observe that
the quality factor cut at $Q_{\rm L}=1.5\,10^{-8}$ clearly separates the
confirmed clusters from the rest. Moreover, the quality factor also
separated the class 3 candidates of PSZ1 from the other PSZ1 sources,
with most of the latter being in the category of low reliability
sources. Some of the class 3 candidates, though pass the cut and are
in the category of highly reliable candidates. Confirmation of their
status by follow-up observation will be an interesting test of the
classification method.

\begin{figure}[!th]
\center
\includegraphics[width=8cm]{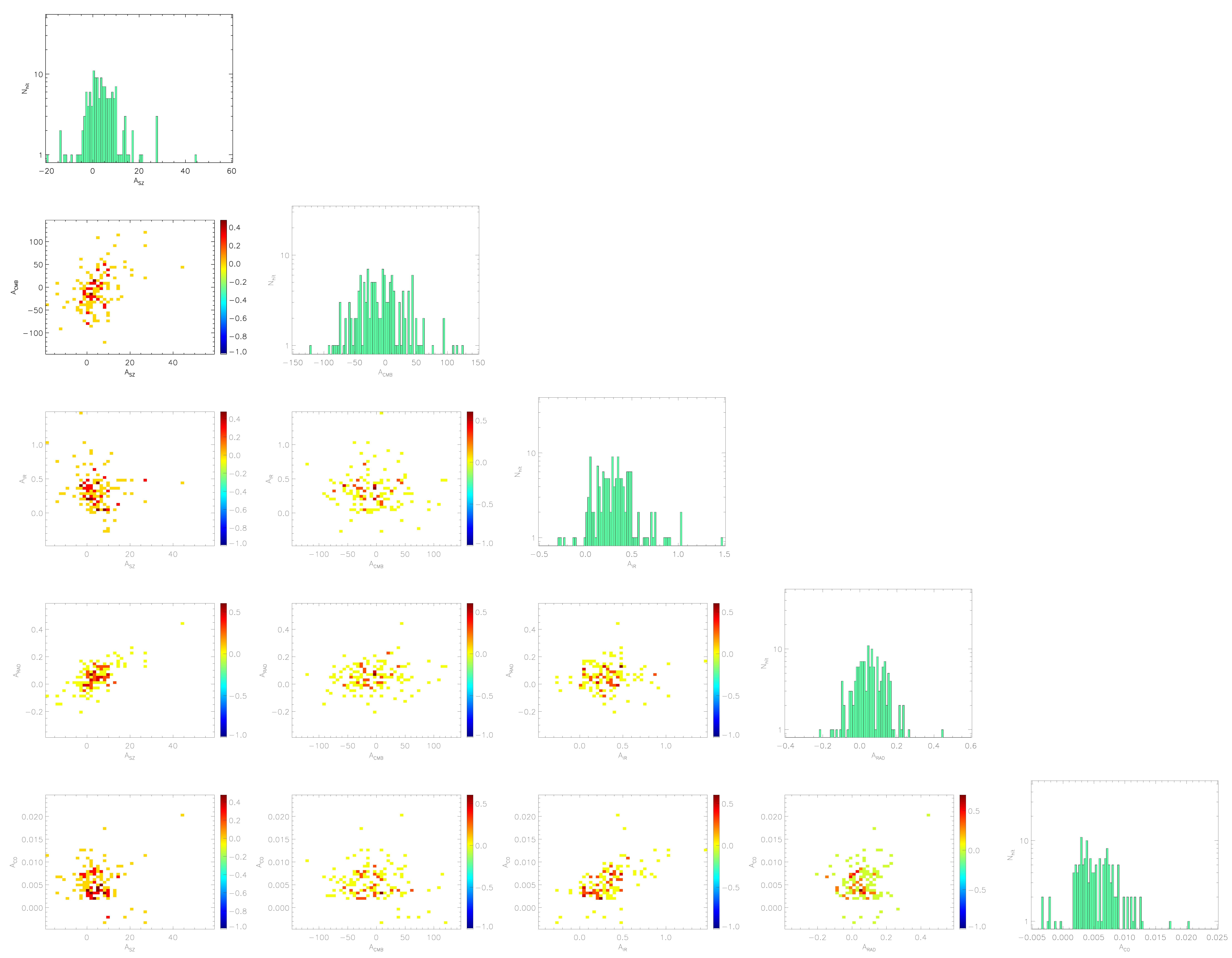}
\caption{Same as Fig.\ref{mcxc} for the high-quality PSZ1 sources
  according, i.e. those with $Q_{\rm L}>1.5
\,10^{-8}$.}
\label{dist_good_QL}
\end{figure}

\begin{figure}[!th]
\center
\includegraphics[width=8cm]{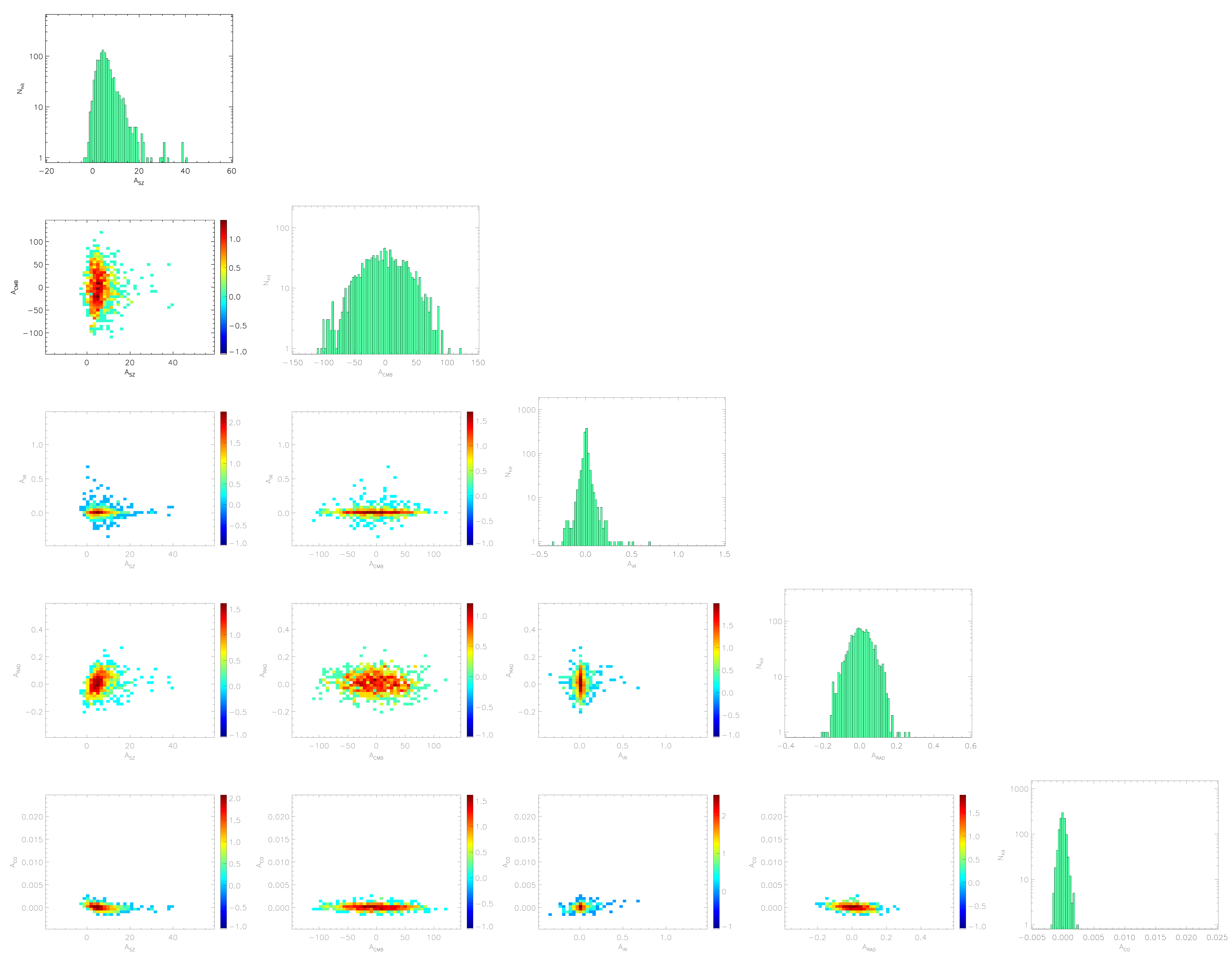}
\caption{Same as Fig.\ref{mcxc} for the low-quality PSZ1 sources those
  with $Q_{\rm L}<1.5 \, 10^{-8}$.}
\label{dist_bad_QL}
\end{figure}

We show the distributions of $A_{\rm SZ}$, $A_{\rm CMB}$, $A_{\rm
  IR}$, $A_{\rm RAD}$, and $A_{\rm CO}$ in Figs.~\ref{dist_good_QL}
and \ref{dist_bad_QL} for the PSZ1 sources with $Q_{\rm L}$ above the
cut $1.5\,10^{-8}$ and for the PSZ1 sources with $Q_{\rm L}< 1.5
\,10^{-8}$. We check that the high-quality sources ($Q_{\rm L}>1.5
\,10^{-8}$) do not show significant contamination by IR, radio nor CO
emissions, while for the sources with $Q_{\rm L}<1.5 \, 10^{-8}$, we
observe a clear contamination, especially the IR-CO plane.


\subsection{Neural network-based quality assessment}

The third approach is based on Artificial Neural Network (ANN,
hereafter), the archetype of supervised machine learning methods. ANNs
are a machine learning methodology based on parallelism and
redundancy. The basic building block of an ANN is the
neuron. Information is passed as inputs to the neuron, which processes
them and produces an output which is a simple
mathematical function of the inputs. The power of the ANN comes from
assembling many neurons into a network. Well-designed networks are
able to learn from a set of training data and to make predictions when
presented with new, possibly incomplete, data.

\begin{figure*}[!th]
\center
\includegraphics[width=16cm]{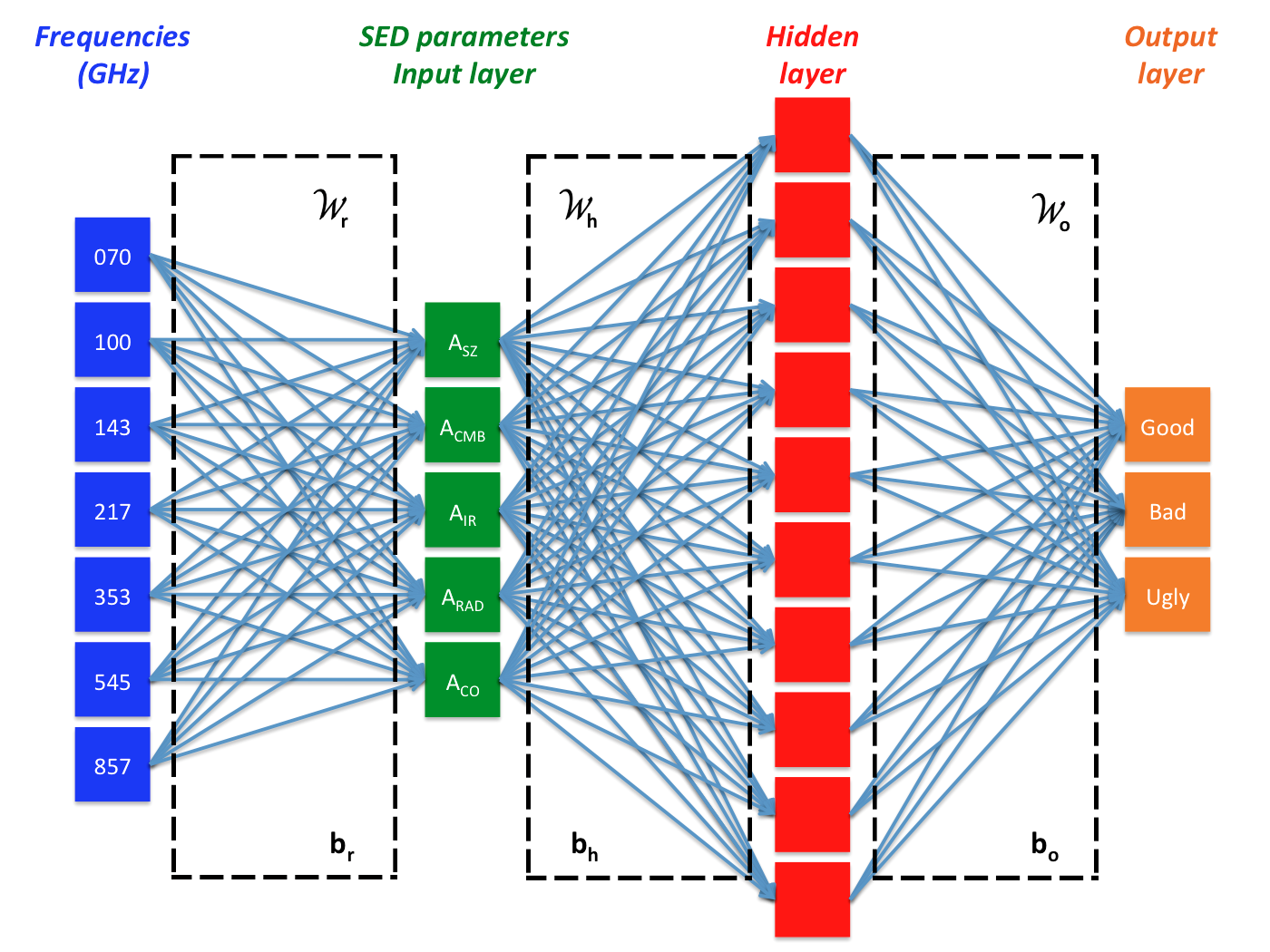}
\caption{Neural network diagram. In our analysis, the input layer is
  composed by the five SED fitted parameters of each source at the
  seven frequencies. The hidden layer is composed by 10 neurons. And
  the output layer contains three values of $Q$ for the categories,
  {\it Good}, {\it Bad}, and {\it Ugly}, standing for reliable,
  unreliable, and noisy sources.}
\label{nnet_diag}
\end{figure*}

We consider a standard three-layer back-propagation ANN to separate
the tSZ detections into three populations of reliable ({\it good}
quality), unreliable/false ({\it bad} quality), and noisy sources
({\it ugly}). A three-layer network consists of a layer of input
neurons, a layer of hidden neurons, and a layer of output neurons. In
such an arrangement each neuron is referred to as a node. The input
layer consists of the five SED parameters and the output nodes
represent the three classes of populations. The layout and number of
nodes represent the architecture of the network.\\
Details on the ANN implementation can be found in Appendix
\ref{ANNET}. We briefly present here the basics of this technique and
illustrate the principle schematically in Fig. \ref{nnet_diag}.

We define
\begin{equation}
Q = g \left( {\cal W}_{\rm o} g\left({\cal W}_{\rm h} \left( {\cal
  W}_{\rm r} F_\nu + {\bf b}_{\rm r} \right) + {\bf b}_{\rm h} \right)
+ {\bf b}_{\rm o} \right),
\label{eqNNET}
\end{equation}
where $g(x) = 1/(1+{\rm exp}(-x))$ is the activation function, $ {\cal
  W}_{\rm r} = ({\cal F}^T {\cal C}^{-1}_{N} {\cal F})^{-1} {\cal F}^T
{\cal C}^{-1}_{N}$, corresponds to a physically-based dimensional
reduction, $ {\cal W}_{\rm h} $ are the weights between input and
hidden layers, $ {\cal W}_{\rm o} $ are the weights between hidden and
output layers, $ {\bf b}_{\rm h} $ are the biases between input and
hidden layers, and $ {\bf b}_{\rm o} $ are the biases between hidden
and output layers.

To train the neural network, we use SED fitted parameters of the
confirmed clusters of PSZ1 catalogue. They are representative of the
{\it Good} high-quality source population. We use the fitted
parameters of the the sample of false detections defined in
Sect. \ref{data}. They are representative of unreliable sources, the
{\it Bad}. We also use the fitted parameters computed in random
position over the sky. They are representative of noise-dominated
population, the {\it Ugly}. We split each catalogue into two 
subsets, one training set and one checking set. The second is used to
estimate the efficiency of the ANN.\\
We defined the error on the classification as,
\begin{equation}
E = \frac{1}{2}\sum_{\rm class} (Q^{(\rm true)}_{\rm class} - Q_{\rm class})^2
\end{equation}
In order to avoid
over-training, we stop the training at the value that minimizes the
error for the checking set. 

\begin{figure}[!th]
\center
\includegraphics[width=8cm]{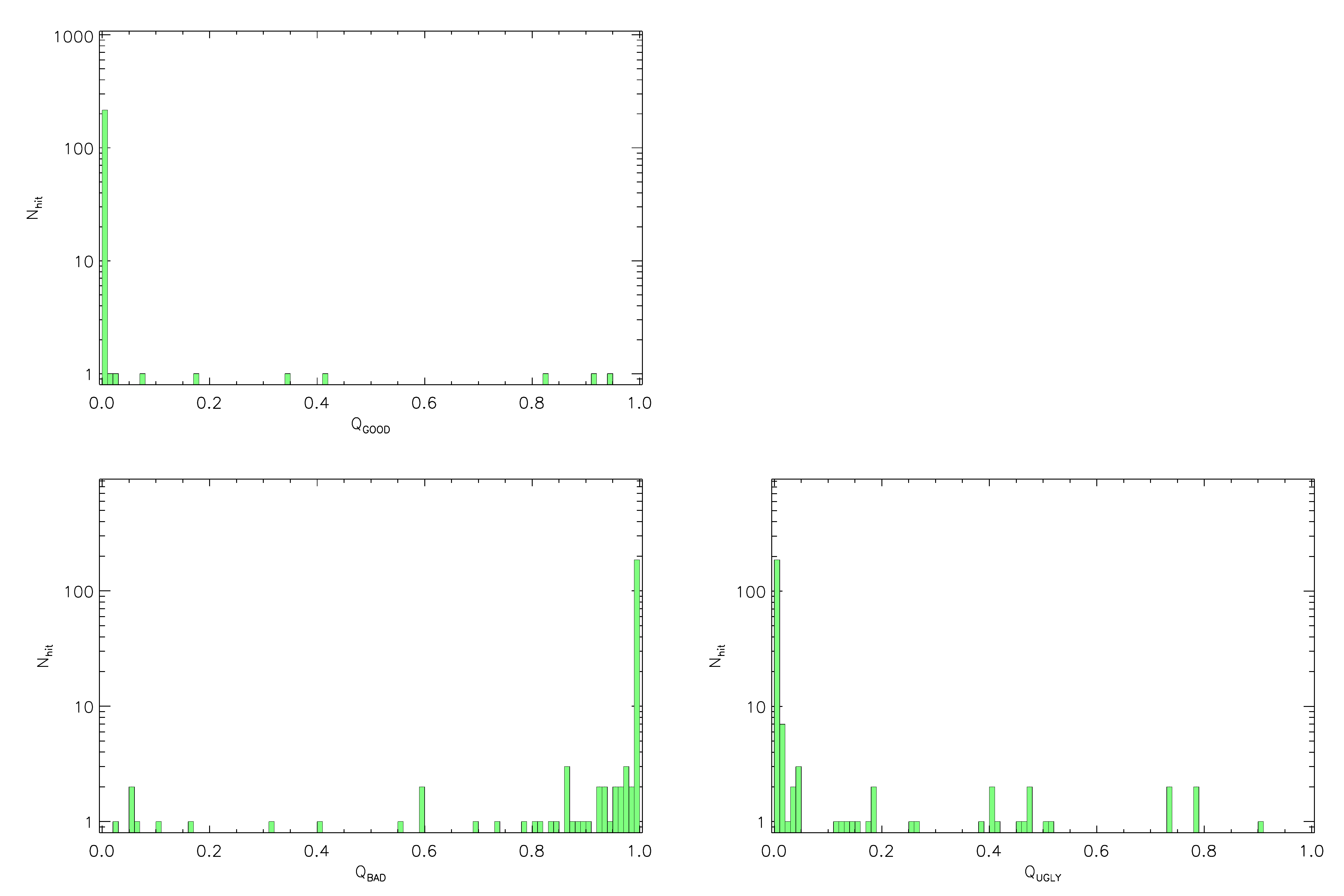}
\caption{Distribution of neural-network-based estimation of $Q_{\rm
    good}$, $Q_{\rm bad}$ and $Q_{\rm ugly}$ for the catalogue of
  false detections (checking-set subsample).}
\label{QUAL_NNET_BAD}
\end{figure}

\begin{figure}[!th]
\center
\includegraphics[width=8cm]{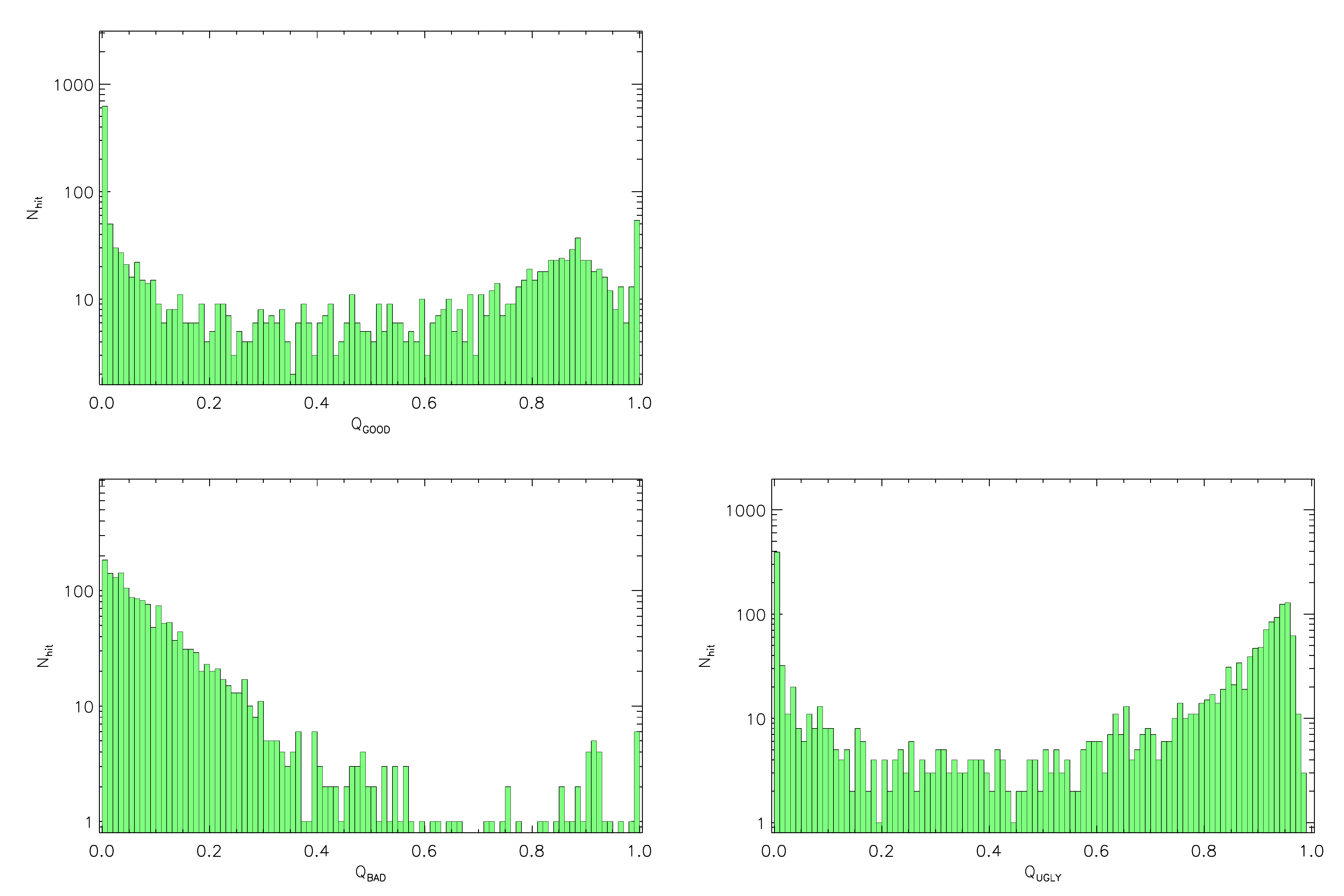}
\caption{Distribution of ANN-based estimation of $Q$ for
  the MCXC catalogue (checking-set subsample).}
\label{QUAL_NNET_MCXC}
\end{figure}

The ANN outputs a value of $Q_{\rm good}$, $Q_{\rm bad}$ and $Q_{\rm
  ugly}$ (as given by Eq. \ref{eqNNET}) for source.  We first show in
Fig. \ref{QUAL_NNET_BAD} the distribution of ANN-based estimation of
the $Q$ values for the catalogue of false detections (checking-set
subsample).  We note that the distribution is dominated by high values
of $Q_{\rm bad}$ and low values of $Q_{\rm good}$. We show by contrast
in Fig. \ref{QUAL_NNET_MCXC} the same distribution of $Q$ values for
actual clusters of galaxies from the MCXC catalogue. In this case, we
note that most of the sources in this catalogue have low values of
$Q_{\rm bad}$. The sources with lowest $Q_{\rm bad}$ values are
clusters exhibiting important contamination from AGNs. We also note a
relatively large number of clusters with high values of $Q_{\rm
  ugly}$. These clusters are associated with the low-mass clusters
that have no significant SZ counterpart in the \Planck\ data.

\begin{figure}[!th]
\center
\includegraphics[width=8cm]{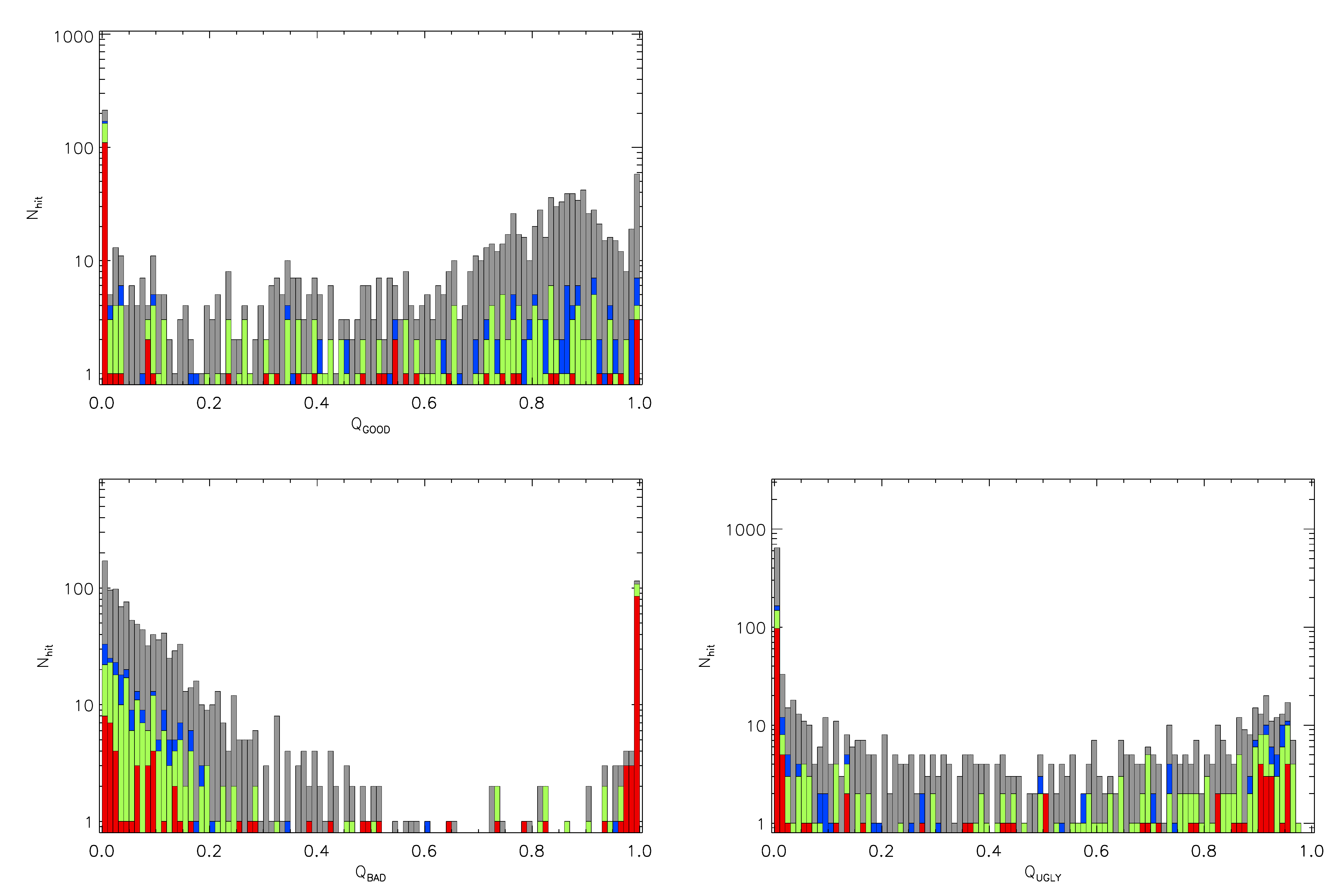}
\caption{Distribution of neural-network-based estimation of the
  quality factor for the PSZ1 catalog (used as training or checking
  sets). Grey, blue, green and red are for confirmed clusters, class
  1, 2 and 3 sources respectively.}
\label{QUAL_NNET_PSZ}
\end{figure}

In Fig.~\ref{QUAL_NNET_PSZ}, we display the piled-up histograms of
$Q_{\rm good}$, $Q_{\rm bad}$ and $Q_{\rm ugly}$ values for sources of
the PSZ1 catalogue. We display in grey, blue, green and red are for
confirmed clusters, class 1, 2 and 3 sources respectively.  We note
that the ANN-based quality factor allows us to separate nicely the
distribution of $Q_{\rm bad}$ into two regimes low and high values
(associated mostly with class 3 PSZ1 sources) thus allowing us to
identify clearly the {\it Bad} sources in the catalogue. The
distribution of $Q_{\rm ugly}$ is flatter and shows that the category
of {\it ugly} noisy sources is evenly distributed including among
confirmed bona fide clusters. The distribution of $Q_{\rm good}$ is
dominated by high values. The lowest end of the distribution is
populated by class 3 sources from the PSZ1. 

\begin{figure}[!th]
\center
\includegraphics[width=8cm]{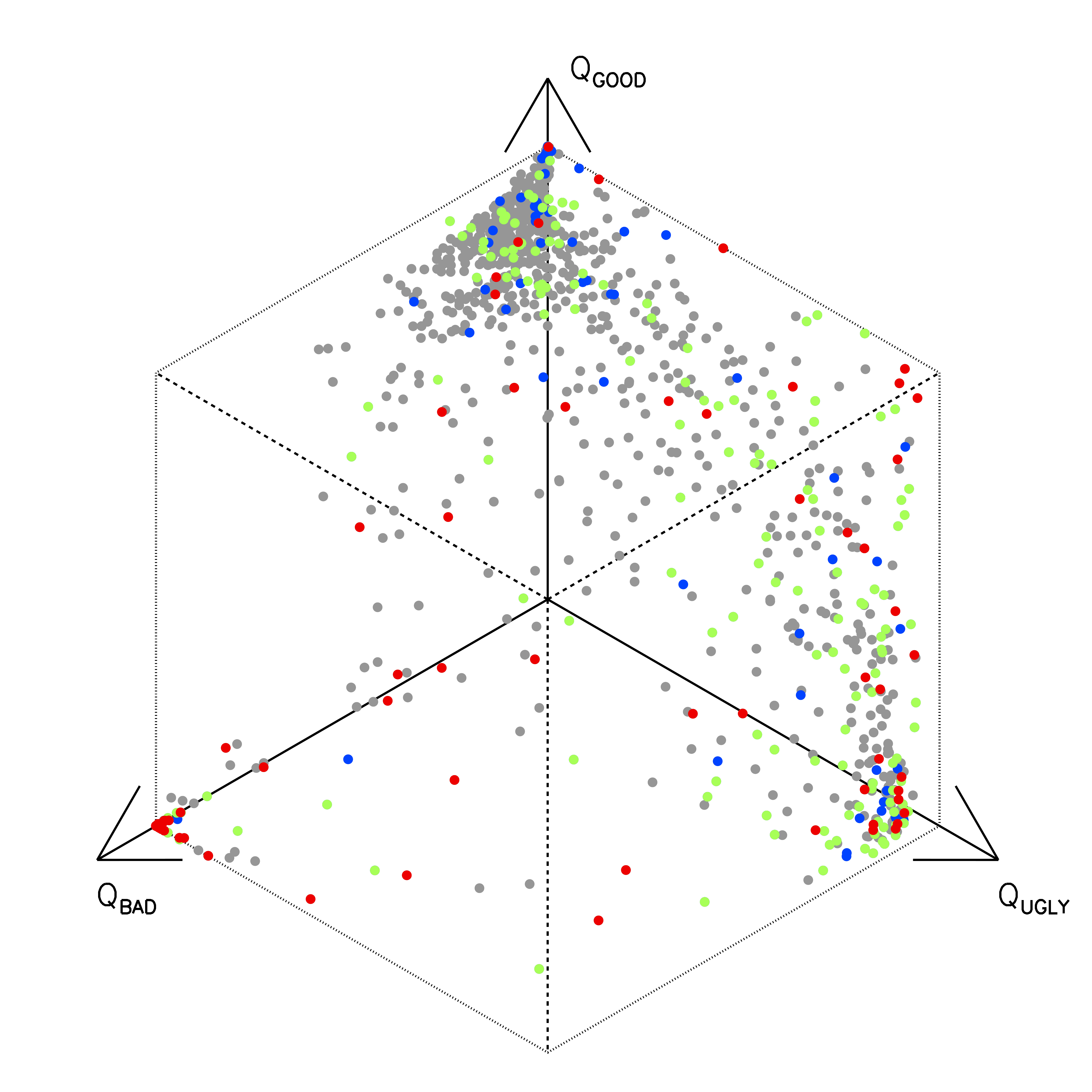}
\includegraphics[width=8cm]{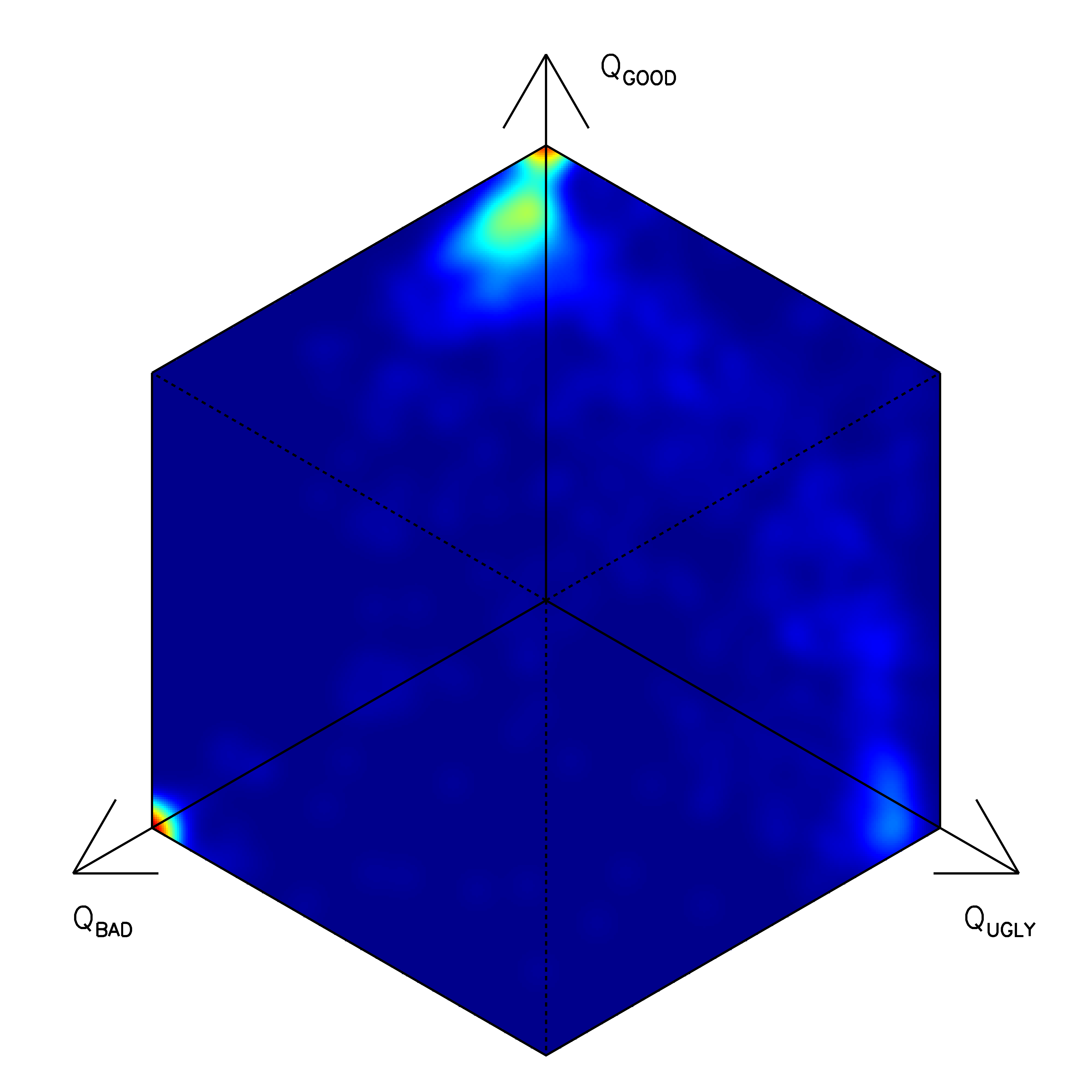}
\caption{Distribution of PSZ1 sources as a function of $Q_{\rm good}$,
  $Q_{\rm bad}$, and $Q_{\rm ugly}$. Top panel : In grey for confirmed
  clusters, in blue for class 1 sources, in green for class 2 sources,
  and in red for class 3 sources. Bottom panel : density of sources as
  a function of $Q_{\rm good}$, $Q_{\rm bad}$, and $Q_{\rm ugly}$.}
\label{NNET_FNAC}
\end{figure}

In Fig.~\ref{NNET_FNAC}, we present the distribution of PSZ1 sources
as a function of $Q_{\rm good}$, $Q_{\rm bad}$, and $Q_{\rm
  ugly}$. The abscissa, $x$, and ordinates, $y$, of each source is
given by,
\begin{align}
x &=  \frac{1}{2}\left( Q_{\rm ugly} - Q_{\rm bad} \right), \nonumber \\
y &=  Q_{\rm good} - \frac{\sqrt{3}}{2} \left( Q_{\rm ugly} + Q_{\rm bad} \right),
\end{align}
We observe clearly three populations of sources, associated with the
{\it good}, {\it bad}, and {\it ugly} noisy categories. The PSZ1
sample is dominated by the {\it good} sources, it also contains about
10\% of {\it bad} sources and a number of {\it ugly} sources that have
low signal-to-noise from the aperture photometry. We observe a clear
separation between the {\it bad} sources and the other ones.These
representation illustrates the efficiency of the neural network to
separate reject {\it bad} most likely spurious SZ sources. \\ The same
representation of $Q_{\rm good}$, $Q_{\rm bad}$, and $Q_{\rm ugly}$
for other samples considered in this study can be found in
Appendix.~\ref{nneteff}. \\

\begin{figure}[!th]
\center
\includegraphics[width=8cm]{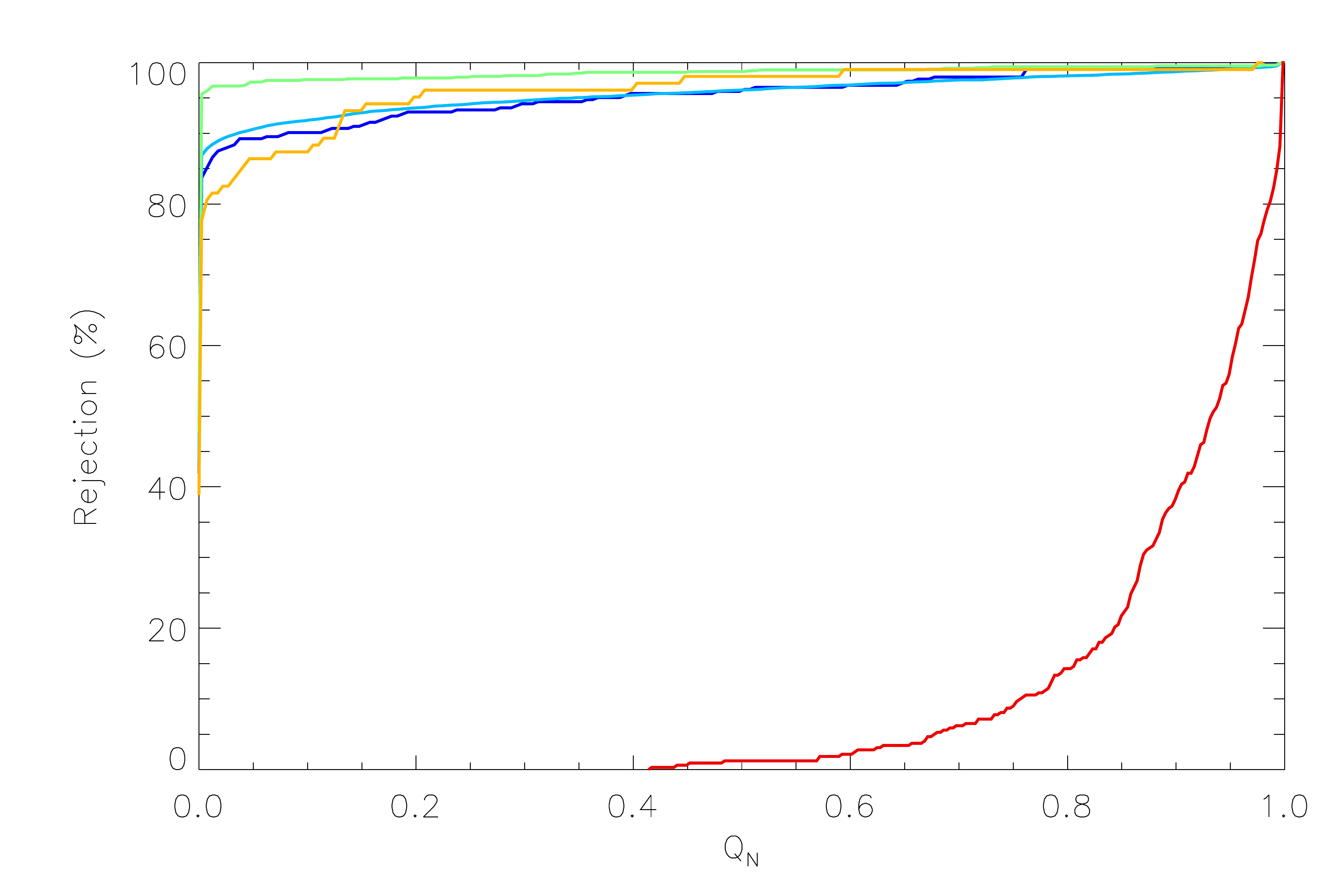}
\caption{For the PSZ1 checking-set sample: Fraction of rejected
  sources as a function of the quality factor $Q_{\rm N}$ cut. In red and
  orange are the true clusters and false detections respectively. We
  also display in blue, cyan and green the radio, IR and CG sources.}
\label{rej_nnet_test}
\end{figure}

We now use the ANN results to find a quantitative way to identify the
{\it bad} sources from the catalogue. We thus define a quality factor
of the SZ detection as, $Q_{\rm N}=1-Q_{\rm bad}$. We display, in
Fig. \ref{rej_nnet_test} and for the PSZ1 checking-set sample, the
fraction of rejected sources as a function of the quality factor $Q_{\rm N}$
cut. In red and orange are the true clusters and false detections
respectively. In blue, cyan and green are the radio, IR and CG
sources.  We see that a cut at $Q_{\rm N}=0.4$ ensures that we remove 95\%
of the {\it bad} sources without affecting the true cluster
distribution. Such a cut allows us to reject 90\% of the IR at 353~GHz
and radio at 30~GHz sources and more that 95\% of the CG sources. The
cut in $Q_{\rm N}$ translates into a $Q_{\rm bad}=0.6$ which marks the
boundary of the $Q_{\rm bad}$ distribution in
Fig. \ref{QUAL_NNET_BAD}.

\begin{figure}[!th]
\center
\includegraphics[width=8cm]{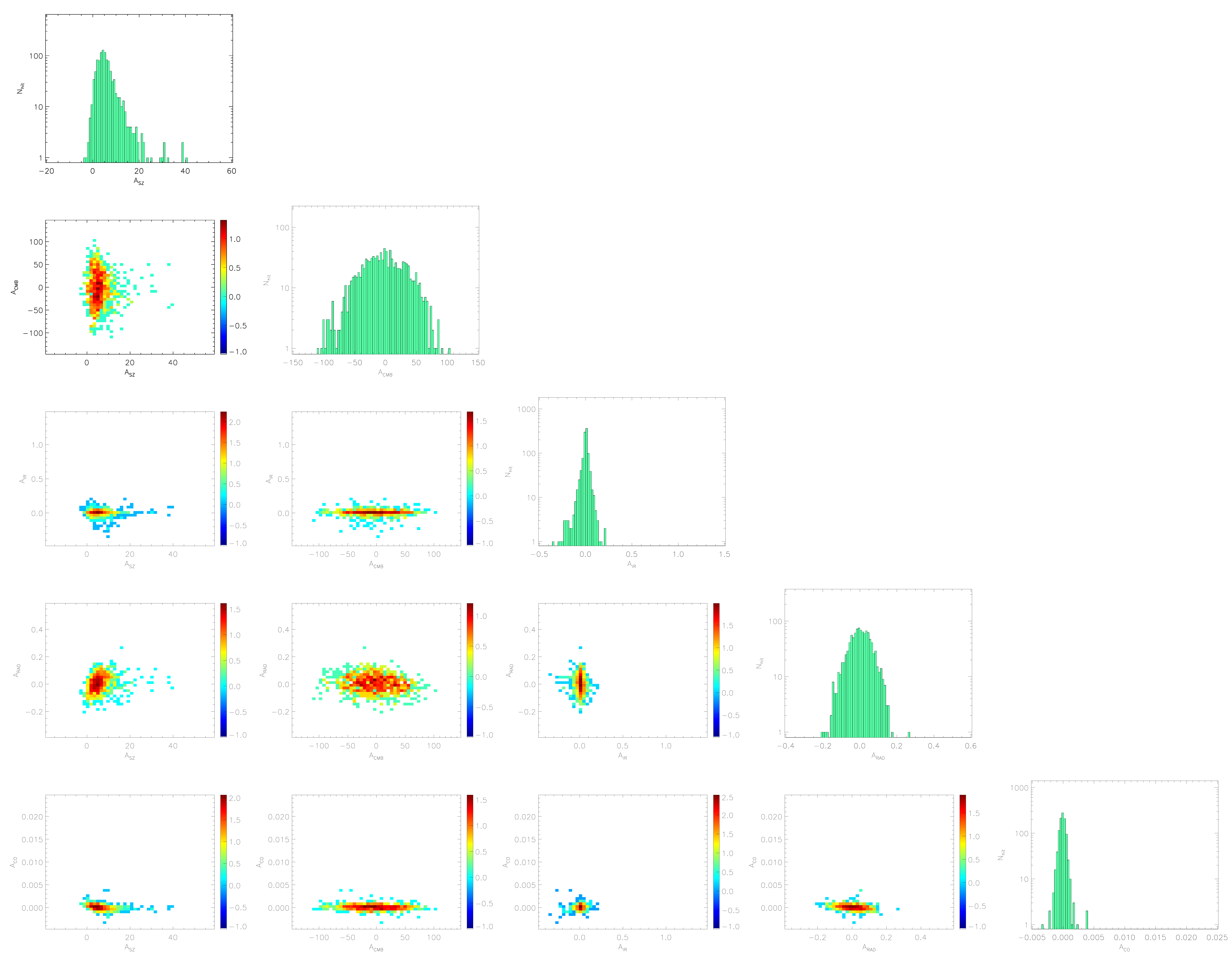}
\caption{Same as Fig. \ref{mcxc} for PSZ1 sources with the ANN-based
  quality factor $Q_{\rm N}>0.4$.}
\label{distrib_ok}
\end{figure}

\begin{figure}[!th]
\center
\includegraphics[width=8cm]{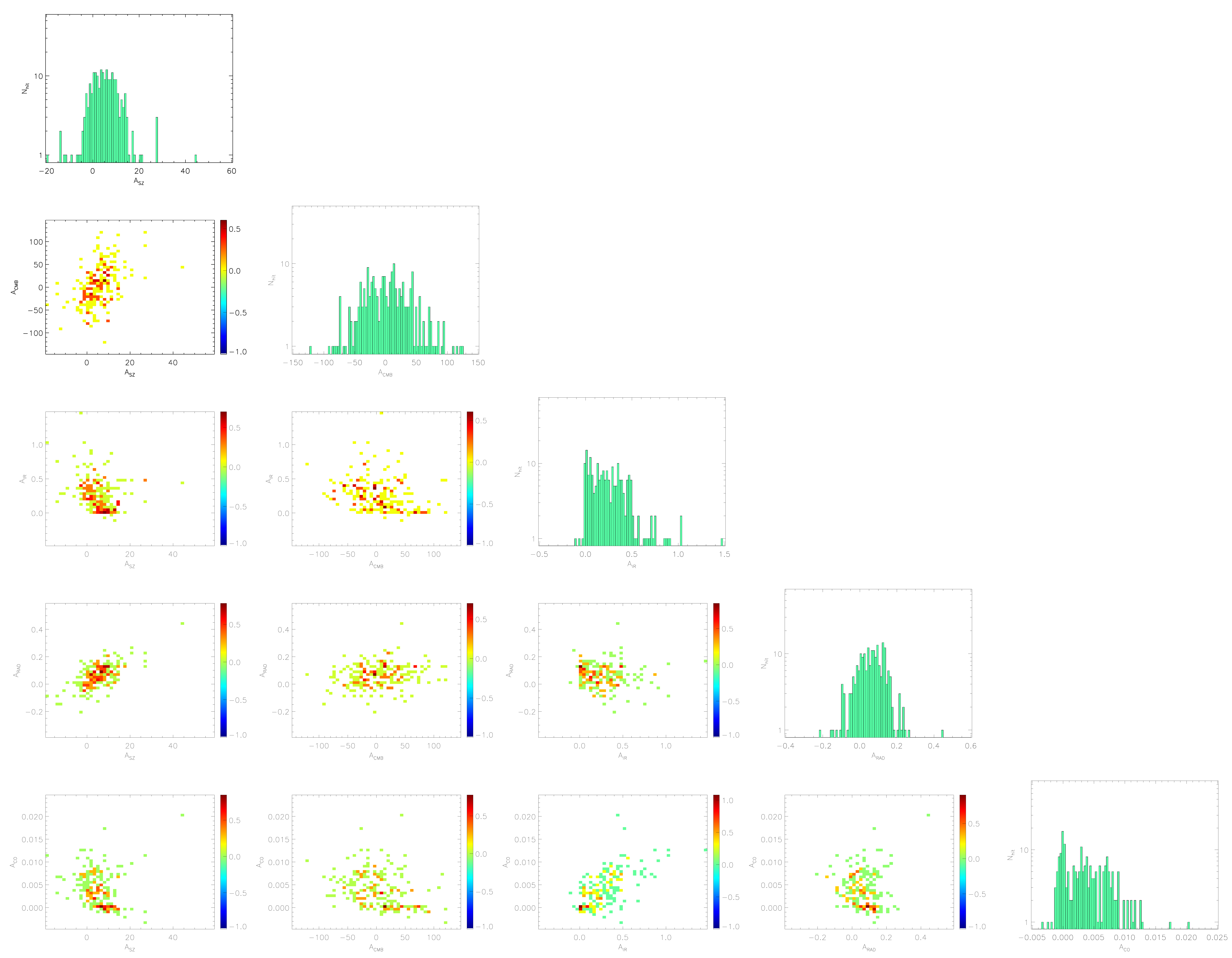}
\caption{Same as Fig. \ref{mcxc} for PSZ1 sources with the ANN-based
  quality factor $Q_{\rm N}<0.4$.}
\label{distrib_nok}
\end{figure}

We show the distributions of $A_{\rm SZ}$, $A_{\rm CMB}$, $A_{\rm
  IR}$, $A_{\rm RAD}$, and $A_{\rm CO}$ in Figs.~\ref{distrib_ok} and
\ref{distrib_nok} for the PSZ1 sources after applying the cut in $Q_{\rm N}$
and for the PSZ1 sources with $Q_{\rm N}<0.4$. We check that the
good-quality sources ($Q_{\rm N}>0.4$) do not show significant contamination
by IR, radio or CO emissions, while for the sources with $Q_{\rm N}<0.4$, we
observe a clear contamination, especially the IR-CO plane.

\section{Discussion} \label{disc}

\begin{figure}[!th]
\center
\includegraphics[width=8cm]{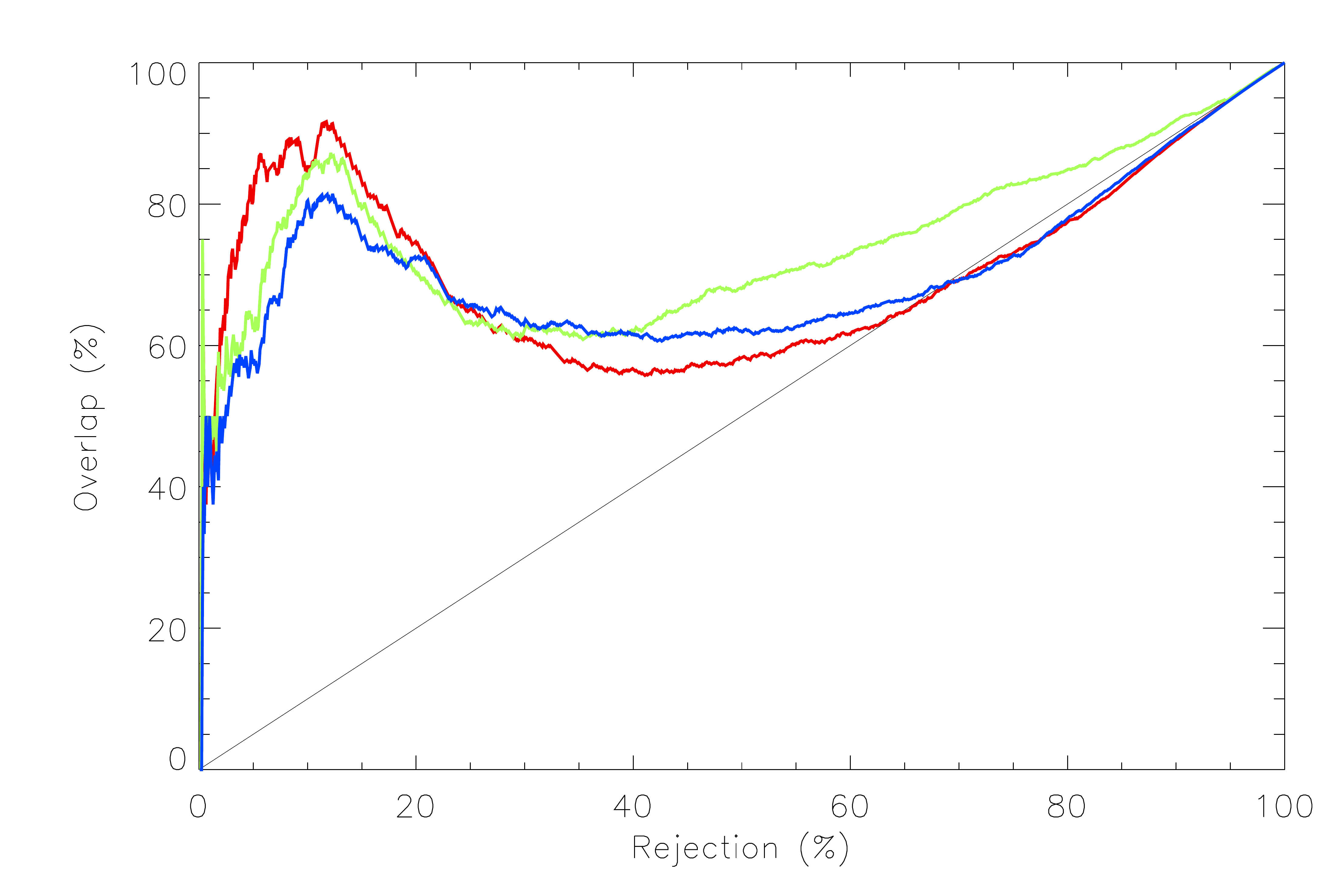}
\caption{Overlap between $Q_{\rm N}$ and $Q_{\rm L}$ in red, $Q_{\rm
    N}$ and $Q_{\rm P}$ in green, and $Q_{\rm L}$ and $Q_{\rm P}$ in
  blue as a function of the rejection percentage for the PSZ1
  catalogue. In black is shown the expected overlap between
  uncorrelated variables.}
\label{sz_overlap_psz}
\end{figure}

We compare the efficiency of the different quality factors to
distinguish between high-quality and low-quality SZ detections. We
first illustrate this comparison by plotting in
Fig. \ref{sz_overlap_psz} the fraction of overlap between $Q_{\rm N}$,
$Q_{\rm L}$, and $Q_{\rm P}$ as a function of the rejection percentage
for the PSZ1 catalogue. We observe that the best agreement between
$Q_{\rm N}$ and $Q_{\rm P}$ is obtained with an overlap of 91\% for a
rejection of 12\%. For higher rejection percentage, we observe that
the overlap decreases.

We also examine the distribution of the PSZ1 sources as a function of
the quality factors $Q_{\rm N}$, $Q_{\rm L}$, and the penalty $Q_{\rm
  P}$ and we show the 2-D scatter plots in the quality-factor
planes. On the one hand, we see that the cuts in $Q_{\rm N}$ and
$Q_{\rm L}$ nicely separate the population of high- and low-quality SZ
sources, with the ANN-based quality assessment seeming more efficient
at identify the {\it bad} sources. Moreover the two cuts preserve the
confirmed clusters as only less that 2\% of these fall in the category
of low-quality sources. We have checked the status of the 22
confirmed clusters that are excluded by the combination of $Q_{\rm N}$ and
$Q_{\rm L}$ cuts.  We find that they are located mostly between
$-30^{o} < b < 30^{o}$, and contaminated by IR, Radio point sources,
or CO and thermal dust emission. 

\begin{figure}[!th]
\center
\includegraphics[width=8cm]{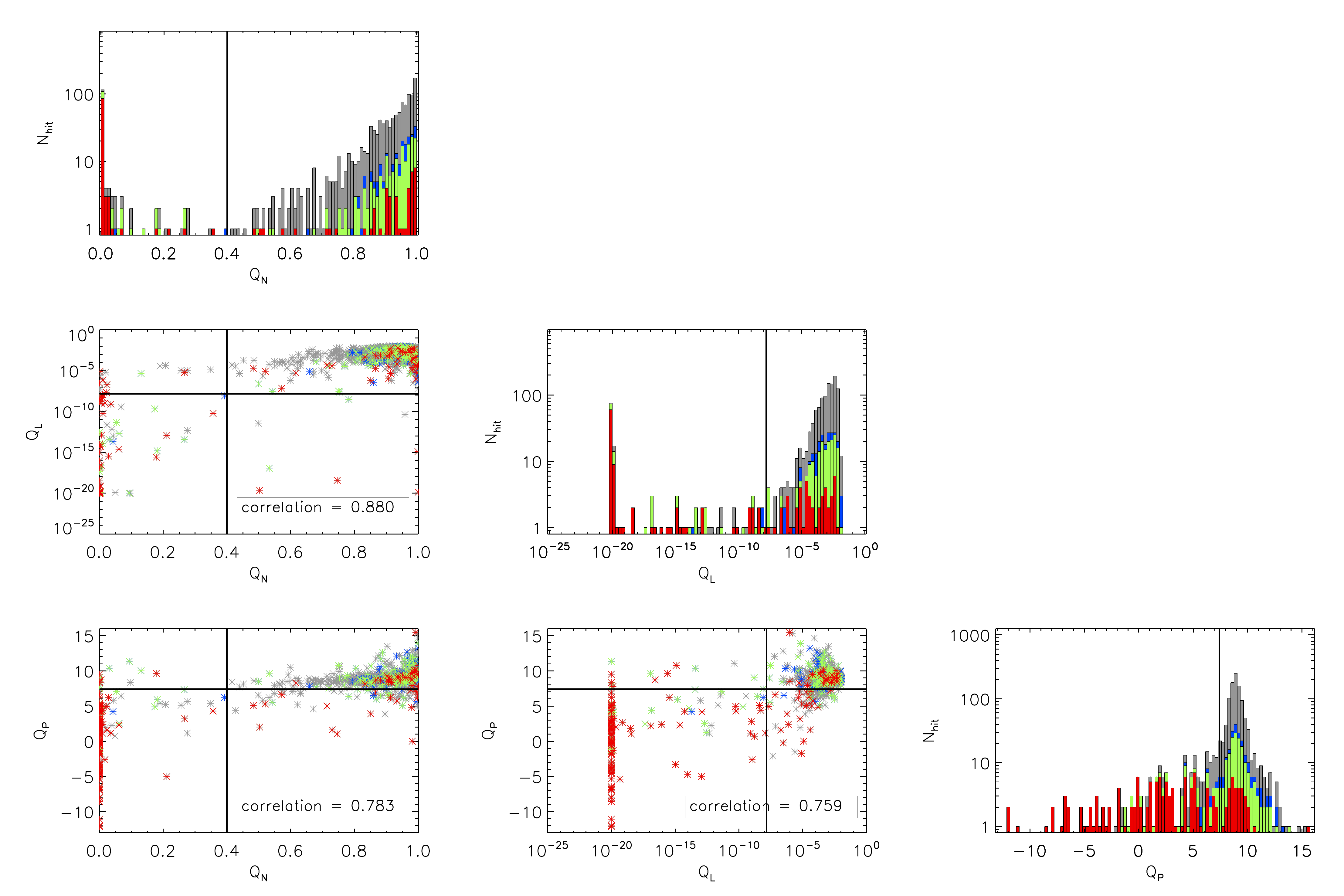}
\caption{Piled-up distribution of the quality factors $Q_{\rm N}$, $Q_{\rm L}$,
  and $Q_{\rm P}$ for the PSZ1 sources (grey: confirmed clusters,
  blue: class1, green: class 2 , red: class 3). The vertical solid
  line represents the cut separating the population of high- and
  low-quality detections. The 2-D scatter plots show the cuts for the
  pair of quality factors under consideration. }
\label{sz_compa_qual_psz}
\end{figure}

\begin{figure}[!th]
\center
\includegraphics[width=8cm]{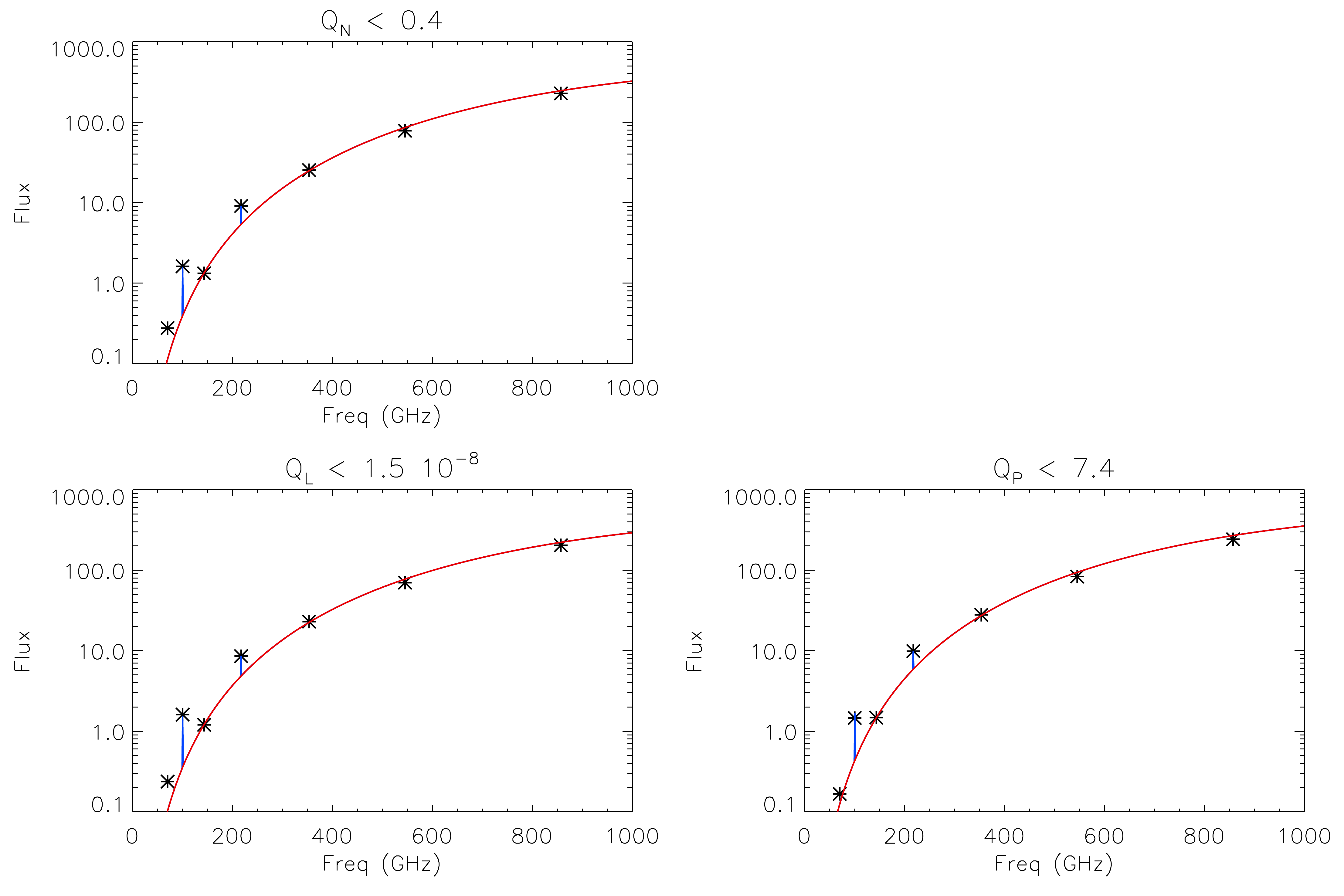}
\caption{Average SED for {\it bad} PSZ1 sources, i.e. with $Q_{\rm
    N}<0.4$, $Q_{\rm L}<1.5\,10^{-8}$ and $Q_{\rm P}<7.4$
  respectively. In red is the best fit for an infra-red SED and in
  blue the contribution from CO rotational lines.}
\label{sz_sed_bad_psz}
\end{figure}

\begin{figure}[!th]
\center
\includegraphics[width=8cm]{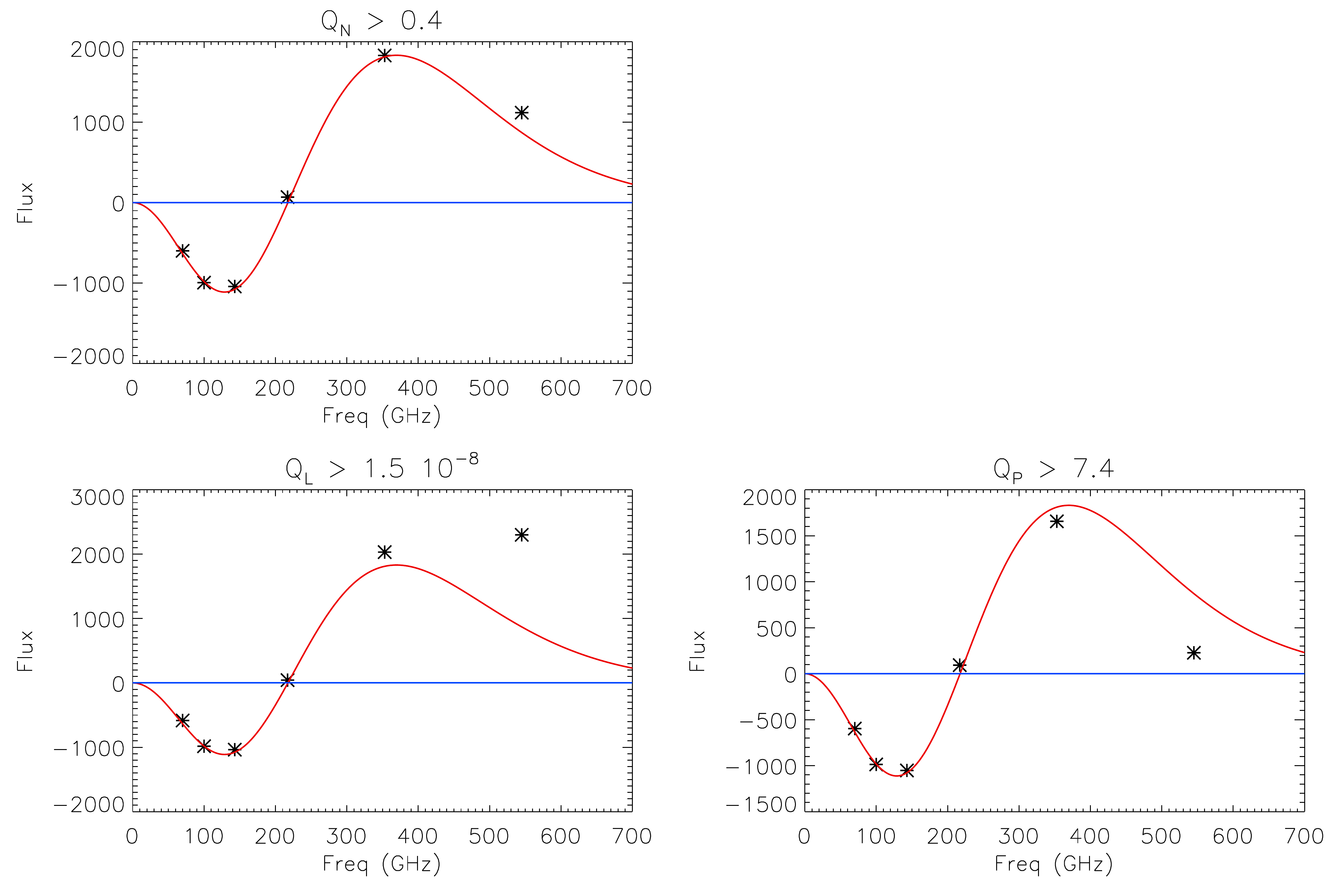}
\caption{Average SED for {\it good} PSZ1 sources, i.e. with $Q_{\rm
    N}>0.4$, $Q_{\rm L}>1.5\,10^{-8}$ and $Q_{\rm P}>7.4$
  respectively. In red is the expected SED for tSZ effect.}
\label{sz_sed_good_szcand}
\end{figure}

Finally, we check the effect of the classification in high- and
low-quality sources through the average SED of the {\it bad} and {\it
  good} sources defined according to the cuts in $Q_{\rm N}$, $Q_{\rm
  L}$, and $Q_{\rm P}$. For the latter we apply a cut at 7.4 which
exclude a few tens of confirmed clusters. A smaller cut would increase
the contamination at high frequencies but reduce the number of
excluded clusters of galaxies. The SED are displayed in
Figs. \ref{sz_sed_bad_psz} and \ref{sz_sed_good_szcand}. We show in
Fig. \ref{sz_sed_bad_psz} that the SED is in perfect agreement with a
dust-like SED. We also see the contamination from CO at 100 and
217~GHz. By contrast, we see in Fig. \ref{sz_sed_good_szcand} that the
average SED compatible with that of the SZ emission. Again the quality
assessment of {\it good} sources from the ANN analysis shows a better
performance as traced by the low contamination level of the SED at the
highest frequencies as compared with the likelihood-based quality
factor.

\begin{figure}[!th]
\center
\includegraphics[width=8cm]{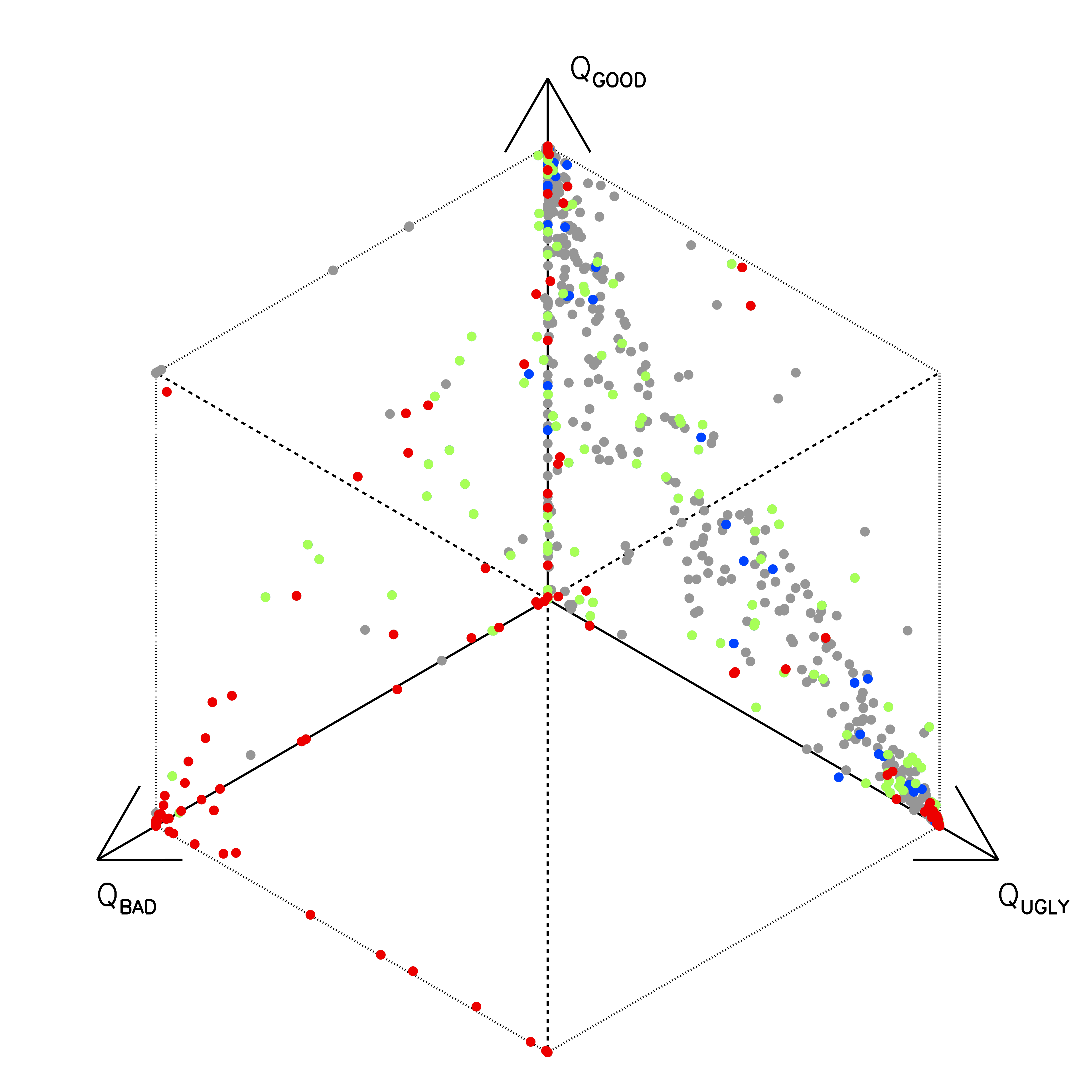}
\includegraphics[width=8cm]{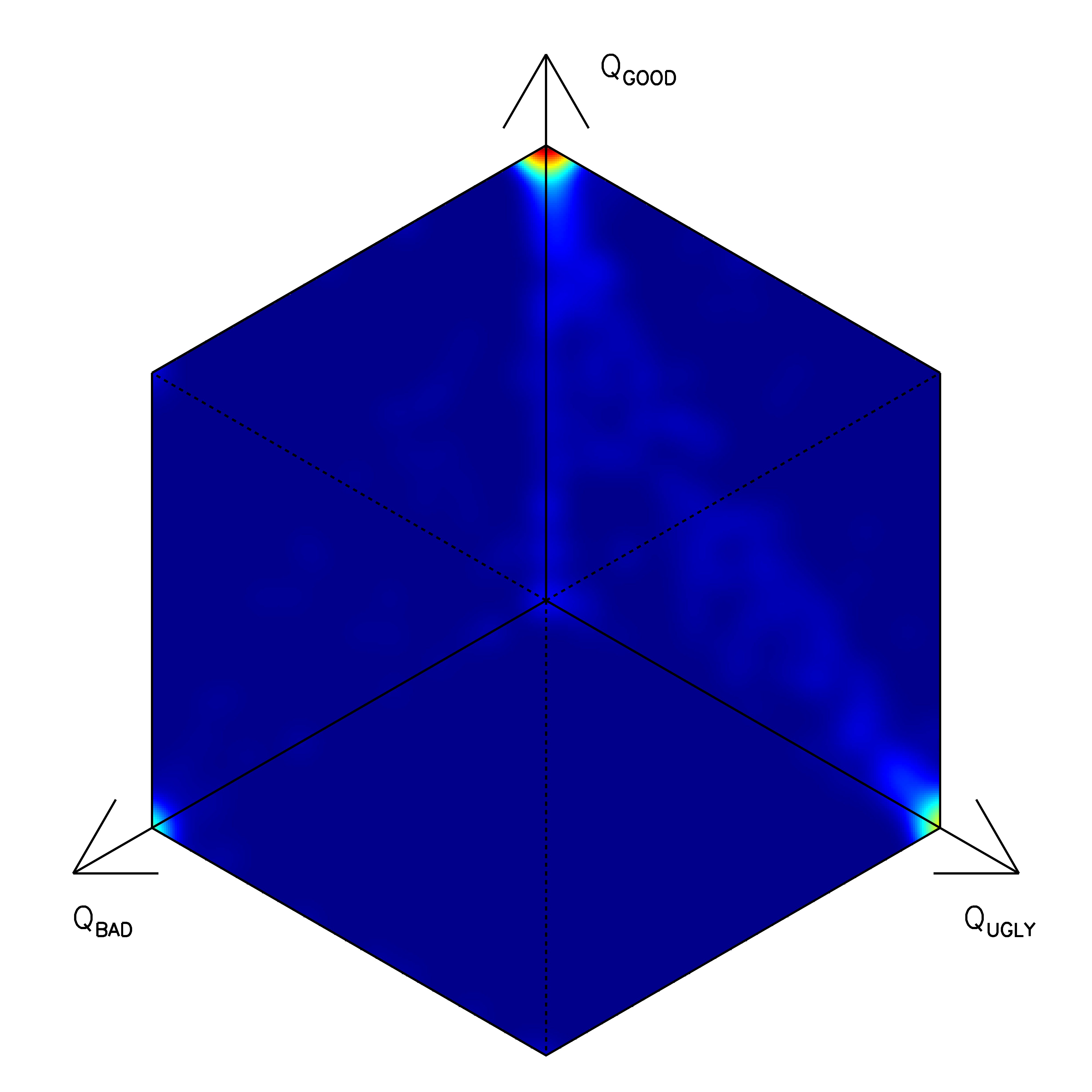}
\caption{{\it Top panel:} Distribution of the PSZ1 sources as a
  function of $Q_{\rm good}$, $Q_{\rm bad}$, and $Q_{\rm ugly}$. In
  grey for confirmed clusters, in blue for class 1 sources, in green
  for class 2 sources, and in red for class 3 sources. {\it Bottom
    panel:} Density of sources for the PSZ1 catalogue as a function of
  $Q_{\rm good}$, $Q_{\rm bad}$, and $Q_{\rm ugly}$.}
\label{NNET_FNAC_TPSZ}
\end{figure}

The classification from the ANN seems to give better results than the
other methods. This is expected from supervised methods where the
training is performed on pre-defined class memberships. We tested the
ANN in a case where no pre-definition of the {\it bad} class is
given. Namely, we used the random positions as the {\it ugly} class
and we trained the network on a sub-sample constructed from the PSZ1
catalogue itself. The results, applied on the checking set, i.e. the
other subsample from PSZ1, are shown in Fig. \ref{NNET_FNAC_TPSZ}. We
note that, without a training on the class of {\it bad} candidates,
the ANN fails at separating efficiently this type of sources. The
network separate the PSZ1 mostly into a population of {\it ugly},
i.e. noisy, and {\it good}. We also not that there is an ensemble of
sources (middle of the lower panel, Fig. \ref{NNET_FNAC_TPSZ}) for
which the network is unable to set a class. \\
Although, the performance of the ANN is decreased as compared to a 
case where the classes are predefined we nevertheless note that this 
method gives very satisfactory results.

Finally, we have investigated the ANN-based quality factors for SZ
sources detected in \Planck\ that proved to unambiguously be false candidates by follow-up in
X-rays based on Director's Discretionary Time on the \xmm\ observatory
\citep{planck2011-5.1b,planck2012-I,planck2012-IV}. No
significant extended X-ray emission was associated with eight SZ
detections in \Planck\ data: PLCK G321.410+19.941, PLCK
G355.247-61.038, PLCK G93.139-19.040, PLCK G320.145-53.631, PLCK
G10.161-11.706, PLCK G201.148-35.245, PLCK G34.92-19.263, PLCK
G120.218+11.093.  We find that all SZ detections have very high
$Q_{\rm ugly}$ factor (0.7 to 1) except the first PLCK G321.410+19.941
which has a slightly smaller value of $Q_{\rm ugly}$ but has $Q_{\rm
  ugly}=0.4$, i.e. is identified as a spurious detection. The quality
factor $Q_{\rm good}$ for all but two detections are below 0.01. Only
PLCK G201.148-35.245 and PLCK G34.92-19.263 have $Q_{\rm good}\sim0.2$
and 0.6 respectively but they both are in the class of noisy {\it
  ugly} sources. \\
The a posteriori quality assessment of the confirmed false SZ 
sources in \Planck\ shows that these spurious detections were mostly 
related to noise fluctuations.


\section{Summary and Conclusions}\label{ccl}

We have addressed the question of classification of population
illustrating the approaches used on the catalogue of SZ sources
detected by Planck. To do so we build an SED model including all
the major sources of signal in the range of frequencies considered for
the dataset. This projection of the data onto an SED has allowed us to
reduce the dimensionality of the problem and to resort to statistical
classification techniques.

We explore three techniques clustering, an unsupervised machine
learning, artificial neural networks, a supervised machine learning,
and likelihood, half way between supervised and unsupervised. Each of
the three methods outputs quantitative quality factors to the SZ
sources. The classification techniques separate statistically the
sources into populations of different quality and reliability.

We apply the techniques to cluster catalogues detected in the X-rays
and in the optical and to catalogues of point sources. Each time, the
statistical classification was able to separate the cases of bona fide
and the cases of sources that are not clusters.  We then applied our
methods to the PSZ1 catalogue. All three classification results
agree. They reproduce rather well the distribution of the PSZ1 sources
in confirmed clusters and class 1, 2, 3 candidates. For each
technique, most of the class 3 objects are put in the least reliable
population.

We show that although all methods agree, the supervised neural
network-based classification shows better performances than the
likelihood approach or the unsupervised clustering method.  This is
exhibited by the clean average SED of the sources in the {\it good}
population. The higher performance is expected since the supervised
methods utilise more information. The performance is even better when
we have an a priori definition of class membership. The classification
then serves to determine whether or not the pre-defined populations
are distinguishable.  We show, however, that the ANN can detect
differences between classes of populations even when the training is
not performed on pre-defined populations.

Finally, we suggest, on the basis of our results that a supervised
learning approach should be the method of choice when classifying
individuals into pre-defined populations. These classification methods
applied in the present study to the assess the quality factor of SZ
detections and separate the populations can easily be adapted and
generalised to other contexts such as the detection of galaxy
clusters, and more generally sources, in the X-rays in SRG/eROSITA and
in {\it Euclid}. An adaptation is ongoing.

\begin{acknowledgements}
 The authors thank A. Beelen, B. Bertincourt, O. Forni, and B. Roukema for useful remarks. We acknowledge the support of the French \emph{Agence Nationale de la Recherche} under grant ANR-11-BD56-015. 
The development of Planck has been supported by: ESA; CNES and CNRS/INSU-IN2P3-INP (France); ASI, CNR, and INAF (Italy); NASA and DoE (USA); STFC and UKSA (UK); CSIC, MICINN and JA (Spain); Tekes, AoF and CSC (Finland); DLR and MPG (Germany); CSA (Canada); DTU Space (Denmark); SER/SSO (Switzerland); RCN (Norway); SFI (Ireland); FCT/MCTES (Portugal); and The development of Planck has been supported by: ESA; CNES and CNRS/INSU-IN2P3-INP (France); ASI, CNR, and INAF (Italy); NASA and DoE (USA); STFC and UKSA (UK); CSIC, MICINN and JA (Spain); Tekes, AoF and CSC (Finland); DLR and MPG (Germany); CSA (Canada); DTU Space (Denmark); SER/SSO (Switzerland); RCN (Norway); SFI (Ireland); FCT/MCTES (Portugal); and PRACE (EU).
\end{acknowledgements}

\bibliographystyle{aa}

\bibliography{sz_likelihood_paper}

\appendix

\section{Neural network}
\label{ANNET}

In this section we details the concept of back propagation neural
network we used in the present analysis.\\ The concept of neural
network, consists in multiple layers of neurons. Each neuron is set to
a value ranging from zero to one. The value of neurons, ${\bf
  x}^{(n)}$, in the layer $n$, is fully determined by the value in the
neurons, ${\bf x}^{(n-1)}$, of layer $n-1$ via a linear combination
using weights, ${\cal W}^{(n)}$. This relation reads,
\begin{equation}
{\bf x}^{(n)}_k = g^{(n)}\left( \sum_i {\cal W}^{(n)}_{ki} {\bf x}^{(n-1)}_i \right),
\end{equation}
where $g^{(n)}(x)$ is the activation function of the neuron. In the
following, we do not mention the bias term, as it can be considered as
an extra neuron added at each "input" layer for which the value is
always set to one.\\

To be trained, such neural network needs a set of input neurons, ${\bf
  x}^{(0)}$, for which one the expected value of the neurons of the
output layer, ${\bf y}$, are known.  Using the neural network, it is
possible to estimate the values of ${\bf y}$ from the values ${\bf
  x}^{(0)}$.  We can define the distance of the estimated value, ${\bf
  \widehat y}$, to the known solution ${\bf y}$ as,
\begin{equation}
E = \frac{1}{2}\sum_i ({\bf y}_i - {\bf \widehat y}_i)^2.
\end{equation}
Then, we aim at minimizing $E$ by adjusting the weights, ${\cal
  W}^{(n)}$, of each layer.\\

To do so, we need to compute the derivative of $E$ as a function of a
given weights ${\cal W}^{(l)}_{ab}$ of the layer $l \neq m$,
\begin{align}
\frac{\partial E}{\partial {\cal W}^{(l)}_{ab}} &= -\sum_i ({\bf y}_i - {\bf \widehat y}_i) \frac{\partial {\bf \widehat y}_i}{\partial {\cal W}^{(l)}_{ab}} \nonumber \\
								&= -\sum_i ({\bf y}_i - {\bf \widehat y}_i) g'^{(m)}_i \sum_j {\cal W}^{(m)}_{ij} \frac{\partial  {\bf x}^{(m-1)}_j}{\partial {\cal W}^{(l)}_{ab}},
\end{align}
where $m$ is the total number of layers and $g'^{(m)}_i$ is the
derivative of $g^{(m)}_i$ with respect to $\sum_j {\cal W}^{(n)}_{ij}
{\bf x}^{(n-1)}_j$.  Following the same approach, and considering $m-1
> l$, it is possible to estimate $ \frac{\partial {\bf
    x}^{(m-1)}_j}{{\cal W}^{(l)}_{ab}}$,
\begin{equation}
\frac{\partial  {\bf x}^{(m-1)}_j}{\partial {\cal W}^{(l)}_{ab}} = g'^{(m-1)}_j \sum_k {\cal W}^{(m-1)}_{jk} \frac{\partial  {\bf x}^{(m-2)}_k}{\partial {\cal W}^{(l)}_{ab}}.
\label{eq1}
\end{equation}
We define the error, ${\bf e}^{(m)}_i =({\bf y}_i - {\bf \widehat
  y}_i) g'^{(m)}_i$, and the back propagated error ${\bf e}^{(n-1)}_i
= g'^{(n-1)}_i \sum_j {\cal W}^{(n)}_{ji} {\bf e}^{(n)}_j$.  Then we
have,
\begin{equation}
\frac{\partial E}{\partial {\cal W}^{(l)}_{ab}} = -\sum_j \frac{\partial  {\bf x}^{(m-1)}_j}{\partial {\cal W}^{(l)}_{ab}} \sum_i {\cal W}^{(m)}_{ij} {\bf e}^{(m)}_i.
\label{eq2}
\end{equation}
Using Eq.~\ref{eq1}, Eq.~\ref{eq2}, and the relation between ${\bf e}^{(n-1)}$ and ${\bf e}^{(n)}$ it comes,
\begin{equation}
\frac{\partial E}{\partial {\cal W}^{(l)}_{ab}} = -\sum_j \frac{\partial  {\bf x}^{(m-2)}_j}{\partial {\cal W}^{(l)}_{ab}} \sum_i {\cal W}^{(m-1)}_{ij} {\bf e}^{(m-1)}_i.
\end{equation}
Then, by iteration we can derive,
\begin{equation}
\frac{\partial E}{\partial {\cal W}^{(l)}_{ab}} = -\sum_j \frac{\partial  {\bf x}^{(l)}_j}{\partial {\cal W}^{(l)}_{ab}} \sum_i {\cal W}^{(l+1)}_{ij} {\bf e}^{(l+1)}_i.
\end{equation}
We need to estimate $ \frac{\partial  {\bf x}^{(l)}_j}{\partial {\cal W}^{(l)}_{ab}}$,
\begin{align}
\frac{\partial  {\bf x}^{(l)}_j}{\partial {\cal W}^{(l)}_{ab}} &=  \frac{\partial  g^{(l-1)}\left( \sum_i {\cal W}^{(l)}_{ji} {\bf x}^{(l-1)}_i \right)}{\partial {\cal W}^{(l)}_{ab}}\nonumber \\
										&=  g'^{(l-1)}_j  \sum_i \frac{\partial {\cal W}^{(l)}_{ji}}{\partial {\cal W}^{(l)}_{ab}} {\bf x}^{(l-1)}_i \nonumber \\
										&=  g'^{(l-1)}_j  \sum_i \delta_{ja} \delta_{ib} {\bf x}^{(l-1)}_i \nonumber \\
										&= g'^{(l-1)}_j  \delta_{ja} {\bf x}^{(l-1)}_b.
\end{align}
Indeed, ${\bf x}^{(l-1)}$ do not depends on ${\cal W}^{(l)}$, as ${\bf x}^{(l-1)}$ is the input layer for weights ${\cal W}^{(l)}$.
Finally, we derive,
\begin{align}
\frac{\partial E}{\partial {\cal W}^{(l)}_{ab}} &= -\sum_j g'^{(l-1)}_j  \delta_{ja} {\bf x}^{(l-1)}_b \sum_i {\cal W}^{(l+1)}_{ij} {\bf e}^{(l+1)}_i \nonumber \\
								&= -{\bf x}^{(l-1)}_b {\bf e}^{(l)}_a.
\end{align}
As a consequence, the gradient of $E$ with respect to ${\cal W}^{(l)}$
can be directly expressed from the input layer values $ {\bf
  x}^{(l-1)}$ and the back propagated error ${\bf e}^{(l)}$.  Then the
weights of the neural network can be adjusted iteratively through a
gradient descent,
\begin{equation}
 {\cal W}^{(l)}_{ab}(t+1) =  {\cal W}^{(l)}_{ab}(t) +  \alpha {\bf x}^{(l-1)}_b {\bf e}^{(l)}_a + \mu \left( {\cal W}^{(l)}_{ab}(t)  - {\cal W}^{(l)}_{ab}(t-1)  \right),
\end{equation}
where $\alpha$ is the learning rate and $\mu$ the momentum, both set
to values ranging from 0 to 1.\\ Low values for $\alpha$ avoid for
oscillations towards the minimum of $E$. High values for $\mu$ avoid
for local minima stabilization. However, extremely low value for both
parameters can slow down the speed of training of the network.

\section{Neural network results for various sample of sources}

\label{nneteff}

\begin{figure*}[!th]
\center
\includegraphics[width=8cm]{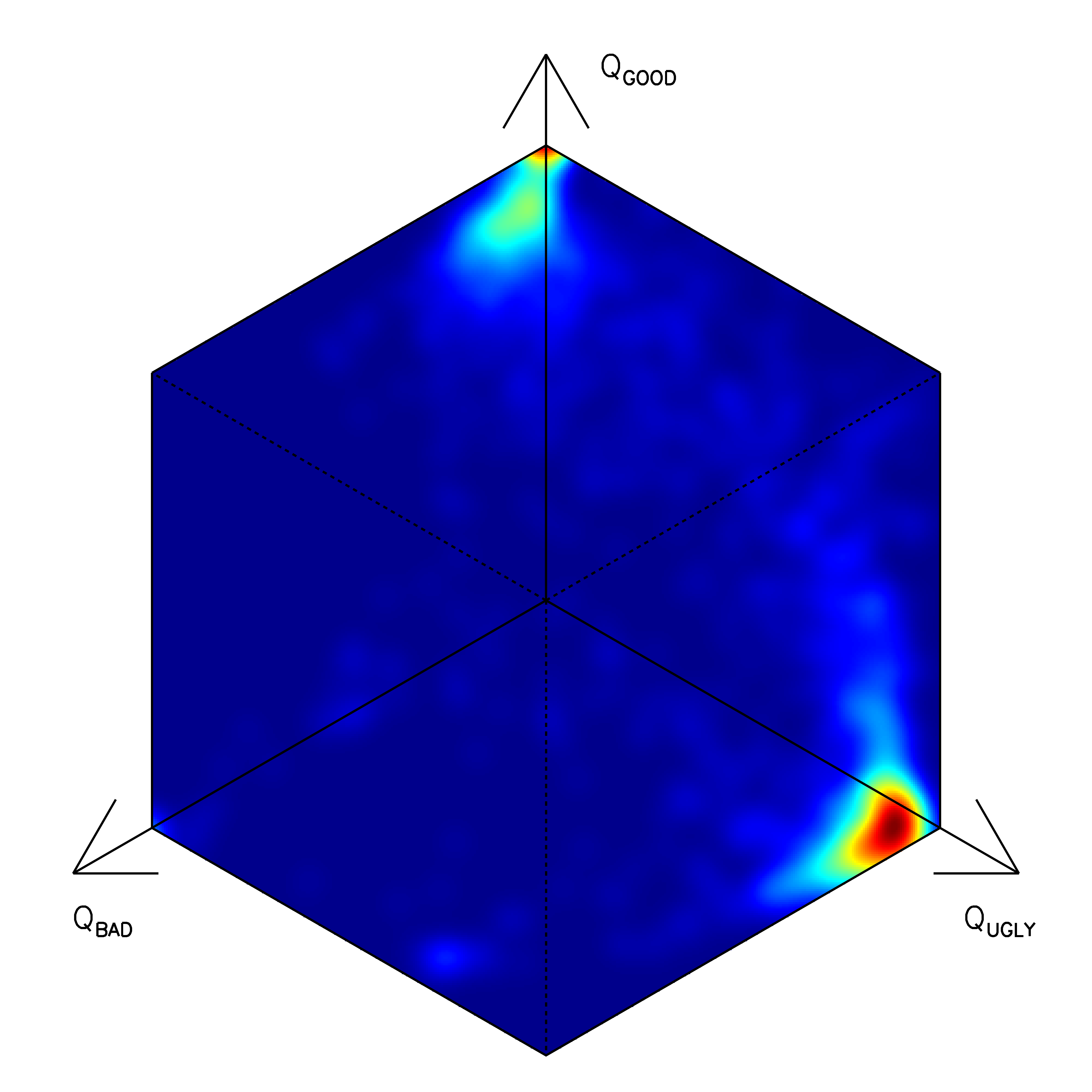}
\includegraphics[width=8cm]{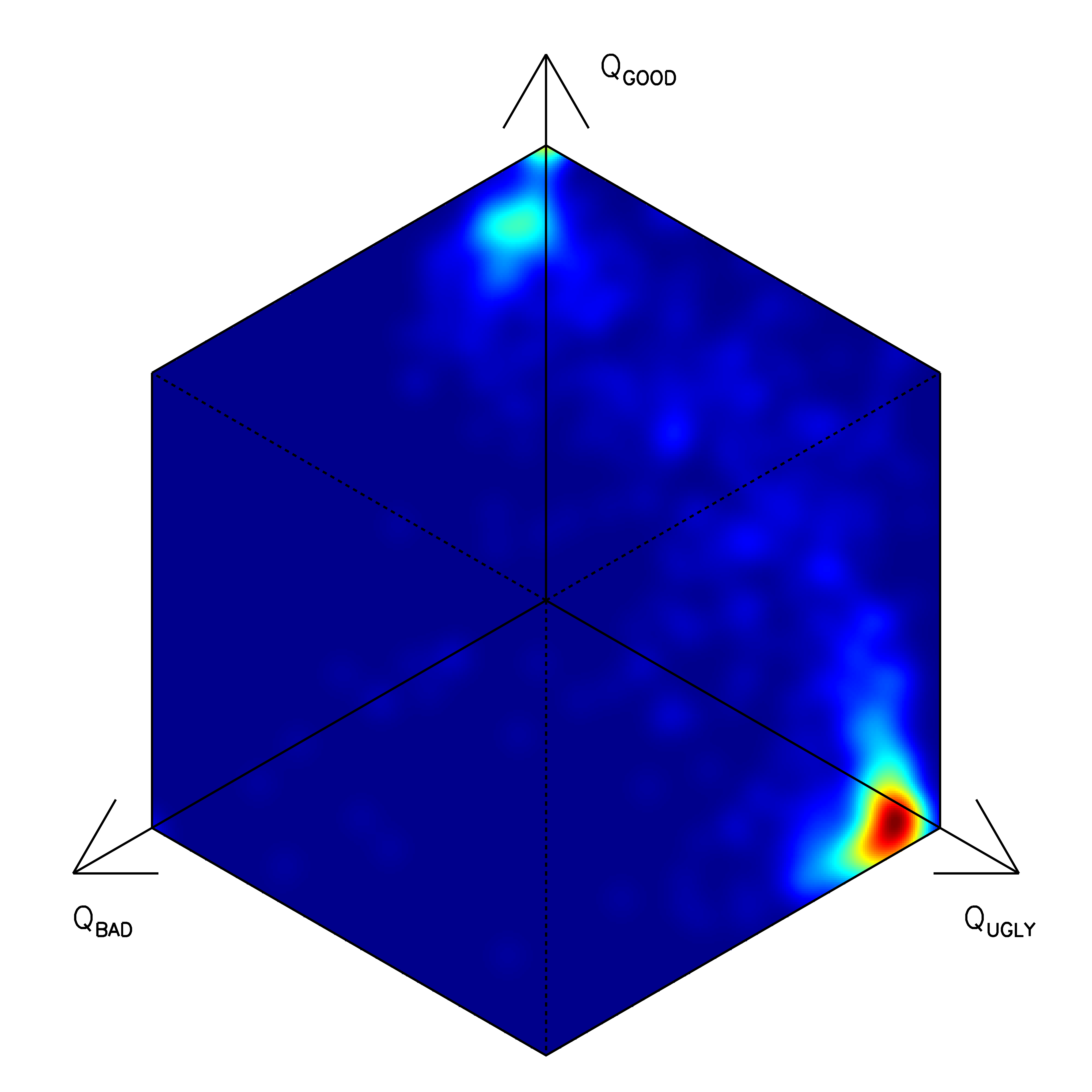}
\includegraphics[width=8cm]{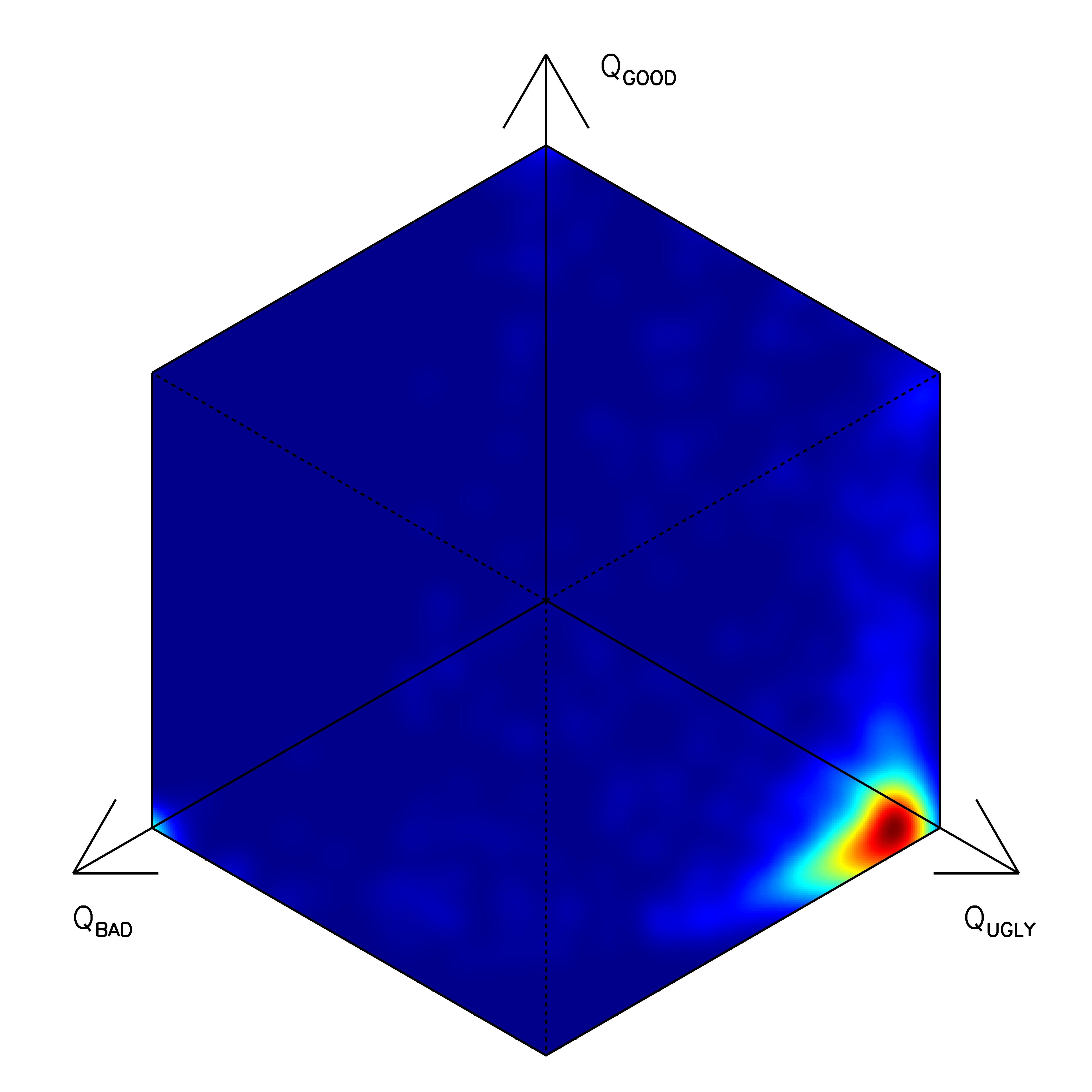}
\includegraphics[width=8cm]{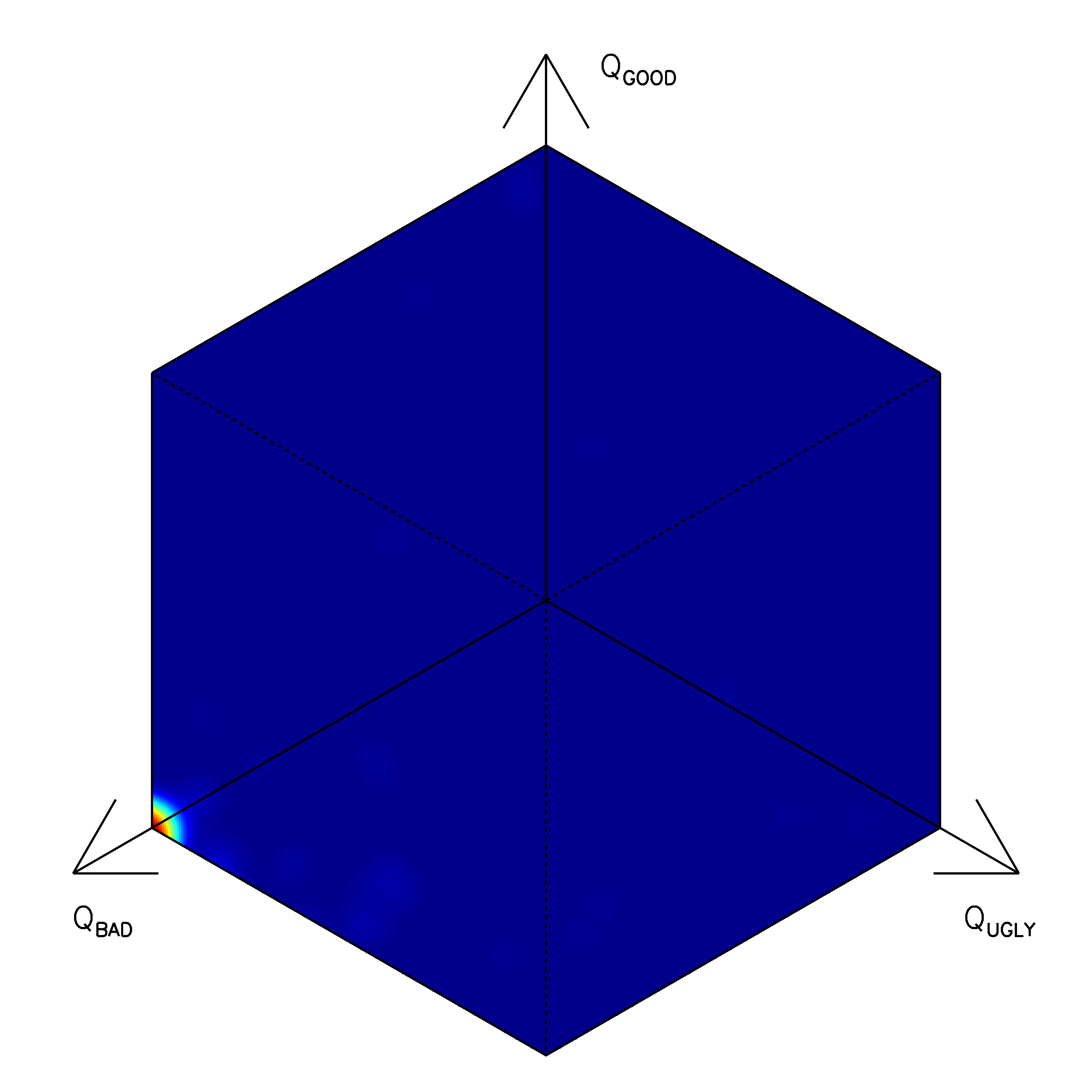}
\includegraphics[width=8cm]{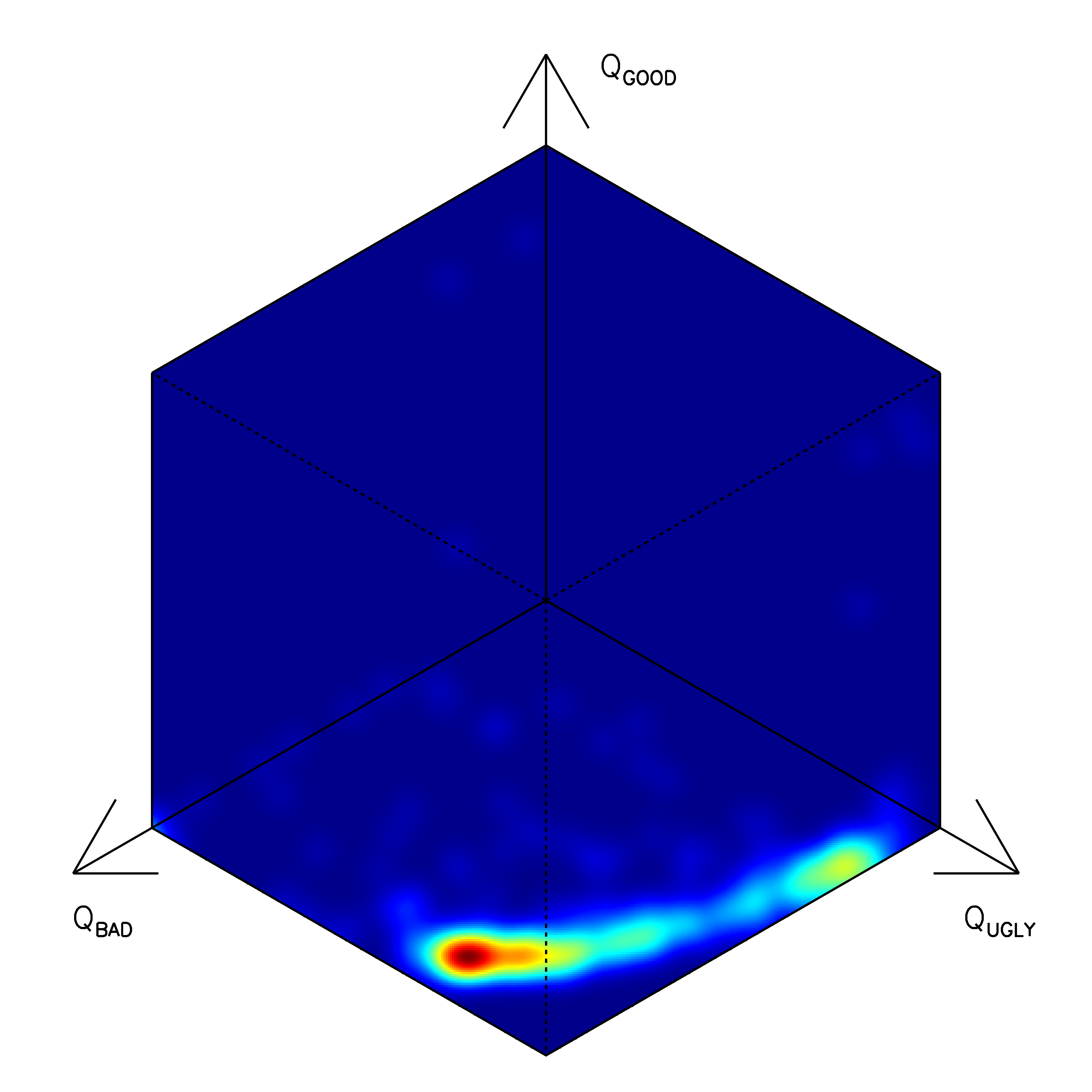}
\includegraphics[width=8cm]{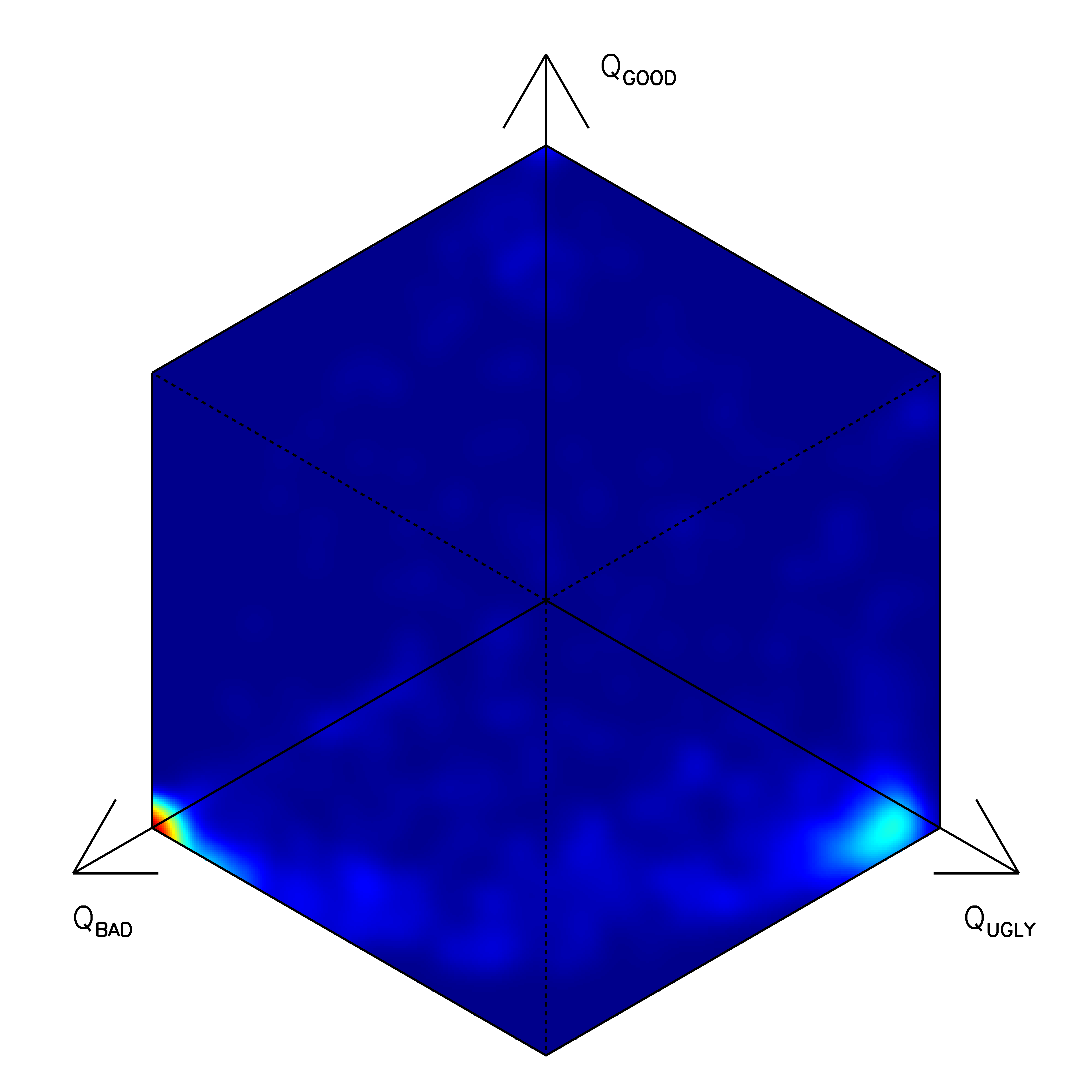}
\caption{From left to right and top to bottom : Density of sources as
  a function of $Q_{\rm good}$, $Q_{\rm bad}$, and $Q_{\rm ugly}$ for
  MCXC, SDSS, random, bad, PCCS at 30 and 353~GHz sources
  respectively.}
\label{NNET_FNAC_ALL}
\end{figure*}

In this section we present the distributions from various samples as a
function of $Q_{\rm good}$, $Q_{\rm bad}$, and $Q_{\rm ugly}$. We test
our neural network on the MCXC catalogue, the catalogue of clusters
from \citet{wen12} based on SDSS data, a set of 2000 random positions
over the sky, a set of bad detections and PCCS sources at 30 and
353~GHz. \\ We observe, in Fig.~\ref{NNET_FNAC_ALL}, that galaxy
clusters are flagged as {\it good} quality sources or {\it ugly}
sources in the case of low signal-to-noise ratio for the tSZ emission.
We observe that random positions over the sky are effectively
classified as {\it ugly}.  We observe that the false detections are
flagged as {\it bad} quality sources.  We note that PCCS sources are
flagged as {\it bad} quality sources or as {\it ugly} for low signal
to noise sources.\\ All these tests demonstrate that the neural
network-based quality assessment is able to accurately separate
classes of sources for which we have a significant signal-to-noise
ratio.

\raggedright
\end{document}

%% file: Planck.tex
\def\setsymbol#1#2{\expandafter\def\csname #1\endcsname{#2}}
\def\getsymbol#1{\csname #1\endcsname}

\def\Planck{\textit{Planck}}

\newbox\tablebox    \newdimen\tablewidth
\def\leaderfil{\leaders\hbox to 5pt{\hss.\hss}\hfil}
\def\tablenote#1 #2\par{\begingroup \parindent=0.8em
    \abovedisplayshortskip=0pt\belowdisplayshortskip=0pt
    \noindent
    $$\hss\vbox{\hsize\tablewidth \hangindent=\parindent \hangafter=1 \noindent
    \hbox to \parindent{$^#1$\hss}\strut#2\strut\par}\hss$$
    \endgroup}

\def\L2{\ifmmode L_2\else $L_2$\fi}

\def\DeltaT{\ifmmode \Delta T\else $\Delta T$\fi}
\def\deltat{\ifmmode \Delta t\else $\Delta t$\fi}
\def\fknee{\ifmmode f_{\rm knee}\else $f_{\rm knee}$\fi}
\def\Fmax{\ifmmode F_{\rm max}\else $F_{\rm max}$\fi}
\def\solar{\ifmmode{\rm M}_{\mathord\odot}\else${\rm M}_{\mathord\odot}$\fi}
\def\Msolar{\ifmmode{\rm M}_{\mathord\odot}\else${\rm M}_{\mathord\odot}$\fi}
\def\Lsolar{\ifmmode{\rm L}_{\mathord\odot}\else${\rm L}_{\mathord\odot}$\fi}

\def\inv{\ifmmode^{-1}\else$^{-1}$\fi}
\def\mo{\ifmmode^{-1}\else$^{-1}$\fi}
\def\sup#1{\ifmmode ^{\rm #1}\else $^{\rm #1}$\fi}
\def\expo#1{\ifmmode \times 10^{#1}\else $\times 10^{#1}$\fi}
\def\,{\thinspace}
\def\lsim{\mathrel{\raise .4ex\hbox{\rlap{$<$}\lower 1.2ex\hbox{$\sim$}}}}
\def\gsim{\mathrel{\raise .4ex\hbox{\rlap{$>$}\lower 1.2ex\hbox{$\sim$}}}}

\def\simprop{\mathrel{\raise .4ex\hbox{\rlap{$\propto$}\lower 1.2ex\hbox{$\sim$}}}}
\def\deg{\ifmmode^\circ\else$^\circ$\fi}
\def\pdeg{\ifmmode $\setbox0=\hbox{$^{\circ}$}\rlap{\hskip.11\wd0 .}$^{\circ}
          \else \setbox0=\hbox{$^{\circ}$}\rlap{\hskip.11\wd0 .}$^{\circ}$\fi}
\def\arcs{\ifmmode {^{\scriptstyle\prime\prime}}
          \else $^{\scriptstyle\prime\prime}$\fi}
\def\arcm{\ifmmode {^{\scriptstyle\prime}}
          \else $^{\scriptstyle\prime}$\fi}
\newdimen\sa  \newdimen\sb
\def\parcs{\sa=.07em \sb=.03em
     \ifmmode \hbox{\rlap{.}}^{\scriptstyle\prime\kern -\sb\prime}\hbox{\kern -\sa}
     \else \rlap{.}$^{\scriptstyle\prime\kern -\sb\prime}$\kern -\sa\fi}
\def\parcm{\sa=.08em \sb=.03em
     \ifmmode \hbox{\rlap{.}\kern\sa}^{\scriptstyle\prime}\hbox{\kern-\sb}
     \else \rlap{.}\kern\sa$^{\scriptstyle\prime}$\kern-\sb\fi}
\def\ra[#1 #2 #3.#4]{#1\sup{h}#2\sup{m}#3\sup{s}\llap.#4}
\def\dec[#1 #2 #3.#4]{#1\deg#2\arcm#3\arcs\llap.#4}
\def\deco[#1 #2 #3]{#1\deg#2\arcm#3\arcs}
\def\rra[#1 #2]{#1\sup{h}#2\sup{m}}

\def\dots{\relax\ifmmode \ldots\else $\ldots$\fi}
\def\WHzsr{\ifmmode $W\,Hz\mo\,sr\mo$\else W\,Hz\mo\,sr\mo\fi}
\def\mHz{\ifmmode $\,mHz$\else \,mHz\fi}
\def\GHz{\ifmmode $\,GHz$\else \,GHz\fi}
\def\mKs{\ifmmode $\,mK\,s$^{1/2}\else \,mK\,s$^{1/2}$\fi}
\def\muKs{\ifmmode \,\mu$K\,s$^{1/2}\else \,$\mu$K\,s$^{1/2}$\fi}
\def\muKRJs{\ifmmode \,\mu$K$_{\rm RJ}$\,s$^{1/2}\else \,$\mu$K$_{\rm RJ}$\,s$^{1/2}$\fi}
\def\muKHz{\ifmmode \,\mu$K\,Hz$^{-1/2}\else \,$\mu$K\,Hz$^{-1/2}$\fi}
\def\MJysr{\ifmmode \,$MJy\,sr\mo$\else \,MJy\,sr\mo\fi}
\def\MJysrmK{\ifmmode \,$MJy\,sr\mo$\,mK$_{\rm CMB}\mo\else \,MJy\,sr\mo\,mK$_{\rm CMB}\mo$\fi}
\def\microns{\ifmmode \,\mu$m$\else \,$\mu$m\fi}

\def\muK{\ifmmode \,\mu$K$\else \,$\mu$\hbox{K}\fi}
\def\microK{\ifmmode \,\mu$K$\else \,$\mu$\hbox{K}\fi}
\def\muW{\ifmmode \,\mu$W$\else \,$\mu$\hbox{W}\fi}
\def\kms{\ifmmode $\,km\,s$^{-1}\else \,km\,s$^{-1}$\fi}
\def\kmsMpc{\ifmmode $\,\kms\,Mpc\mo$\else \,\kms\,Mpc\mo\fi}
\setsymbol{LFI:center:frequency:70GHz:units}{70.3\,GHz}
\setsymbol{LFI:center:frequency:44GHz:units}{44.1\,GHz}
\setsymbol{LFI:center:frequency:30GHz:units}{28.5\,GHz}

\setsymbol{LFI:center:frequency:70GHz}{70.3}
\setsymbol{LFI:center:frequency:44GHz}{44.1}
\setsymbol{LFI:center:frequency:30GHz}{28.5}

\setsymbol{LFI:center:frequency:LFI18:Rad:M:units}{71.7\GHz}
\setsymbol{LFI:center:frequency:LFI19:Rad:M:units}{67.5\GHz}
\setsymbol{LFI:center:frequency:LFI20:Rad:M:units}{69.2\GHz}
\setsymbol{LFI:center:frequency:LFI21:Rad:M:units}{70.4\GHz}
\setsymbol{LFI:center:frequency:LFI22:Rad:M:units}{71.5\GHz}
\setsymbol{LFI:center:frequency:LFI23:Rad:M:units}{70.8\GHz}
\setsymbol{LFI:center:frequency:LFI24:Rad:M:units}{44.4\GHz}
\setsymbol{LFI:center:frequency:LFI25:Rad:M:units}{44.0\GHz}
\setsymbol{LFI:center:frequency:LFI26:Rad:M:units}{43.9\GHz}
\setsymbol{LFI:center:frequency:LFI27:Rad:M:units}{28.3\GHz}
\setsymbol{LFI:center:frequency:LFI28:Rad:M:units}{28.8\GHz}
\setsymbol{LFI:center:frequency:LFI18:Rad:S:units}{70.1\GHz}
\setsymbol{LFI:center:frequency:LFI19:Rad:S:units}{69.6\GHz}
\setsymbol{LFI:center:frequency:LFI20:Rad:S:units}{69.5\GHz}
\setsymbol{LFI:center:frequency:LFI21:Rad:S:units}{69.5\GHz}
\setsymbol{LFI:center:frequency:LFI22:Rad:S:units}{72.8\GHz}
\setsymbol{LFI:center:frequency:LFI23:Rad:S:units}{71.3\GHz}
\setsymbol{LFI:center:frequency:LFI24:Rad:S:units}{44.1\GHz}
\setsymbol{LFI:center:frequency:LFI25:Rad:S:units}{44.1\GHz}
\setsymbol{LFI:center:frequency:LFI26:Rad:S:units}{44.1\GHz}
\setsymbol{LFI:center:frequency:LFI27:Rad:S:units}{28.5\GHz}
\setsymbol{LFI:center:frequency:LFI28:Rad:S:units}{28.2\GHz}

\setsymbol{LFI:center:frequency:LFI18:Rad:M}{71.7}
\setsymbol{LFI:center:frequency:LFI19:Rad:M}{67.5}
\setsymbol{LFI:center:frequency:LFI20:Rad:M}{69.2}
\setsymbol{LFI:center:frequency:LFI21:Rad:M}{70.4}
\setsymbol{LFI:center:frequency:LFI22:Rad:M}{71.5}
\setsymbol{LFI:center:frequency:LFI23:Rad:M}{70.8}
\setsymbol{LFI:center:frequency:LFI24:Rad:M}{44.4}
\setsymbol{LFI:center:frequency:LFI25:Rad:M}{44.0}
\setsymbol{LFI:center:frequency:LFI26:Rad:M}{43.9}
\setsymbol{LFI:center:frequency:LFI27:Rad:M}{28.3}
\setsymbol{LFI:center:frequency:LFI28:Rad:M}{28.8}
\setsymbol{LFI:center:frequency:LFI18:Rad:S}{70.1}
\setsymbol{LFI:center:frequency:LFI19:Rad:S}{69.6}
\setsymbol{LFI:center:frequency:LFI20:Rad:S}{69.5}
\setsymbol{LFI:center:frequency:LFI21:Rad:S}{69.5}
\setsymbol{LFI:center:frequency:LFI22:Rad:S}{72.8}
\setsymbol{LFI:center:frequency:LFI23:Rad:S}{71.3}
\setsymbol{LFI:center:frequency:LFI24:Rad:S}{44.1}
\setsymbol{LFI:center:frequency:LFI25:Rad:S}{44.1}
\setsymbol{LFI:center:frequency:LFI26:Rad:S}{44.1}
\setsymbol{LFI:center:frequency:LFI27:Rad:S}{28.5}
\setsymbol{LFI:center:frequency:LFI28:Rad:S}{28.2}

\setsymbol{LFI:white:noise:sensitivity:70GHz:units}{134.7\muKs}
\setsymbol{LFI:white:noise:sensitivity:44GHz:units}{164.7\muKs}
\setsymbol{LFI:white:noise:sensitivity:30GHz:units}{143.4\muKs}

\setsymbol{LFI:white:noise:sensitivity:70GHz}{134.7}
\setsymbol{LFI:white:noise:sensitivity:44GHz}{164.7}
\setsymbol{LFI:white:noise:sensitivity:30GHz}{143.4}

\setsymbol{LFI:white:noise:sensitivity:LFI18:Rad:M:units}{512.0\muKs}
\setsymbol{LFI:white:noise:sensitivity:LFI19:Rad:M:units}{581.4\muKs}
\setsymbol{LFI:white:noise:sensitivity:LFI20:Rad:M:units}{590.8\muKs}
\setsymbol{LFI:white:noise:sensitivity:LFI21:Rad:M:units}{455.2\muKs}
\setsymbol{LFI:white:noise:sensitivity:LFI22:Rad:M:units}{492.0\muKs}
\setsymbol{LFI:white:noise:sensitivity:LFI23:Rad:M:units}{507.7\muKs}
\setsymbol{LFI:white:noise:sensitivity:LFI24:Rad:M:units}{462.2\muKs}
\setsymbol{LFI:white:noise:sensitivity:LFI25:Rad:M:units}{413.6\muKs}
\setsymbol{LFI:white:noise:sensitivity:LFI26:Rad:M:units}{478.6\muKs}
\setsymbol{LFI:white:noise:sensitivity:LFI27:Rad:M:units}{277.7\muKs}
\setsymbol{LFI:white:noise:sensitivity:LFI28:Rad:M:units}{312.3\muKs}
\setsymbol{LFI:white:noise:sensitivity:LFI18:Rad:S:units}{465.7\muKs}
\setsymbol{LFI:white:noise:sensitivity:LFI19:Rad:S:units}{555.6\muKs}
\setsymbol{LFI:white:noise:sensitivity:LFI20:Rad:S:units}{623.2\muKs}
\setsymbol{LFI:white:noise:sensitivity:LFI21:Rad:S:units}{564.1\muKs}
\setsymbol{LFI:white:noise:sensitivity:LFI22:Rad:S:units}{534.4\muKs}
\setsymbol{LFI:white:noise:sensitivity:LFI23:Rad:S:units}{542.4\muKs}
\setsymbol{LFI:white:noise:sensitivity:LFI24:Rad:S:units}{399.2\muKs}
\setsymbol{LFI:white:noise:sensitivity:LFI25:Rad:S:units}{392.6\muKs}
\setsymbol{LFI:white:noise:sensitivity:LFI26:Rad:S:units}{418.6\muKs}
\setsymbol{LFI:white:noise:sensitivity:LFI27:Rad:S:units}{302.9\muKs}
\setsymbol{LFI:white:noise:sensitivity:LFI28:Rad:S:units}{285.3\muKs}

\setsymbol{LFI:white:noise:sensitivity:LFI18:Rad:M}{512.0}
\setsymbol{LFI:white:noise:sensitivity:LFI19:Rad:M}{581.4}
\setsymbol{LFI:white:noise:sensitivity:LFI20:Rad:M}{590.8}
\setsymbol{LFI:white:noise:sensitivity:LFI21:Rad:M}{455.2}
\setsymbol{LFI:white:noise:sensitivity:LFI22:Rad:M}{492.0}
\setsymbol{LFI:white:noise:sensitivity:LFI23:Rad:M}{507.7}
\setsymbol{LFI:white:noise:sensitivity:LFI24:Rad:M}{462.2}
\setsymbol{LFI:white:noise:sensitivity:LFI25:Rad:M}{413.6}
\setsymbol{LFI:white:noise:sensitivity:LFI26:Rad:M}{478.6}
\setsymbol{LFI:white:noise:sensitivity:LFI27:Rad:M}{277.7}
\setsymbol{LFI:white:noise:sensitivity:LFI28:Rad:M}{312.3}
\setsymbol{LFI:white:noise:sensitivity:LFI18:Rad:S}{465.7}
\setsymbol{LFI:white:noise:sensitivity:LFI19:Rad:S}{555.6}
\setsymbol{LFI:white:noise:sensitivity:LFI20:Rad:S}{623.2}
\setsymbol{LFI:white:noise:sensitivity:LFI21:Rad:S}{564.1}
\setsymbol{LFI:white:noise:sensitivity:LFI22:Rad:S}{534.4}
\setsymbol{LFI:white:noise:sensitivity:LFI23:Rad:S}{542.4}
\setsymbol{LFI:white:noise:sensitivity:LFI24:Rad:S}{399.2}
\setsymbol{LFI:white:noise:sensitivity:LFI25:Rad:S}{392.6}
\setsymbol{LFI:white:noise:sensitivity:LFI26:Rad:S}{418.6}
\setsymbol{LFI:white:noise:sensitivity:LFI27:Rad:S}{302.9}
\setsymbol{LFI:white:noise:sensitivity:LFI28:Rad:S}{285.3}

\setsymbol{LFI:knee:frequency:70GHz:units}{29.5\mHz}
\setsymbol{LFI:knee:frequency:44GHz:units}{56.2\mHz}
\setsymbol{LFI:knee:frequency:30GHz:units}{113.7\mHz}

\setsymbol{LFI:knee:frequency:70GHz}{29.5}
\setsymbol{LFI:knee:frequency:44GHz}{56.2}
\setsymbol{LFI:knee:frequency:30GHz}{113.7}

\setsymbol{LFI:knee:frequency:LFI18:Rad:M:units}{16.3\mHz}
\setsymbol{LFI:knee:frequency:LFI19:Rad:M:units}{15.1\mHz}
\setsymbol{LFI:knee:frequency:LFI20:Rad:M:units}{18.7\mHz}
\setsymbol{LFI:knee:frequency:LFI21:Rad:M:units}{37.2\mHz}
\setsymbol{LFI:knee:frequency:LFI22:Rad:M:units}{12.7\mHz}
\setsymbol{LFI:knee:frequency:LFI23:Rad:M:units}{34.6\mHz}
\setsymbol{LFI:knee:frequency:LFI24:Rad:M:units}{46.2\mHz}
\setsymbol{LFI:knee:frequency:LFI25:Rad:M:units}{24.9\mHz}
\setsymbol{LFI:knee:frequency:LFI26:Rad:M:units}{67.6\mHz}
\setsymbol{LFI:knee:frequency:LFI27:Rad:M:units}{187.4\mHz}
\setsymbol{LFI:knee:frequency:LFI28:Rad:M:units}{122.2\mHz}
\setsymbol{LFI:knee:frequency:LFI18:Rad:S:units}{17.7\mHz}
\setsymbol{LFI:knee:frequency:LFI19:Rad:S:units}{22.0\mHz}
\setsymbol{LFI:knee:frequency:LFI20:Rad:S:units}{8.7\mHz}
\setsymbol{LFI:knee:frequency:LFI21:Rad:S:units}{25.9\mHz}
\setsymbol{LFI:knee:frequency:LFI22:Rad:S:units}{15.8\mHz}
\setsymbol{LFI:knee:frequency:LFI23:Rad:S:units}{129.8\mHz}
\setsymbol{LFI:knee:frequency:LFI24:Rad:S:units}{100.9\mHz}
\setsymbol{LFI:knee:frequency:LFI25:Rad:S:units}{38.9\mHz}
\setsymbol{LFI:knee:frequency:LFI26:Rad:S:units}{58.9\mHz}
\setsymbol{LFI:knee:frequency:LFI27:Rad:S:units}{104.4\mHz}
\setsymbol{LFI:knee:frequency:LFI28:Rad:S:units}{40.7\mHz}

\setsymbol{LFI:knee:frequency:LFI18:Rad:M}{16.3}
\setsymbol{LFI:knee:frequency:LFI19:Rad:M}{15.1}
\setsymbol{LFI:knee:frequency:LFI20:Rad:M}{18.7}
\setsymbol{LFI:knee:frequency:LFI21:Rad:M}{37.2}
\setsymbol{LFI:knee:frequency:LFI22:Rad:M}{12.7}
\setsymbol{LFI:knee:frequency:LFI23:Rad:M}{34.6}
\setsymbol{LFI:knee:frequency:LFI24:Rad:M}{46.2}
\setsymbol{LFI:knee:frequency:LFI25:Rad:M}{24.9}
\setsymbol{LFI:knee:frequency:LFI26:Rad:M}{67.6}
\setsymbol{LFI:knee:frequency:LFI27:Rad:M}{187.4}
\setsymbol{LFI:knee:frequency:LFI28:Rad:M}{122.2}
\setsymbol{LFI:knee:frequency:LFI18:Rad:S}{17.7}
\setsymbol{LFI:knee:frequency:LFI19:Rad:S}{22.0}
\setsymbol{LFI:knee:frequency:LFI20:Rad:S}{8.7}
\setsymbol{LFI:knee:frequency:LFI21:Rad:S}{25.9}
\setsymbol{LFI:knee:frequency:LFI22:Rad:S}{15.8}
\setsymbol{LFI:knee:frequency:LFI23:Rad:S}{129.8}
\setsymbol{LFI:knee:frequency:LFI24:Rad:S}{100.9}
\setsymbol{LFI:knee:frequency:LFI25:Rad:S}{38.9}
\setsymbol{LFI:knee:frequency:LFI26:Rad:S}{58.9}
\setsymbol{LFI:knee:frequency:LFI27:Rad:S}{104.4}
\setsymbol{LFI:knee:frequency:LFI28:Rad:S}{40.7}

\setsymbol{LFI:slope:70GHz:units}{$-1.03$\mHz}
\setsymbol{LFI:slope:44GHz:units}{$-0.89$\mHz}
\setsymbol{LFI:slope:30GHz:units}{$-0.87$\mHz}

\setsymbol{LFI:slope:70GHz}{$-1.03$}
\setsymbol{LFI:slope:44GHz}{$-0.89$}
\setsymbol{LFI:slope:30GHz}{$-0.87$}

\setsymbol{LFI:slope:LFI18:Rad:M:units}{$-1.04$\mHz}
\setsymbol{LFI:slope:LFI19:Rad:M:units}{$-1.09$\mHz}
\setsymbol{LFI:slope:LFI20:Rad:M:units}{$-0.69$\mHz}
\setsymbol{LFI:slope:LFI21:Rad:M:units}{$-1.56$\mHz}
\setsymbol{LFI:slope:LFI22:Rad:M:units}{$-1.01$\mHz}
\setsymbol{LFI:slope:LFI23:Rad:M:units}{$-0.96$\mHz}
\setsymbol{LFI:slope:LFI24:Rad:M:units}{$-0.83$\mHz}
\setsymbol{LFI:slope:LFI25:Rad:M:units}{$-0.91$\mHz}
\setsymbol{LFI:slope:LFI26:Rad:M:units}{$-0.95$\mHz}
\setsymbol{LFI:slope:LFI27:Rad:M:units}{$-0.87$\mHz}
\setsymbol{LFI:slope:LFI28:Rad:M:units}{$-0.88$\mHz}
\setsymbol{LFI:slope:LFI18:Rad:S:units}{$-1.15$\mHz}
\setsymbol{LFI:slope:LFI19:Rad:S:units}{$-1.00$\mHz}
\setsymbol{LFI:slope:LFI20:Rad:S:units}{$-0.95$\mHz}
\setsymbol{LFI:slope:LFI21:Rad:S:units}{$-0.92$\mHz}
\setsymbol{LFI:slope:LFI22:Rad:S:units}{$-1.01$\mHz}
\setsymbol{LFI:slope:LFI23:Rad:S:units}{$-0.95$\mHz}
\setsymbol{LFI:slope:LFI24:Rad:S:units}{$-0.73$\mHz}
\setsymbol{LFI:slope:LFI25:Rad:S:units}{$-1.16$\mHz}
\setsymbol{LFI:slope:LFI26:Rad:S:units}{$-0.79$\mHz}
\setsymbol{LFI:slope:LFI27:Rad:S:units}{$-0.82$\mHz}
\setsymbol{LFI:slope:LFI28:Rad:S:units}{$-0.91$\mHz}

\setsymbol{LFI:slope:LFI18:Rad:M}{$-1.04$}
\setsymbol{LFI:slope:LFI19:Rad:M}{$-1.09$}
\setsymbol{LFI:slope:LFI20:Rad:M}{$-0.69$}
\setsymbol{LFI:slope:LFI21:Rad:M}{$-1.56$}
\setsymbol{LFI:slope:LFI22:Rad:M}{$-1.01$}
\setsymbol{LFI:slope:LFI23:Rad:M}{$-0.96$}
\setsymbol{LFI:slope:LFI24:Rad:M}{$-0.83$}
\setsymbol{LFI:slope:LFI25:Rad:M}{$-0.91$}
\setsymbol{LFI:slope:LFI26:Rad:M}{$-0.95$}
\setsymbol{LFI:slope:LFI27:Rad:M}{$-0.87$}
\setsymbol{LFI:slope:LFI28:Rad:M}{$-0.88$}
\setsymbol{LFI:slope:LFI18:Rad:S}{$-1.15$}
\setsymbol{LFI:slope:LFI19:Rad:S}{$-1.00$}
\setsymbol{LFI:slope:LFI20:Rad:S}{$-0.95$}
\setsymbol{LFI:slope:LFI21:Rad:S}{$-0.92$}
\setsymbol{LFI:slope:LFI22:Rad:S}{$-1.01$}
\setsymbol{LFI:slope:LFI23:Rad:S}{$-0.95$}
\setsymbol{LFI:slope:LFI24:Rad:S}{$-0.73$}
\setsymbol{LFI:slope:LFI25:Rad:S}{$-1.16$}
\setsymbol{LFI:slope:LFI26:Rad:S}{$-0.79$}
\setsymbol{LFI:slope:LFI27:Rad:S}{$-0.82$}
\setsymbol{LFI:slope:LFI28:Rad:S}{$-0.91$}

\setsymbol{LFI:FWHM:70GHz:units}{13\parcm01}
\setsymbol{LFI:FWHM:44GHz:units}{27\parcm92}
\setsymbol{LFI:FWHM:30GHz:units}{32\parcm65}

\setsymbol{LFI:FWHM:70GHz}{13.01}
\setsymbol{LFI:FWHM:44GHz}{27.92}
\setsymbol{LFI:FWHM:30GHz}{32.65}

\setsymbol{LFI:FWHM:LFI18:units}{13\parcm39}
\setsymbol{LFI:FWHM:LFI19:units}{13\parcm01}
\setsymbol{LFI:FWHM:LFI20:units}{12\parcm75}
\setsymbol{LFI:FWHM:LFI21:units}{12\parcm74}
\setsymbol{LFI:FWHM:LFI22:units}{12\parcm87}
\setsymbol{LFI:FWHM:LFI23:units}{13\parcm27}
\setsymbol{LFI:FWHM:LFI24:units}{22\parcm98}
\setsymbol{LFI:FWHM:LFI25:units}{30\parcm46}
\setsymbol{LFI:FWHM:LFI26:units}{30\parcm31}
\setsymbol{LFI:FWHM:LFI27:units}{32\parcm65}
\setsymbol{LFI:FWHM:LFI28:units}{32\parcm66}

\setsymbol{LFI:FWHM:LFI18}{13.39}
\setsymbol{LFI:FWHM:LFI19}{13.01}
\setsymbol{LFI:FWHM:LFI20}{12.75}
\setsymbol{LFI:FWHM:LFI21}{12.74}
\setsymbol{LFI:FWHM:LFI22}{12.87}
\setsymbol{LFI:FWHM:LFI23}{13.27}
\setsymbol{LFI:FWHM:LFI24}{22.98}
\setsymbol{LFI:FWHM:LFI25}{30.46}
\setsymbol{LFI:FWHM:LFI26}{30.31}
\setsymbol{LFI:FWHM:LFI27}{32.65}
\setsymbol{LFI:FWHM:LFI28}{32.66}

\setsymbol{LFI:FWHM:uncertainty:LFI18:units}{0.170\arcm}
\setsymbol{LFI:FWHM:uncertainty:LFI19:units}{0.174\arcm}
\setsymbol{LFI:FWHM:uncertainty:LFI20:units}{0.170\arcm}
\setsymbol{LFI:FWHM:uncertainty:LFI21:units}{0.156\arcm}
\setsymbol{LFI:FWHM:uncertainty:LFI22:units}{0.164\arcm}
\setsymbol{LFI:FWHM:uncertainty:LFI23:units}{0.171\arcm}
\setsymbol{LFI:FWHM:uncertainty:LFI24:units}{0.652\arcm}
\setsymbol{LFI:FWHM:uncertainty:LFI25:units}{1.075\arcm}
\setsymbol{LFI:FWHM:uncertainty:LFI26:units}{1.131\arcm}
\setsymbol{LFI:FWHM:uncertainty:LFI27:units}{1.266\arcm}
\setsymbol{LFI:FWHM:uncertainty:LFI28:units}{1.287\arcm}

\setsymbol{LFI:FWHM:uncertainty:LFI18}{0.170}
\setsymbol{LFI:FWHM:uncertainty:LFI19}{0.174}
\setsymbol{LFI:FWHM:uncertainty:LFI20}{0.170}
\setsymbol{LFI:FWHM:uncertainty:LFI21}{0.156}
\setsymbol{LFI:FWHM:uncertainty:LFI22}{0.164}
\setsymbol{LFI:FWHM:uncertainty:LFI23}{0.171}
\setsymbol{LFI:FWHM:uncertainty:LFI24}{0.652}
\setsymbol{LFI:FWHM:uncertainty:LFI25}{1.075}
\setsymbol{LFI:FWHM:uncertainty:LFI26}{1.131}
\setsymbol{LFI:FWHM:uncertainty:LFI27}{1.266}
\setsymbol{LFI:FWHM:uncertainty:LFI28}{1.287}

\setsymbol{HFI:center:frequency:100GHz:units}{100\,GHz}
\setsymbol{HFI:center:frequency:143GHz:units}{143\,GHz}
\setsymbol{HFI:center:frequency:217GHz:units}{217\,GHz}
\setsymbol{HFI:center:frequency:353GHz:units}{353\,GHz}
\setsymbol{HFI:center:frequency:545GHz:units}{545\,GHz}
\setsymbol{HFI:center:frequency:857GHz:units}{857\,GHz}

\setsymbol{HFI:center:frequency:100GHz}{100}
\setsymbol{HFI:center:frequency:143GHz}{143}
\setsymbol{HFI:center:frequency:217GHz}{217}
\setsymbol{HFI:center:frequency:353GHz}{353}
\setsymbol{HFI:center:frequency:545GHz}{545}
\setsymbol{HFI:center:frequency:857GHz}{857}

\setsymbol{HFI:Ndetectors:100GHz}{8}
\setsymbol{HFI:Ndetectors:143GHz}{11}
\setsymbol{HFI:Ndetectors:217GHz}{12}
\setsymbol{HFI:Ndetectors:353GHz}{12}
\setsymbol{HFI:Ndetectors:545GHz}{3}
\setsymbol{HFI:Ndetectors:857GHz}{4}

\setsymbol{HFI:FWHM:Maps:100GHz:units}{9\parcm88}
\setsymbol{HFI:FWHM:Maps:143GHz:units}{7\parcm18}
\setsymbol{HFI:FWHM:Maps:217GHz:units}{4\parcm87}
\setsymbol{HFI:FWHM:Maps:353GHz:units}{4\parcm65}
\setsymbol{HFI:FWHM:Maps:545GHz:units}{4\parcm72}
\setsymbol{HFI:FWHM:Maps:857GHz:units}{4\parcm39}
\setsymbol{HFI:FWHM:Maps:100GHz}{9.88}
\setsymbol{HFI:FWHM:Maps:143GHz}{7.18}
\setsymbol{HFI:FWHM:Maps:217GHz}{4.87}
\setsymbol{HFI:FWHM:Maps:353GHz}{4.65}
\setsymbol{HFI:FWHM:Maps:545GHz}{4.72}
\setsymbol{HFI:FWHM:Maps:857GHz}{4.39}

\setsymbol{HFI:beam:ellipticity:Maps:100GHz}{1.15}
\setsymbol{HFI:beam:ellipticity:Maps:143GHz}{1.01}
\setsymbol{HFI:beam:ellipticity:Maps:217GHz}{1.06}
\setsymbol{HFI:beam:ellipticity:Maps:353GHz}{1.05}
\setsymbol{HFI:beam:ellipticity:Maps:545GHz}{1.14}
\setsymbol{HFI:beam:ellipticity:Maps:857GHz}{1.19}

\setsymbol{HFI:FWHM:Mars:100GHz:units}{9\parcm37}
\setsymbol{HFI:FWHM:Mars:143GHz:units}{7\parcm04}
\setsymbol{HFI:FWHM:Mars:217GHz:units}{4\parcm68}
\setsymbol{HFI:FWHM:Mars:353GHz:units}{4\parcm43}
\setsymbol{HFI:FWHM:Mars:545GHz:units}{3\parcm80}
\setsymbol{HFI:FWHM:Mars:857GHz:units}{3\parcm67}

\setsymbol{HFI:FWHM:Mars:100GHz}{9.37}
\setsymbol{HFI:FWHM:Mars:143GHz}{7.04}
\setsymbol{HFI:FWHM:Mars:217GHz}{4.68}
\setsymbol{HFI:FWHM:Mars:353GHz}{4.43}
\setsymbol{HFI:FWHM:Mars:545GHz}{3.80}
\setsymbol{HFI:FWHM:Mars:857GHz}{3.67}

\setsymbol{HFI:beam:ellipticity:Mars:100GHz}{1.18}
\setsymbol{HFI:beam:ellipticity:Mars:143GHz}{1.03}
\setsymbol{HFI:beam:ellipticity:Mars:217GHz}{1.14}
\setsymbol{HFI:beam:ellipticity:Mars:353GHz}{1.09}
\setsymbol{HFI:beam:ellipticity:Mars:545GHz}{1.25}
\setsymbol{HFI:beam:ellipticity:Mars:857GHz}{1.03}

\setsymbol{HFI:CMB:relative:calibration:100GHz}{$\lsim 1\%$}
\setsymbol{HFI:CMB:relative:calibration:143GHz}{$\lsim 1\%$}
\setsymbol{HFI:CMB:relative:calibration:217GHz}{$\lsim 1\%$}
\setsymbol{HFI:CMB:relative:calibration:353GHz}{$\lsim 1\%$}
\setsymbol{HFI:CMB:relative:calibration:545GHz}{}
\setsymbol{HFI:CMB:relative:calibration:857GHz}{}

\setsymbol{HFI:CMB:absolute:calibration:100GHz}{$\lsim 2\%$}
\setsymbol{HFI:CMB:absolute:calibration:143GHz}{$\lsim 2\%$}
\setsymbol{HFI:CMB:absolute:calibration:217GHz}{$\lsim 2\%$}
\setsymbol{HFI:CMB:absolute:calibration:353GHz}{$\lsim 2\%$}
\setsymbol{HFI:CMB:absolute:calibration:545GHz}{}
\setsymbol{HFI:CMB:absolute:calibration:857GHz}{}

\setsymbol{HFI:FIRAS:gain:calibration:accuracy:statistical:100GHz}{}
\setsymbol{HFI:FIRAS:gain:calibration:accuracy:statistical:143GHz}{}
\setsymbol{HFI:FIRAS:gain:calibration:accuracy:statistical:217GHz}{}
\setsymbol{HFI:FIRAS:gain:calibration:accuracy:statistical:353GHz}{2.5\%}
\setsymbol{HFI:FIRAS:gain:calibration:accuracy:statistical:545GHz}{1\%}
\setsymbol{HFI:FIRAS:gain:calibration:accuracy:statistical:857GHz}{0.5\%}

\setsymbol{HFI:FIRAS:gain:calibration:accuracy:systematic:100GHz}{}
\setsymbol{HFI:FIRAS:gain:calibration:accuracy:systematic:143GHz}{}
\setsymbol{HFI:FIRAS:gain:calibration:accuracy:systematic:217GHz}{}
\setsymbol{HFI:FIRAS:gain:calibration:accuracy:systematic:353GHz}{}
\setsymbol{HFI:FIRAS:gain:calibration:accuracy:systematic:545GHz}{7\%}
\setsymbol{HFI:FIRAS:gain:calibration:accuracy:systematic:857GHz}{7\%}

\setsymbol{HFI:FIRAS:zero:point:accuracy:100GHz:units}{0.8\MJysr}
\setsymbol{HFI:FIRAS:zero:point:accuracy:143GHz:units}{}
\setsymbol{HFI:FIRAS:zero:point:accuracy:217GHz:units}{}
\setsymbol{HFI:FIRAS:zero:point:accuracy:353GHz:units}{1.4\MJysr}
\setsymbol{HFI:FIRAS:zero:point:accuracy:545GHz:units}{2.2\MJysr}
\setsymbol{HFI:FIRAS:zero:point:accuracy:857GHz:units}{1.7\MJysr}

\setsymbol{HFI:FIRAS:zero:point:accuracy:100GHz}{0.8}
\setsymbol{HFI:FIRAS:zero:point:accuracy:143GHz}{}
\setsymbol{HFI:FIRAS:zero:point:accuracy:217GHz}{}
\setsymbol{HFI:FIRAS:zero:point:accuracy:353GHz}{1.4}
\setsymbol{HFI:FIRAS:zero:point:accuracy:545GHz}{2.2}
\setsymbol{HFI:FIRAS:zero:point:accuracy:857GHz}{1.7}

\setsymbol{HFI:unit:conversion:100GHz:units}{0.2415\MJysrmK}
\setsymbol{HFI:unit:conversion:143GHz:units}{0.3694\MJysrmK}
\setsymbol{HFI:unit:conversion:217GHz:units}{0.4811\MJysrmK}
\setsymbol{HFI:unit:conversion:353GHz:units}{0.2883\MJysrmK}
\setsymbol{HFI:unit:conversion:545GHz:units}{0.05826\MJysrmK}
\setsymbol{HFI:unit:conversion:857GHz:units}{0.002238\MJysrmK}

\setsymbol{HFI:unit:conversion:100GHz}{0.2415}
\setsymbol{HFI:unit:conversion:143GHz}{0.3694}
\setsymbol{HFI:unit:conversion:217GHz}{0.4811}
\setsymbol{HFI:unit:conversion:353GHz}{0.2883}
\setsymbol{HFI:unit:conversion:545GHz}{0.05826}
\setsymbol{HFI:unit:conversion:857GHz}{0.002238}

\setsymbol{HFI:colour:correction:alpha=-2:V1.01:100GHz}{0.9893}
\setsymbol{HFI:colour:correction:alpha=-2:V1.01:143GHz}{0.9759}
\setsymbol{HFI:colour:correction:alpha=-2:V1.01:217GHz}{1.0007}
\setsymbol{HFI:colour:correction:alpha=-2:V1.01:353GHz}{1.0028}
\setsymbol{HFI:colour:correction:alpha=-2:V1.01:545GHz}{1.0019}
\setsymbol{HFI:colour:correction:alpha=-2:V1.01:857GHz}{0.9889}

\setsymbol{HFI:colour:correction:alpha=0:V1.01:100GHz}{1.0008}
\setsymbol{HFI:colour:correction:alpha=0:V1.01:143GHz}{1.0148}
\setsymbol{HFI:colour:correction:alpha=0:V1.01:217GHz}{0.9909}
\setsymbol{HFI:colour:correction:alpha=0:V1.01:353GHz}{0.9888}
\setsymbol{HFI:colour:correction:alpha=0:V1.01:545GHz}{0.9878}
\setsymbol{HFI:colour:correction:alpha=0:V1.01:857GHz}{1.0014}